\documentclass[aps,prd,10pt,preprintnumbers,showpacs,twocolumn,superscriptaddress,nofootinbib,amsmath,amssymb]{revtex4-1}
\usepackage{cases}
\usepackage{graphicx}
\usepackage{xcolor}
\usepackage[colorlinks]{hyperref}
\newcommand{\epow}[1]{e^{#1}}
\newcommand{\oder}[2]{\frac{\partial #1}{\partial #2}}
\newcommand{\dd}{\mathrm{d}}

\begin{document}

\title{Dynamical instability of polytropic spheres in spacetimes with a cosmological constant}

\author{Camilo Posada}\email{camilo.posada@physics.slu.cz}
\author{Jan Hlad\'ik}\email{jan.hladik@physics.slu.cz}
\author{Zden\v{e}k Stuchl\'ik}\email{zdenek.stuchlik@physics.slu.cz}
\affiliation{Institute of Physics and Research Centre of Theoretical Physics and Astrophysics, Silesian University in Opava, Bezru\v{c}ovo n\'{a}m. 13, CZ-746 01 Opava, Czech Republic}

\begin{abstract}
The dynamical instability of relativistic polytropic spheres, embedded in a spacetime with a repulsive cosmological constant, is studied in the framework of general relativity. We apply the methods used in our preceding paper to study the trapping polytropic spheres with $\Lambda = 0$, namely, the critical point method and the infinitesimal and adiabatic radial perturbations method developed by Chandrasekhar. We compute numerically the critical adiabatic index, as a function of the parameter $\sigma = p_{\mathrm{c}}/(\rho_{\mathrm{c}} c^2)$, for several values of the cosmological parameter $\lambda$ giving the ratio of the vacuum energy density to the central energy density of the polytrope. We also determine the critical values for the parameter $\sigma_{\mathrm{cr}}$, for the onset of instability, by using both approaches. We found that for large values of the parameter $\lambda$, the differences between the values  of $\sigma_{\mathrm{cr}}$ calculated by the critical point method differ from those obtained via the radial perturbations method.  Our results, given by both applied methods, indicate that large values of the cosmological parameter $\lambda$ have relevant effects on the dynamical stability of the polytropic configurations.
\end{abstract}

\maketitle

\section{Introduction}\label{intro}
There are indications that contrary to the inflationary era, in the recent era the dark energy could correspond to the vacuum energy related to a repulsive cosmological constant $\Lambda > 0$ implying important consequences in astrophysical phenomena~\citep{Stuchlik:2005dv,Stuchlik:2020rls}. The estimate of the so-called relic cosmological constant, governing the acceleration of the recent stage of the Universe expansion, reads $\Lambda \sim 10^{-52}~\mathrm{m}^{-2}$~\citep{Ade:2015xua}. Although the relic cosmological constant is extremely small, its role in astrophysical phenomena could be quite significant, being limited by the so-called static radius introduced in~\cite{Stuchlik:1983BAC} and discussed in~\cite{Stuchlik:1984BAC,Stuchlik:1999qk,Stuchlik:2004dt,Far:2016:Uni,Stuchlik:2018qyz}. The static radius represents an upper limit on the existence of both Keplerian~\citep{Stuchlik:2004wk} and toroidal fluids~\citep{Stuchlik:2000AA,Slany:2005vd,Stuchlik:2008ea,Stuchlik:2009jv}, accretion disks, the limit on gravitationally bound galaxy systems~\citep{Stuchlik:2011zz,Stuchlik:2012,Schee:2013wqa}, and even the limit on gravitationally bound polytropic configurations that could represent a model of dark matter halos~\citep{Stuchlik:2016xiq,Novotny:2017cep,Stuchlik:2017qiz}. Test fields around black holes in spacetimes with $\Lambda > 0$ were treated in~\cite{Konoplya:2011qq,Toshmatov:2017qrq}, indicating possible instabilities.

In spacetimes with a cosmological constant $\Lambda$, the interior Schwarzschild solution with uniform distribution of energy density was found in~\cite{Stuchlik:2000xe} for starlike configurations and extended to more general situations in~\cite{Boehmer:2003iv,Boehmer:2003uz}. The effect of $\Lambda$ on gravitational instabilities for isothermal spheres in the Newtonian limit was considered in~\cite{Axenides:2013hrq}. The role of the cosmological constant on the so-called electrovacuum solutions was studied in~\cite{Posada:2013eqa}. 

In spacetimes with a positive cosmological constant, the polytropic spheres represented by a polytropic index $n$, a parameter $\sigma$, giving the ratio of pressure and energy density at the center, and vacuum constant index $\lambda$, giving the ratio of the vacuum energy to the central energy density, were exhaustively discussed in~\cite{Stuchlik:2016xiq,Novotny:2017cep}. It has been shown that in some special cases of polytropes with sufficiently large values of the polytropic index $n$, extremely extended configurations representing a dark matter halo could have gravitationally unstable central parts that could collapse leading to the formation of a supermassive black hole~\citep{Stuchlik:2017qiz}.

The polytropic spheres are well-known models of compact objects, as they represent extremely dense nuclear matter inside neutron or quark stars. For example, they describe the fluid configurations constituted from relativistic ($n = 3$) and nonrelativistic ($n = 3/2$) Fermi gas \citep{Shapiro:1983du}, considered as basic approximations of neutron stars matter. Of course, in realistic models describing the neutron stars interior, the equations of state (EOSs) of nuclear matter are considered. However, in a recently developed approach, such realistic EOSs are represented by sequences of polytropes with appropriately tuned parameters~\citep{Alvarez-Castillo:2017qki}.

The stability of the polytropic spheres can be addressed from two different approaches. One of them is related to the energetic considerations, or critical point method \citep{Zeldovich:1971,Shapiro:1983du}, developed by Tooper~\cite{Tooper:1964}. A second approach deals with the dynamical theory of infinitesimal, and adiabatic, radial oscillations pioneered in a seminal paper by Chandrasekhar~\cite{Chandrasekhar:1964zz}. Using his ‘pulsation equation', Chandrasekhar established the conditions of stability, against radial oscillations, for homogenous stars and polytropic spheres. The main conclusion of this study is that the critical adiabatic index $\gamma_{\mathrm{cr}}$, for the onset of instability, increases due to relativistic effects from the Newtonian value $\gamma = 4/3$. The radial oscillations method has been widely used in different contexts~\citep{Gleiser:1988ih,Moustakidis:2016ndw,Posada:2018goy}. The role of the cosmological constant on the radial stability of the uniform energy density stars, using Chandrasekhar's method, was studied in~\cite{Stuchlik:2005rag,Boehmer:2005kk}. These authors concluded that a large value of the vacuum constant index $\lambda$ increases significantly the critical adiabatic index from its value with $\lambda=0$.

The purpose of this paper is to study in detail the role of the cosmological constant in the stability of the polytropic fluid spheres against radial oscillations, that is expected to be relevant for extremely extended noncompact configurations modeling galactic dark matter halos~\citep{Stuchlik:2016xiq}. For that purpose, we are reconsidering the analysis carried out in~\cite{Stuchlik:2005rag} in two ways: first, we will apply the methods introduced in our preceding paper~\citep{Hladik:2020xfw} to study the stability of polytropic spheres with $\Lambda = 0$, namely, the shooting method and trial functions to solve the Sturm--Liouville equation; and the  critical point method based on the energetic approach. Second, we will extend the analysis to a larger family of spheres in the range of polytropic indexes $0.5 \leq \, n\, \leq 3$, for several values of the vacuum constant index $\lambda \in [10^{-9}, 10^{-1}]$.

The paper is organized as follows. In Sec.~\ref{sect:2}, we review the equations of structure of the relativistic polytropes in the presence of a cosmological constant. In Sec.~\ref{sect:3}, we summarize the general properties and gravitational energy for polytropic spheres with $\Lambda$.  In Sec.~\ref{sect:4}, Chandrasekhar's procedure and the associated Sturm--Liouville eigenvalue problem, including the cosmological term, are outlined. In Sec.~\ref{sect:5}, we present our methods and results. In Sect.~\ref{sect:6}, we discuss our conclusions.


\section{Structure equations of relativistic polytropic spheres with a cosmological constant}\label{sect:2}
We will consider throughout the paper perturbations which preserve spherical symmetry. This condition guarantees that motions along the radial direction will ensue. Thus, our starting point is a spherically symmetric spacetime in the standard Schwarzschild coordinates
\begin{equation}\label{spher_metric}
    \dd s^2 = -\epow{2\Phi}(c\,\dd t)^2 + \epow{2\Psi}\dd r^2+r^2(\dd\theta^2+\sin^2 \theta\, \dd\phi^2)\, ,
\end{equation}
where $\Phi(r,t)$ and $\Psi(r,t)$ are functions of $t$ and the radial coordinate $r$. The energy-momentum tensor for a spherically symmetric configuration takes the form
\begin{equation}\label{em_tensor}
    T_{\mu}^{\hphantom{\nu}\nu} = (\epsilon + p) u_{\mu}u^{\nu} + p\delta_{\mu}^{\hphantom{\mu}\nu}\, ,
\end{equation}
where $\epsilon=\rho c^2$ is the energy density (written as the product of the mass density $\rho$ times the speed of light squared), $p$ is the fluid pressure, and $u^{\mu} = \dd x^{\mu}/\dd\tau$ is its four-velocity.

Following our preceding paper~\citep{Hladik:2020xfw}, we consider the models of static polytropic fluid spheres proposed by Tooper~\cite{Tooper:1964}, which are governed by the EOS
\begin{equation} \label{polytropic}
    p = K\rho^{1 + (1/n)}\, ,
\end{equation}
where $n$ is the polytropic index and $K$ is a constant related to the characteristics of a specific configuration. It is conventional to introduce the parameter
\begin{equation}\label{sigma}
    \sigma\equiv\frac{p_{\mathrm{c}}}{\rho_{\mathrm{c}} c^2}\, ,
\end{equation}
which denotes the ratio of pressure to energy density at the centre of the configuration. The radial profiles of the mass density and pressure of the polytropic spheres are given by the relations
\begin{equation}\label{polyeos}
    \rho = \rho_{\mathrm{c}}\theta^n\, ,\qquad p = p_{\mathrm{c}} \theta^{n+1}\, ,
\end{equation}
where $\theta(x)$ is function of the dimensionless radius,
\begin{equation}\label{polyx}
    x \equiv \frac{r}{L}\, ,\qquad L \equiv \left[\frac{\sigma(n + 1)c^2}{4\pi G\rho_{\mathrm{c}}}\right]^{\frac{1}{2}}\, .
\end{equation}
Here $L$ is the characteristic length scale of the polytropic sphere and it is determined by the polytropic index $n$, the parameter $\sigma$, and the central density $\rho_\mathrm{c}$. From Eq.~\eqref{polyeos}, we obtain immediately the boundary condition $\theta(r = 0) = 1$.

We consider a static configuration in equilibrium, immersed in a cosmological background. The relevant components of Einstein's equations for this problem are $(t)(t)$ and $(r)(r)$, which in the presence of a cosmological constant $\Lambda$ are given by~\citep{Tolman:1934}
\begin{equation}\label{GttLambda}
    \frac{\dd}{\dd r}(r \epow{-2\Psi}) = 1 - \frac{8\pi G}{c^4} T_{0}^{0}r^2 - \Lambda r^2\, ,
\end{equation}
\begin{equation}\label{GrrLambda}
    \frac{2\dd\Phi}{\dd r} = \frac{\epow{2\Psi} - 1}{r} - \frac{8\pi G}{c^4} T_{1}^{1}r - \Lambda r\, .
\end{equation}
Equation~\eqref{GttLambda} can be recast into the form
\begin{equation}\label{GttLambda2}
    \epow{2\Psi} = \left[1 - \frac{2Gm(r)}{c^2r} - \frac{1}{3}\Lambda r^2\right]^{-1}\, ,
\end{equation}
where
\begin{equation}\label{MisnerMass}
    m(r) = 4\pi\int_{0}^{r}\rho(r)r^2\, \dd r
\end{equation}
is the mass inside the radius $r$.  Using Eqs.~\eqref{GttLambda2} and~\eqref{MisnerMass}, we transform Eq.~\eqref{GrrLambda} into
\begin{equation}\label{GrrLambda2}
    \frac{\dd\Phi}{\dd r} = \frac{(G/c^2)m(r) - \frac{\Lambda}{3} r^2 + (4\pi G/c^4)pr^3}{r^2\left[1 - \frac{2Gm(r)}{c^2r} - \frac{\Lambda}{3}r^2\right]}\, .
\end{equation}
Using the energy-momentum ‘conservation' $T^{\mu\nu}_{\hphantom{\mu}\hphantom{\mu};\nu} = 0$, we can write Eq.~\eqref{GrrLambda2} as a relation between pressure and energy density in the form~\citep{Boehmer:2005kk,Stuchlik:2000xe}
\begin{equation}\label{TOVLambda}
    \frac{\dd p}{\dd r} = -(\epsilon + p)\frac{(G/c^2)m(r) + \left[(4\pi G/c^4)p - \frac{\Lambda}{3}\right]r^3}{r^2\left[1 - \frac{2Gm(r)}{c^2r} - \frac{\Lambda}{3}r^2\right]}\, ,
\end{equation}
which is the Tolman-Oppenheimer-Volkoff (TOV) equation, including the effect of the cosmological constant.

Once an EOS $p = p(\rho)$ is given, Eqs.~\eqref{MisnerMass} and~\eqref{TOVLambda} can be integrated subject to the boundary conditions $m(0) = 0$ and $p(R) = 0$, where $R$ is the radius of the star. In the exterior of the configuration, Eq.~\eqref{MisnerMass} gives the total mass $M = m(R)$, and the spacetime is described by the Kottler metric~\citep{Kottler:1918AnP}. The mass relation Eq.~\eqref{MisnerMass} can be written in terms of the parameter $\sigma$ as follows:
\begin{equation}\label{MisnerMass2}
    \sigma(n + 1)\,\dd\theta = -(\sigma\theta + 1)\,\dd\Phi\, .
\end{equation}
Solving Eq.~\eqref{MisnerMass2} with the condition that the interior and exterior metrics are smoothly matched at the surface $r = R$, we have
\begin{equation}\label{gtt_interior}
    \epow{2\Phi} = (1 + \sigma\theta)^{-2(n + 1)}\left(1 - \frac{2GM}{c^2R} - \frac{\Lambda}{3}R^2\right)\, ,
\end{equation}
which is a function of $\theta$ and $\sigma$. In order to find a relation for the function $\theta$, we rewrite Eq.~\eqref{MisnerMass2} in the form
\begin{equation}\label{dPhidr}
    \frac{\dd\Phi}{\dd r} = -\frac{\sigma(n + 1)}{1 + \sigma\theta}\frac{\dd\theta}{\dd r}\, .
\end{equation}
Substituting Eq.~\eqref{dPhidr} into Eq.~\eqref{GrrLambda} we have
\begin{multline}\label{GrrLambda3}
    \frac{\sigma(n + 1)r}{1 + \sigma\theta}\left[1 - \frac{2Gm(r)}{c^2r} - \frac{\Lambda}{3}r^2\right]\left(\frac{\dd\theta}{\dd r}\right)\\
        + \frac{Gm(r)}{c^2r} - \frac{\Lambda}{3}r^2 = -\frac{G\sigma\theta}{c^2}\left(\frac{\dd m}{\dd r}\right)\, .
\end{multline}
Similarly, the mass relation Eq.~\eqref{MisnerMass} can be written in terms of $\theta$ as follows:
\begin{equation}\label{MisnerMass3}
    \frac{\dd m}{\dd r} = 4\pi r^2\rho_\mathrm{c}\,\theta^{n}\, .
\end{equation}
To facilitate the numerical computations, it is convenient to write Eq.~\eqref{GrrLambda2} in a dimensionless form by using Eq.~\eqref{polyx} and the quantities
\begin{gather}\label{vx}
    v(x) \equiv \frac{m(r)}{4\pi L^3 \rho_\mathrm{c}} = \frac{m(r)}{\mathcal{M}}\, ,\\
    \lambda \equiv \frac{\rho_{\mathrm{vac}}}{\rho_\mathrm{c}}\, ,\label{lambda}
\end{gather}
where $\mathcal{M}$ is a characteristic mass scale of the polytrope
\begin{equation}\label{mass_scale}
    \mathcal{M} = 4\pi L^3 \rho_\mathrm{c} = \frac{c^2}{G}\sigma L(n + 1)\, ,
\end{equation}
and $\lambda$ indicates the vacuum constant index giving the ratio between the vacuum energy density and the central energy density of the polytropic sphere. The cosmological constant $\Lambda$ is related to the energy density of the vacuum by
\begin{equation}\label{vacuum}
    \Lambda = \frac{8\pi G}{c^2}\rho_{\mathrm{vac}}\, .
\end{equation}
Thus, the index $\lambda$ in Eq.~\eqref{lambda} is connected with $\Lambda$ through the relation
\begin{equation}\label{vacuum2}
    \lambda = \frac{\Lambda c^4}{8\pi G\rho_\mathrm{c}}\, .
\end{equation}
In terms of Eqs.~\eqref{polyx}, \eqref{vx}, and \eqref{mass_scale}, Eq. \eqref{GrrLambda2} takes the final form
\begin{equation}\label{dthetadx}
    \frac{\dd\theta}{\dd x} = \left[\left(\frac{2\lambda}{3} - \sigma\theta^{n + 1}\right)x - \frac{v}{x^2}\right]\left(1 + \sigma\theta\right)g_{rr}\, ,
\end{equation}
\begin{equation}\label{dvdx}
    \frac{\dd v}{\dd x} = x^2 \theta^n\, ,
\end{equation}
where
\begin{equation}\label{grr}
    g_{rr}\equiv\left[1 - 2\sigma(n + 1)\left(\frac{v}{x} + \frac{\lambda}{3}x^2\right)\right]^{-1}\, .
\end{equation}
These equations, subject to the boundary conditions
\begin{equation}
    \theta(0) = 1\, ,\quad v(0)=0\,
\end{equation}
can be solved numerically to give the radius $x=x_1$ of the configuration as the first solution $\theta(x) = 0$.


\section{General properties of the solutions}\label{sect:3}

\subsection{Structural parameters}
Except for the case $n = 0$, corresponding to a cons\-tant density configuration~\citep{Stuchlik:2000xe, Boehmer:2005kk}, the structure equations~\eqref{dthetadx} and \eqref{dvdx} do not admit analytic solutions in a closed form for $\sigma \neq 0$. Thus, one must turn to numerical integration. Considering that a configuration in equilibrium has positive density and monotonically decreasing pressure (see however \citep{Posada:2018goy}), we will concentrate in the range of values of $x$ such that $\theta>0$.

Assuming that $\lambda$, $n$, $\sigma$, and $\rho_\mathrm{c}$ are given, we start the numerical integration at the center of the sphere $x = 0$, where $\theta(0) = 1$ and $v(0) = 1$, and advance by small steps until the first zero $\theta(x_1) = 0$ is found at $x_1$, if it exists. Using this value in Eq.~\eqref{polyx}, we can determine the radius of the polytrope as
\begin{equation}\label{radius}
    R = L x_1\, .
\end{equation}
The mass of the configuration is determined by the solution of $v(x)$ at the surface
\begin{equation}\label{Mass}
    M = 4\pi L^3 \rho_{\mathrm{c}} v(x_1) = \frac{c^2 }{G}\sigma L(n + 1)v(x_1)\, .
\end{equation}
From Eqs.~\eqref{radius} and \eqref{Mass}, we can obtain the mass-radius relation
\begin{equation}\label{mass_ratio}
    \mathcal{C} \equiv \frac{GM}{c^2R} = \frac{\sigma (n+1)v(x_1)}{x_1}\, ,
\end{equation}
which gives the ratio between the gravitational radius $r_{g} \equiv 2GM/c^2$ and the coordinate radius $R$, once $\sigma$ has been specified. The $g_{tt}$ and $g_{rr}$ metric components can be written in terms of the function $\theta$ and the parameter $\sigma$ as
\begin{equation}\label{gtt_theta}
    \epow{2\Phi} = \frac{1 - 2\sigma(n + 1)\left[\frac{v(x_1)}{x_1}+\frac{\lambda}{3}x_1^2\right]}{(1 + \sigma\theta)^{2(n+1)}}\, ,
\end{equation}
\begin{equation}\label{grr_theta}
    \epow{-2\Psi} = 1 - 2\sigma(n + 1)\left[\frac{v(x)}{x} + \frac{\lambda}{3}x^2\right]\, .
\end{equation}
The exterior region is described by the Kottler metric, or Schwarzschild--de Sitter spacetime, which represents a Schwarzschild mass embedded into an asymptotically de Sitter spacetime. In the Schwarzschild coordinates, the exterior metric takes the form~\citep{Tolman:1934}
\begin{equation}\label{kottler}
    \epow{2\Phi} = \epow{-2\Psi} = 1 - \frac{2GM}{c^2 r} - \frac{\Lambda}{3}r^2\, .
\end{equation}

\subsection{Gravitational energy }
In the relativistic theory, the total energy $E$ of certain fluid sphere, which includes the internal energy and the gravitational potential energy, is $Mc^2$ where $M$ corresponds to the mass producing the gravitational field
\begin{equation}
    E =  Mc^2 = 4\pi \int_{0}^{R}\epsilon r^2\, \dd r\, .
\end{equation}
The proper energy and proper mass of a spherical \emph{gas} is defined by
\begin{equation}\label{M0g}
    E_{0\mathrm{g}} =  M_{0g}c^2 = 4\pi \int_{0}^{R}(\rho_{g}c^2) \epow{\Psi}\, r^2\,\dd r\, ,
\end{equation}
where $M_{0\mathrm{g}}$ equals (approximately) the rest mass density of baryons in the configuration and $\rho_\mathrm{g}c^2$ is the rest energy density of the gas particles. For polytropic spheres, the gas density can be written in terms of the total mass density as \citep{Tooper:1964}
\begin{equation}\label{gas}
    \rho_{g}=  \frac{\rho_\mathrm{c}\theta^n}{(1+\sigma\theta)^n}.
\end{equation}
\noindent In our analysis of stability using the energy considerations, an important quantity is the ratio
\begin{multline}\label{ratio}
    \frac{E_{0\mathrm{g}}}{E} = \frac{1}{v(x_{1})}\\
    \times\int\limits_{0}^{x_{1}}\frac{\theta^{n}x^2}{\left(1 + \sigma\theta\right)^{n}\left[1 - 2\sigma(n + 1)\left(\frac{v}{x} + \frac{\lambda}{3}x^2\right)\right]^{1/2}}\,\dd x\,,
\end{multline}
\noindent which gives the proper energy of the gas in units of the total energy $E=Mc^2$. Note that Eq.~\eqref{ratio} generalizes the expression given in \cite{Tooper:1964} to the case of a nonvanishing cosmological constant.

These two quantities define the binding energy of the system, namely, $E_{\mathrm{b}} = E_{0\mathrm{g}} - E$, which corresponds to the difference in energy between an initial state with zero internal energy where the particles that compose the system are dispersed, and a final state where the particles are bounded by gravitational interaction.


\section{Radial oscillations of relativistic spheres in spacetimes with a cosmological constant}\label{sect:4}
In this section we discuss the theory of infinitesimal, and adiabatic, radial oscillations of relativistic spheres developed by Chandrasekhar~\cite{Chandrasekhar:1964zz}, and its extension to the case of a nonvanishing cosmological constant~\citep{Stuchlik:2005rag,Boehmer:2005kk}.

We consider ‘pulsations' which preserve spherical symmetry; therefore, these do not affect the exterior gravitational field. In other words, there is no gravitational mo\-no\-po\-le radiation.  Thus,  we are considering a situation with the line element, Eq.~\eqref{spher_metric}, and a mass distribution described by the energy-momentum tensor, Eq.~\eqref{em_tensor}

The pulsation dynamics will be determined by the Einstein equations, including the cosmological constant, together with the energy-momentum conservation, ba\-ryon number conservation, and the laws of thermodynamics. The relevant components of the Einstein equations with $\Lambda$ are given in~\citep{Tolman:1934,Boehmer:2005kk}.

To obtain the radial pulsation equation for spherical fluids immersed into a cosmological background, the metric coefficients $\Psi(r,t)$ and $\Phi(r,t)$ together with the fluid variables $\rho(r,t)$, $p(r,t)$ and the number density of baryons $n(r,t)$, as measured in fluid's rest frame, are perturbed generally in the form
\begin{equation}
    q(r,t) = q_0(r) + \delta q(r,t)\, ,
\end{equation}
where the canonical variable $q \equiv (\Phi,\Psi,\epsilon,p,n)$ indicates the metric and physical quantities, and the subscript $0$ refers to the variables at equilibrium.  At first order in the perturbations, the components of the energy-momentum tensor Eq.~\eqref{em_tensor} are given by
\begin{align}
    T_{0}^{0} &= -\epsilon_0\, ,\\
    T_{i}^{i} &= p,\quad\quad i = 1,\, 2,\, 3 \quad(\text{no summation})\, , \\
    T_{0}^{1} &= -(\epsilon_{0} + p_{0})v\, , \\
    T_{1}^{0} &= (\epsilon_{0} + p_{0})v \epow{2(\Psi_0-\Phi_0)}\, ,
\end{align}
where $v = dr/dx^{0}$. The pulsation is represented by the radial, or ‘Lagrangian', displacement $\xi$ of the fluid from the equilibrium position $\xi = \xi(r,t)$, defined as
\begin{equation}
    \frac{u^r}{u^t} = \frac{\partial\xi}{\partial t} \equiv \dot{\xi}\, .
\end{equation}
The derivation of the expressions for the linear perturbations in the quantities $q(r,t)$ follows the same lines as for the case $\Lambda = 0$ discussed in~\cite{Chandrasekhar:1964zz,Misner:1974qy}; therefore, we just summarize the main results here. All the relevant equations must be linearized relative to the displacement from the static equilibrium configuration. We have to obtain the dynamic equation for evolution of the fluid displacement $\xi(t,r)$, and a set of initial-value equations, expressing the perturbation functions $\delta\Phi$, $\delta\Psi$, $\delta\epsilon$, $\delta p$, $\delta n$ in terms of the displacement function $\xi(t,r)$.

No nuclear reactions are assumed during small radial perturbations; therefore, the dynamics of the energy density and the pressure perturbations is governed by the baryon conservation law
\begin{equation}\label{nbaryon}
    \left(nu^{\mu}\right)_{;\mu} = 0\, .
\end{equation}
Using Eq.~\eqref{nbaryon}, we obtain the initial value equation for the pressure perturbation
\begin{equation}
    \delta p = - \gamma p_0 \frac{\epow{\Phi_0}}{r^2}\left(r^2 \epow{-\Phi_0}\xi\right)' - \xi(p_0)'\, ,
\end{equation}
where $'=\partial/\partial r$, and we introduce
\begin{equation}\label{gamma}
    \gamma \equiv \left(p\,\frac{\partial n}{\partial p}\right)^{-1} \left[n-(\epsilon+p)\frac{\partial n}{\partial\epsilon}\right]\, ,
\end{equation}
as the adiabatic index that governs the linear perturbations of pressure inside the star~\citep{Chandrasekhar:1964zz,Shapiro:1983du}. In general, this $\gamma$ is not necessarily the same as the adiabatic index associated to the EOS (see discussion in~\citep{Hladik:2020xfw}).

The initial-value equation for the Lagrangian perturbation of the energy density $\delta\rho$, and the metric functions $\delta\Phi$ and $\delta\Psi$, takes a similar form as for the case $\Lambda = 0$,
\begin{equation}
    \delta\epsilon = -\frac{\epow{\Phi_0}}{r^2}(\epsilon_0 + p_0)\left(r^2\epow{-\Phi_0}\xi\right)' - \xi(\epsilon_0)'\, ,
\end{equation}
\begin{equation}
    \delta\Psi = - \xi\left(\Psi_0 + \Phi_0 \right)'\, ,
\end{equation}
\begin{equation}
    (\delta\Phi)' = \left[\frac{\delta p}{(\epsilon_0+p_0)} - \left(\Phi_0' + \frac{1}{r}\right)\xi\right]\left(\Psi_0 + \Phi_0 \right)'\, .
\end{equation}
It is conventional to assume that all the perturbations have a time dependence of the form $\epow{i\omega t}$, where $\omega$ is a characteristic frequency to be determined.  Thus, using the previous initial-value equations for the perturbations, and introducing the ‘renormalized displacement function' $\zeta$~\citep{Misner:1974qy},
\begin{equation}
    \zeta\equiv r^2 \epow{-\Phi_0}\xi\, ,
\end{equation}
we obtain the Sturm-Liouville dynamic pulsation equation with a cosmological constant
\begin{equation}\label{eSL}
    \frac{\dd}{\dd r}\left(P\frac{\dd\zeta}{\dd r}\right) + \left(Q+\omega^2 W\right)\zeta=0\, ,
\end{equation}
where the functions $P(r)$, $Q(r)$, and $W(r)$ are defined as
\begin{equation}\label{Pr}
    P(r)\equiv  \frac{\gamma p_0}{r^2}\,\epow{3\Phi_0+\Psi_0}\, ,
\end{equation}
\begin{multline}\label{Qr}
    Q(r)\equiv \frac{\epow{3\Phi_0+\Psi_0}}{r^2}\Bigg[\frac{(p_0')^2}{\epsilon_0 + p_0} - \frac{4p'_0}{r}\\
        -\left(\frac{8\pi G}{c^4}p_0 - \Lambda\right)(\epsilon_0 + p_0) \epow{2\Psi_0}\Bigg]\, ,
\end{multline}
\begin{equation}\label{Wr}
    W(r)\equiv\frac{\epsilon_0 + p_0}{r^2}\,\epow{\Phi_0+3\Psi_0}\, .
\end{equation}
The boundary conditions must guarantee that the displacement function is not resulting in a divergent behavior of the energy density and pressure perturbations at the center of the sphere. On the other hand, the variations of the pressure must satisfy the condition $p(R) = 0$ at the surface of the configuration. Therefore, we have
\begin{align}
    & \frac{\zeta}{r^3}&\text{is finite, or zero, as} & \quad r \rightarrow 0\, , \label{bc1}\\
    & \left(\frac{\gamma p_0\epow\Phi}{r^2}\right)\zeta' \rightarrow 0  & \text{as} & \quad r\rightarrow R\, . \label{bc2}
\end{align}
The Sturm-Liouville equation~\eqref{eSL}, together with the boundary conditions Eqs.~\eqref{bc1} and \eqref{bc2}, determine the eigenvalues $\omega_i$ (frequencies) and the pulsation eigenfunctions $\zeta_i(r)$ which satisfy
\begin{equation}\label{orthog}
    \int_{0}^{R}\epow{\Phi_0+3\Psi_0}(\epsilon_0 + p_0)\zeta_{i}\zeta_{j}r^2\,\dd r = 0, \quad i\neq j\, .
\end{equation}
The Sturm-Liouville eigenvalue problem can be written in the variational form, as described in~\cite{Misner:1974qy}, because the extremal values of
\begin{equation}\label{e36}
    \omega^2 = \frac{\displaystyle\int_0^R\left(P\zeta'^2 - Q\zeta^2\right)\,\dd r}{\displaystyle\int_0^R W\zeta^2\,\dd r}
\end{equation}
determine the eigenfrequencies $\omega_i$. The absolute minimum value of Eq.~\eqref{e36} corresponds to the squared frequency of the fundamental mode of the radial pulsations. If $\omega^2$ is positive (negative), the configuration is stable (unstable) against radial oscillations. Moreover, if the fundamental mode is stable ($\omega_0^2 > 0$), all higher radial modes will also be stable. For this reason, a sufficient condition for the dynamical instability is the vanishing of the right-hand side of Eq.~\eqref{e36} for certain trial function satisfying the boundary conditions.

The Sturm-Liouville pulsation equation can be used to determine the dynamical stability of spherical configurations of perfect fluid. Given certain EoS, the critical adiabatic index $\gamma_{\mathrm{cr}}$, given by the marginally stable condition $\omega^2 = 0$, can be determined by integration of the Sturm-Liouville equation. Using Eq.~\eqref{e36}, we can deduce a general formula to find the critical adiabatic index, which reads
\begin{equation}\label{gammacr}
    \gamma_{\mathrm{cr}} =\frac{\displaystyle\int_{0}^{R}Q(r)\zeta^2\,\dd r}{\displaystyle\int_{0}^{R}\frac{p_0}{r^2}\epow{3\Phi_0+\Psi_0}(\zeta')^2\,\dd r}\, .
\end{equation}
Thus, for $\gamma < \gamma_{\mathrm{cr}}$ dynamical instability will ensue and the configuration will collapse. For the case of a homogeneous star in the presence of a cosmological constant, B\"ohmer and Harko~\cite{Boehmer:2005kk} showed that the condition for radial stability reads
\begin{equation}\label{gcBH}
  \gamma > \gamma_{\mathrm{cr}} =\frac{\frac{4}{3}-l}{1-3l} + \frac{19}{42}\left(1-\frac{21}{19}l\right)\left(\frac{r_g}{R}\right)+\mathcal{O}\left(\frac{r_g}{R}\right)^2\, ,
\end{equation}
where $l = \Lambda/(12\pi G\rho_\mathrm{c})$, $r_{g}$ is the gravitational radius, and $R$ indicates the radius of the star. When $\Lambda = 0$, Eq.~\eqref{gcBH} reduces to the value found by Chandrasekhar~\cite{Chandrasekhar:1964zz}.

\subsection{Sturm-Liouville equation for polytropic spheres with a cosmological constant}\label{sec:3.3}
Using the relevant expressions discussed in Sec.~\ref{sect:2}, together with the variational form Eq.~\eqref{e36}, we arrive to the Sturm-Liouville eigenvalue equation for the dynamical stability of relativistic polytropic spheres in the presence of a cosmological constant,
\begin{multline}\label{pulsationx}
    \omega^2L^2\int_0^{x_1}\theta^n(1 + \sigma\theta)\left(\frac{\zeta}{x}\right)^2\epow{\Phi+3\Psi}\,\dd x = \\
        \quad \sigma\int_0^{x_1}\frac{\gamma\,\theta^{n + 1}}{x^2}\left(\oder{\zeta}{x}\right)^2\epow{3\Phi+\Psi}\,\dd x -\\
    (n + 1)\int_0^{x_1}\frac{\theta^n \epow{3\Phi+\Psi}}{x^2}\biggl\{\left(\oder{\theta}{x}\right)\frac{4}{x}\left[\frac{\sigma(n + 1)x}{4(1 + \sigma\theta)}\left(\oder{\theta}{x}\right) - 1\right] \\
        -2(1 + \sigma\theta)\left(\sigma\theta^{n + 1} - \lambda\right)\epow{2\Psi}\biggr\} \zeta^2\,\dd x\, ,
\end{multline}
which constitutes a characteristic eigenvalue problem for the frequency $\omega^2$ and the amplitude $\zeta(x)$ (we have suppressed the subscript zero as no longer needed). For the polytropic spheres considered in Sec.~\ref{sect:2}, the adiabatic index $\gamma$ is given by
\begin{equation}\label{gamma_poly}
    \gamma = \left(1 + \frac{1}{n}\right)(1 + \sigma\theta)\, ,
\end{equation}
which, in general, is a function of the radial coordinate. In his study of the dynamical stability of relativistic polytropes, Chandrasekhar~\cite{Chandrasekhar:1964zz} assumed $\gamma$ to be a constant. Thus, under this assumption, $\gamma$ can be taken out of the integral in Eq.~\eqref{pulsationx} and one can integrate given certain trial function. In a more general approach, for any equilibrium configuration, one can consider $\gamma$ in Eq.~\eqref{pulsationx} as an effective adiabatic index~\citep{Merafina:1989},
\begin{equation}\label{effective}
    \langle \gamma \rangle = \frac{\displaystyle \int\limits_0^{x_1}\,\frac{\gamma\,\theta^{n + 1}}{x^2}\left(\oder{\zeta}{x}\right)^2 \epow{3\Phi+\Psi}\,\dd x}{\displaystyle \int\limits_0^{x_1}\frac{\theta^{n+1}}{x^2}\left(\oder{\zeta}{x}\right)^2 \epow{3\Phi+\Psi}\,\dd x}\, .
\end{equation}
Thus, the condition for stability can be established as
\begin{equation}
    \langle \gamma \rangle > \gamma_\mathrm{cr}\, .
\end{equation}
The mass relation Eq.~\eqref{MisnerMass2} for the gradients of $p$ and $\Phi$ is transferred into the form
\begin{equation}
    \oder{\Phi}{x} = -\frac{\sigma(n + 1)}{1 + \sigma\theta}\oder{\theta}{x}\, .
\end{equation}
In terms of the variables introduced in Eqs.~\eqref{sigma}--\eqref{polyx}, the Sturm-Liouville equation~\eqref{eSL} takes the form
\begin{equation}\label{eSLx}
    \frac{\dd}{\dd x}\left[P(x)\frac{\dd\zeta}{\dd x}\right] + L^2\left[Q(x) + \omega^2\,W(x)\right]\zeta(x)=0\, ,
\end{equation}
where the functions $P$, $Q$, and $W$ given by Eqs.~\eqref{Pr}, \eqref{Qr}, and \eqref{Wr} are now
\begin{equation}\label{Px}
    P(x) = \frac{\langle\gamma\rangle\sigma\rho_\mathrm{c}\theta^{n + 1}}{L^2 x^2}\epow{3\Phi+\Psi}\, ,
\end{equation}
\begin{multline}\label{Qx}
    Q(x) = \frac{\sigma\rho_\mathrm{c}(n + 1)\theta^{n}\epow{3\Phi+\Psi}}{L^4 x^2}\left[\frac{\sigma(n + 1)}{(1 + \sigma\theta)}\left(\frac{\dd\theta}{\dd x}\right)^2 - \right.\\
        \left. \frac{4}{x}\left(\frac{\dd\theta}{\dd x}\right) - 2(1 + \sigma\theta)\left(\sigma\theta^{n + 1} - \lambda\right)\epow{2\Psi}\right]\, ,
\end{multline}
\begin{equation}\label{Wx}
    W(x) = \frac{\rho_\mathrm{c}\,\theta^{n}(1+\sigma\theta)}{L^2 x^2}\epow{\Phi+3\Psi}\, .
\end{equation}
In the next section, we will discuss the methods we used to solve the eigenvalue problem Eq.~\eqref{eSLx}, subject to the boundary conditions Eqs.~\eqref{bc1} and \eqref{bc2}, in order to study the radial stability of polytropic spheres in the presence of a cosmological constant.


\section{Radial stability of relativistic polytropes in the presence a cosmological constant}\label{sect:5}

\subsection{Numerical methods}\label{sect:5A}
Clearly, for general polytropes, the critical value of the adiabatic index related to the dynamical stability can be determined by numerical integration only. Several methods to solve the eigenvalue problem Eq.~\eqref{pulsationx} have been described in the literature (see, e.g.,~\cite{Bardeen:1966} and references therein). Following~\cite{Chandrasekhar:1964zz}, we computed the critical values of the adiabatic index $\gamma_\mathrm{cr}$, for the onset of instability, by integrating numerically Eq.~\eqref{eSLx} in the case $\omega^2 = 0$. For that purpose we followed two different methods: the shooting method and trial functions.

In the shooting method \citep{Press:1992zz}, one integrates Eq.~\eqref{eSLx} from the center up to the surface of the configuration with some trial value of $\gamma$. The value for which the solution satisfies (within a prescribed error) the boundary conditions, Eqs.~\eqref{bc1} and \eqref{bc2}, correspond to the critical adiabatic index $\gamma_\mathrm{cr}$.

In order to apply the shooting method to Eq.~\eqref{eSLx}, it is convenient to transform it to a set of two ordinary differential equations. We follow the convention used in \cite{Kokkotas:2000up} where Eq.~\eqref{eSLx} can be split in the following form:
\begin{gather}
    \frac{\dd \zeta}{\dd x} = \frac{\eta}{P(x)}\, , \label{ode1}\\
    \frac{\dd \eta}{\dd x} = -L^2\,\left[\omega^2\,W(x) + Q(x)\right]\zeta\, , \label{ode2}
\end{gather}
which satisfy the following behaviour near the origin:
\begin{gather}
    \zeta(x) = \frac{\eta_{0}}{3P(0)}x^3 + \mathcal{O}(x^5)\, , \label{origin1}\\
    \eta(r) = \eta_{0}\, , \label{origin2}
\end{gather}
where $\eta_{0}$ is an arbitrary constant, which we choose to be the unity.

The second method is based on using trial functions to integrate Eq.~\eqref{pulsationx}. Following~\cite{Chandrasekhar:1964zz,Boehmer:2005kk}, we chose the following functions
\begin{equation}\label{Chandratrial}
    \xi_1 = x \epow{\Phi/2}\, ,\qquad \xi_2 = x\, ,
\end{equation}
yielding
\begin{equation}
    \zeta_1 = x^3\epow{-\Phi/2}\, ,\qquad \zeta_2 = x^3 \epow{-\Phi}\, .
\end{equation}
We perform a detailed study of the stability for the whole range of the polytropes subject to the condition of causality due to the restriction on the parameter $\sigma$~\citep{Tooper:1964},
\begin{equation}\label{causal}
    \sigma < \sigma_\mathrm{causal} \equiv \frac{n}{n + 1}\, .
\end{equation}
However, we will see that for certain combinations of the parameters $(n,\lambda)$ the range of $\sigma$ is also limited by the condition of having finite size configurations. The limits on the existence of relativistic polytropic spheres were discussed in~\cite{Stuchlik:2016xiq}, in dependence of the polytropic index $n$ and the parameter $\sigma$. Condition~\eqref{causal} is obtained from the relation
\begin{equation}\label{vsc}
    v_\mathrm{sc} = c \left[\frac{\sigma(n + 1)}{n}\right]^{1/2}\, ,
\end{equation}
which corresponds to the speed of sound at the center of the sphere. Thus, it might seem that Eq.~\eqref{vsc} implies the restriction given by Eq.~\eqref{causal}. However, note that Eq.~\eqref{vsc} gives the phase velocity which is not the same as the group velocity; therefore, the condition~\eqref{causal} might not be definitive.

Applying the methods described above, we have computed critical values of the adiabatic index $\gamma_{\mathrm{cr}}$ for polytropes with characteristic values of the index $n$, for several values of the cosmological parameter $\lambda$. Using these results for $\gamma_\mathrm{cr}$, we also determined constraints on the parameter $\sigma$ in order to construct stable configurations. We present our results in the next section.

\subsection{Results obtained via Chandrasekhar's pulsation equation}\label{sect:5B}

As a first step in our analysis, we solved numerically the equations of structure Eqs.~\eqref{dthetadx} and \eqref{dvdx} for relativistic polytropes in the presence of a cosmological constant. The integrations were carried out using the adaptive Runge-Kutta-Fehlberg method~\citep{Press:1992zz}. We provide some profiles of the dimensionless radius $x_{1}$, as a function of $\sigma$, for several values of the vacuum constant index $\lambda$. We studied a whole family of polytropic spheres with index $n \leq\, 3$, and we restrict the values of $\sigma$ by the causality limit Eq.~\eqref{causal}. For comparison, we have included in the same plot (dashed lines) the profiles with $\lambda = 0$.

\begin{figure*}[ht]
    \centering \includegraphics[width=0.4\linewidth]{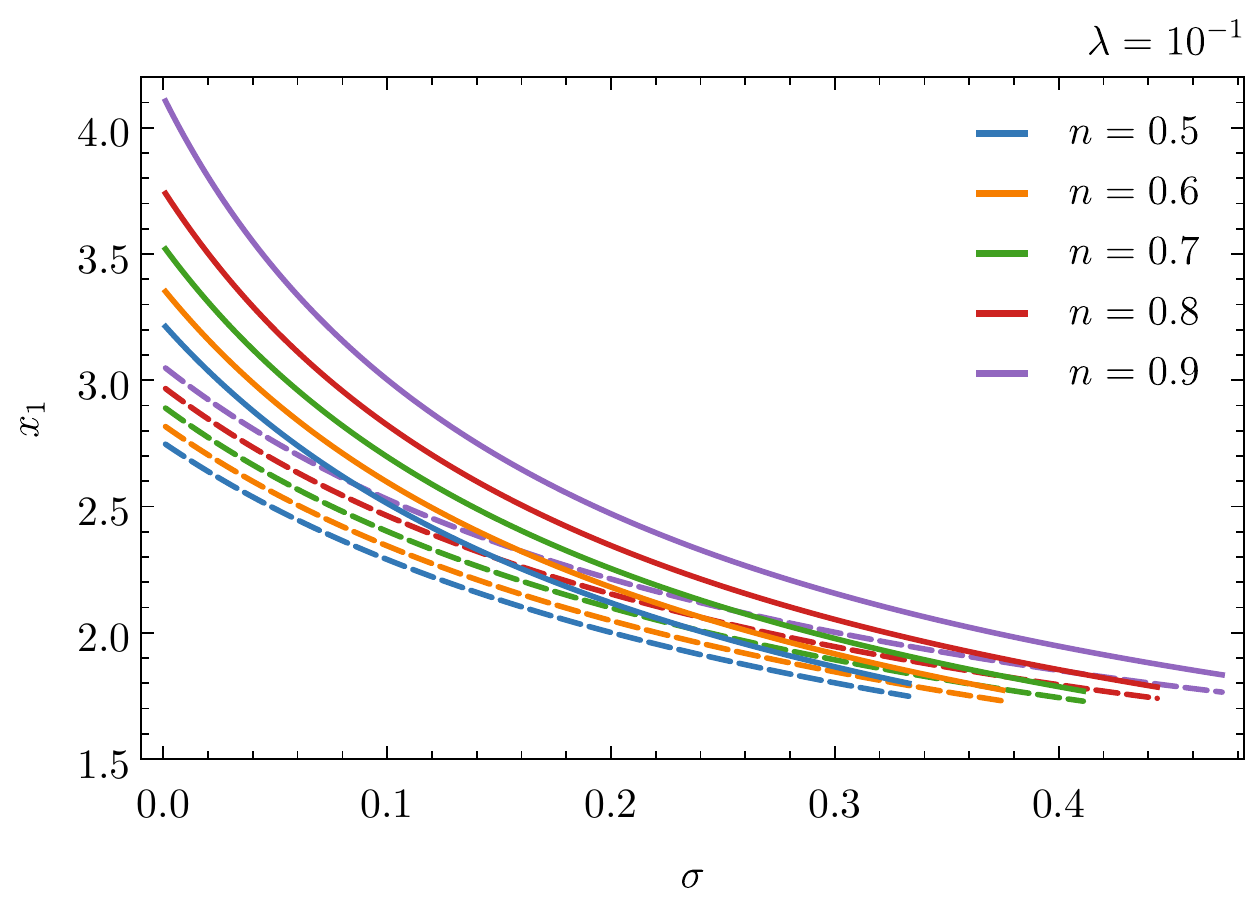} \quad \includegraphics[width=0.4\linewidth]{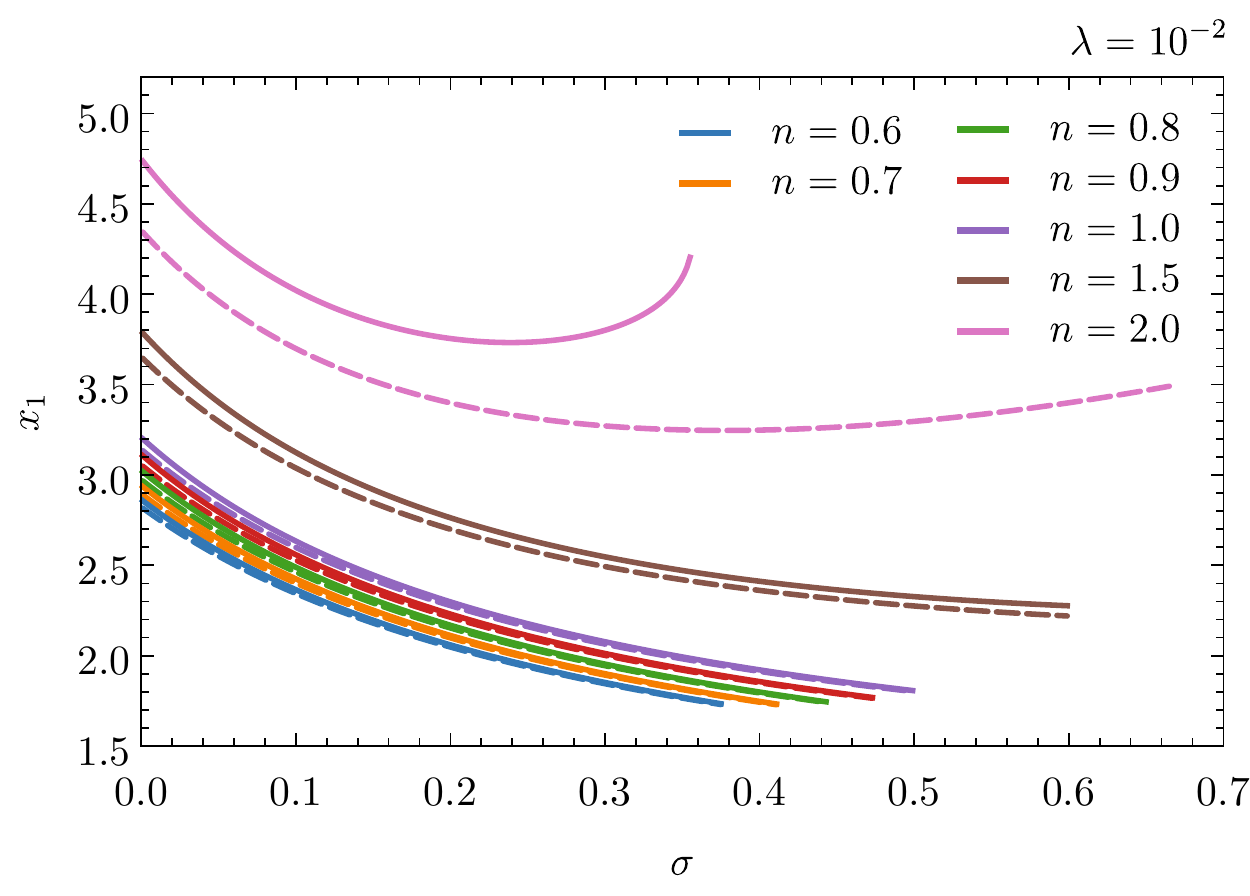}\\
    \centering \includegraphics[width=0.4\linewidth]{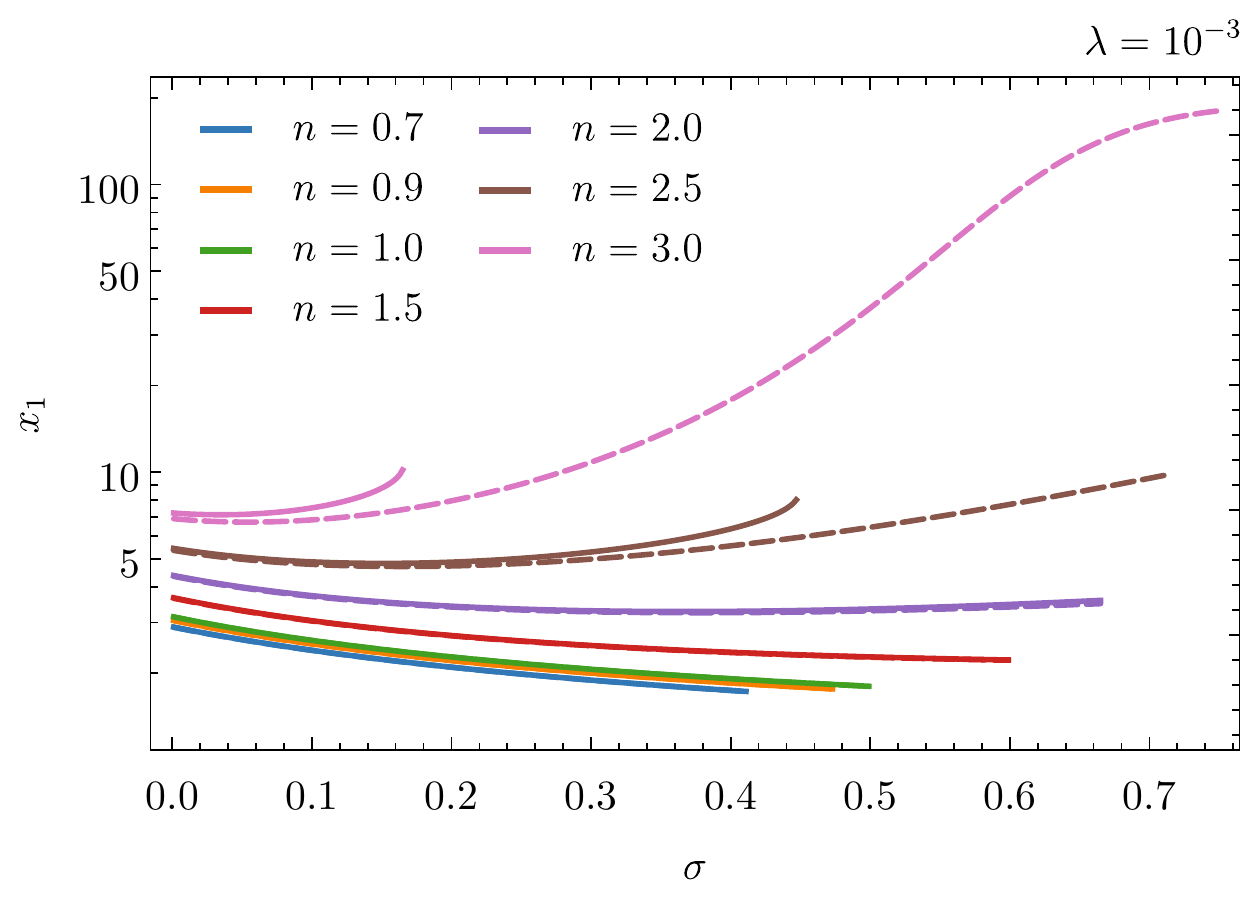} \quad \includegraphics[width=0.4\linewidth]{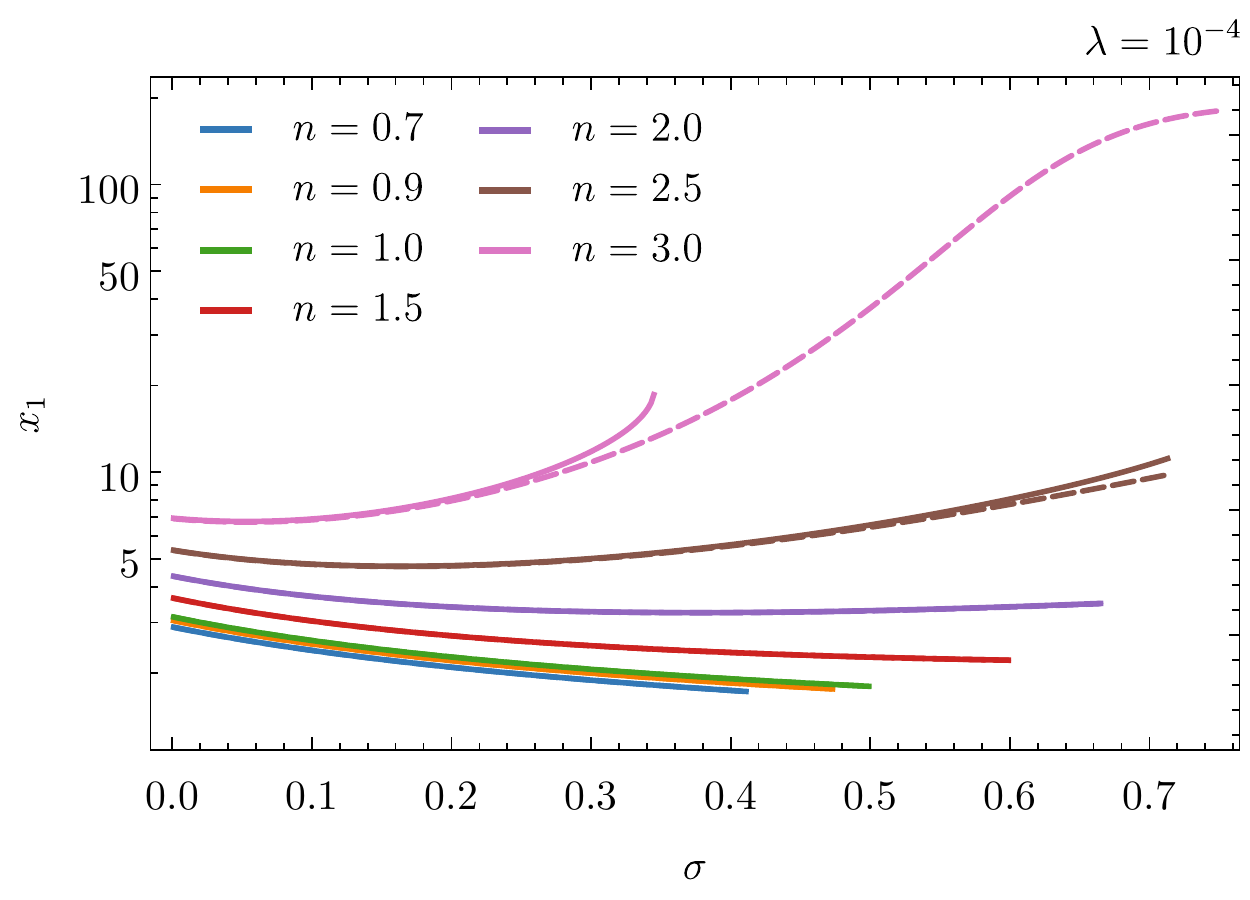}\\
    \centering \includegraphics[width=0.4\linewidth]{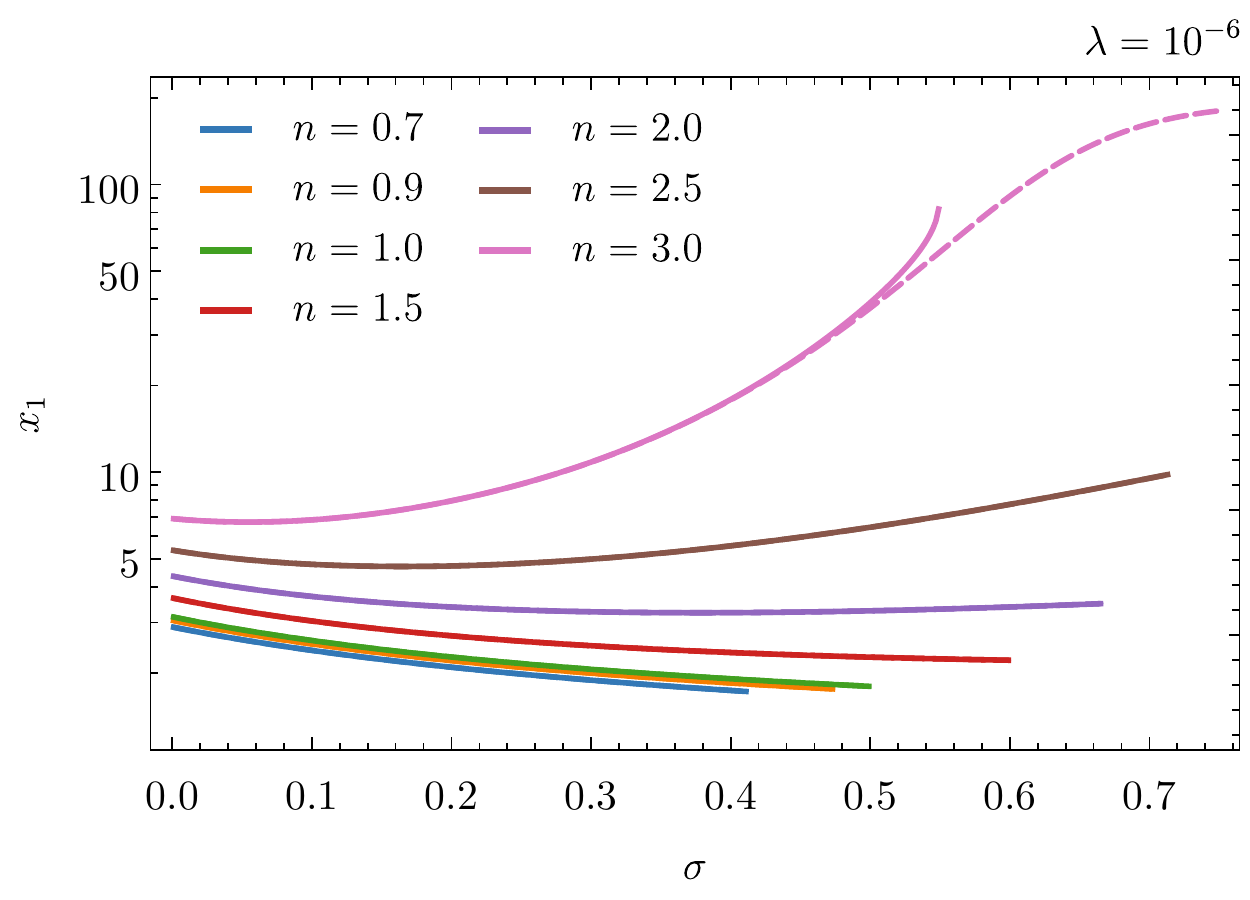} \quad \includegraphics[width=0.4\linewidth]{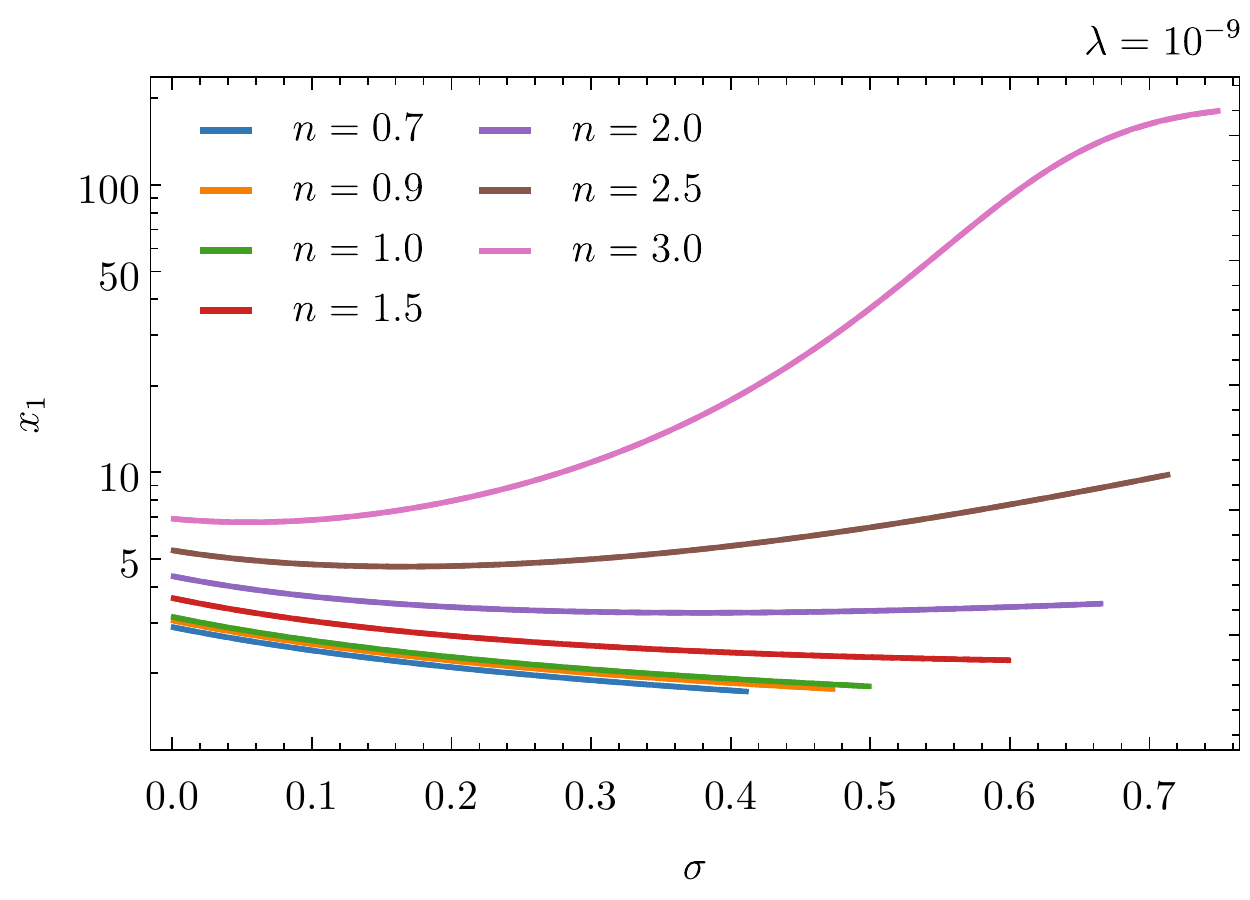}
    \caption{\label{fig1} Profiles of the dimensionless radius $x_{1}$ for relativistic polytropic spheres in dependence of the parameters: $n\in[0.5, 3]$, $\sigma\in[0, n/(n + 1)]$, and $\lambda\in[10^{-9}$, $10^{-1}]$. The dashed lines indicate the corresponding polytropic configuration (in same color) for $\lambda = 0$. Note that for certain combinations of the indexes $(n,\lambda\neq 0)$, the structure equations do not yield finite configurations for $\sigma\in[0,n/(n + 1)]$.}
\end{figure*}

In Fig.~\ref{fig1}, we present our results which are in very good agreement with those reported in~\cite{Stuchlik:2016xiq}. Note that for certain combinations of the parameters $(n,\sigma)$, the extension of the configuration increases as compared to its corresponding value in the case $\lambda = 0$. This is expected as a consequence of the repulsive effect of a positive cosmological constant. Moreover,  the presence of $\lambda$ sets strong constraints on the existence of polytropic configurations. For instance, for the case $\lambda = 10^{-2}$, polytropic configurations with $n > 2.4$ do not exist. For the case $\lambda = 10^{-1}$, existing polytropes are restricted to $n < 1$.

We also found that the vacuum constant index $\lambda$ constraints the allowed values of the parameter $\sigma$ for certain configurations. For instance, for the combination $(n = 2,\,\lambda = 10^{-2})$, the maximum allowed value of the parameter $\sigma$ we found was $\sigma_{\mathrm{max}} \simeq 0.3563$. Note that this value is lower than the value restricted by causality given by Eq.~\eqref{causal}. In Fig.~\ref{fig1}, we show the results for various combinations of the parameters $(n,\sigma)$. For the case of $\lambda = 10^{-9}$, deviations from the $\lambda = 0$ case are negligible.

\begin{figure*}[ht]
    \centering \includegraphics[width=0.44\linewidth]{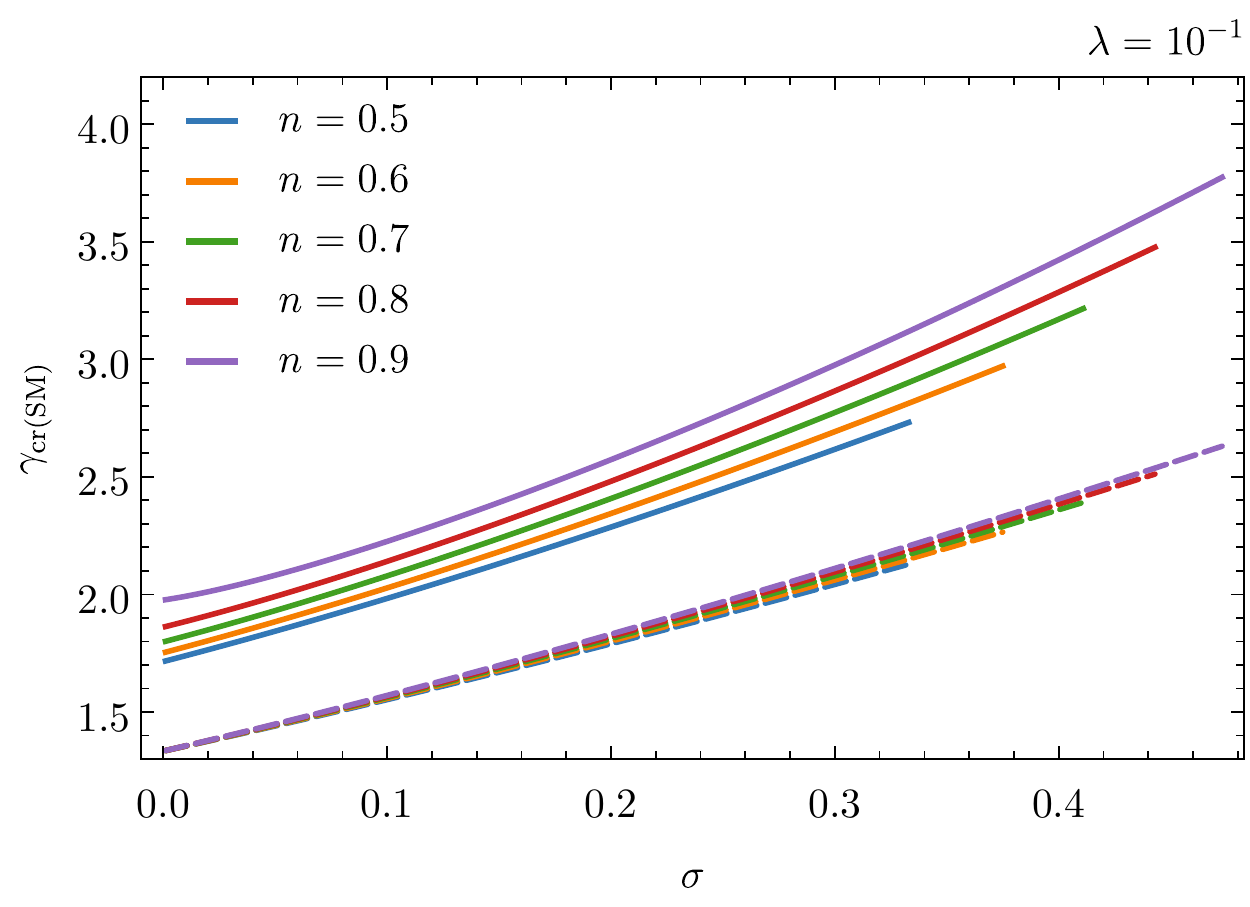} \quad \includegraphics[width=0.44\linewidth]{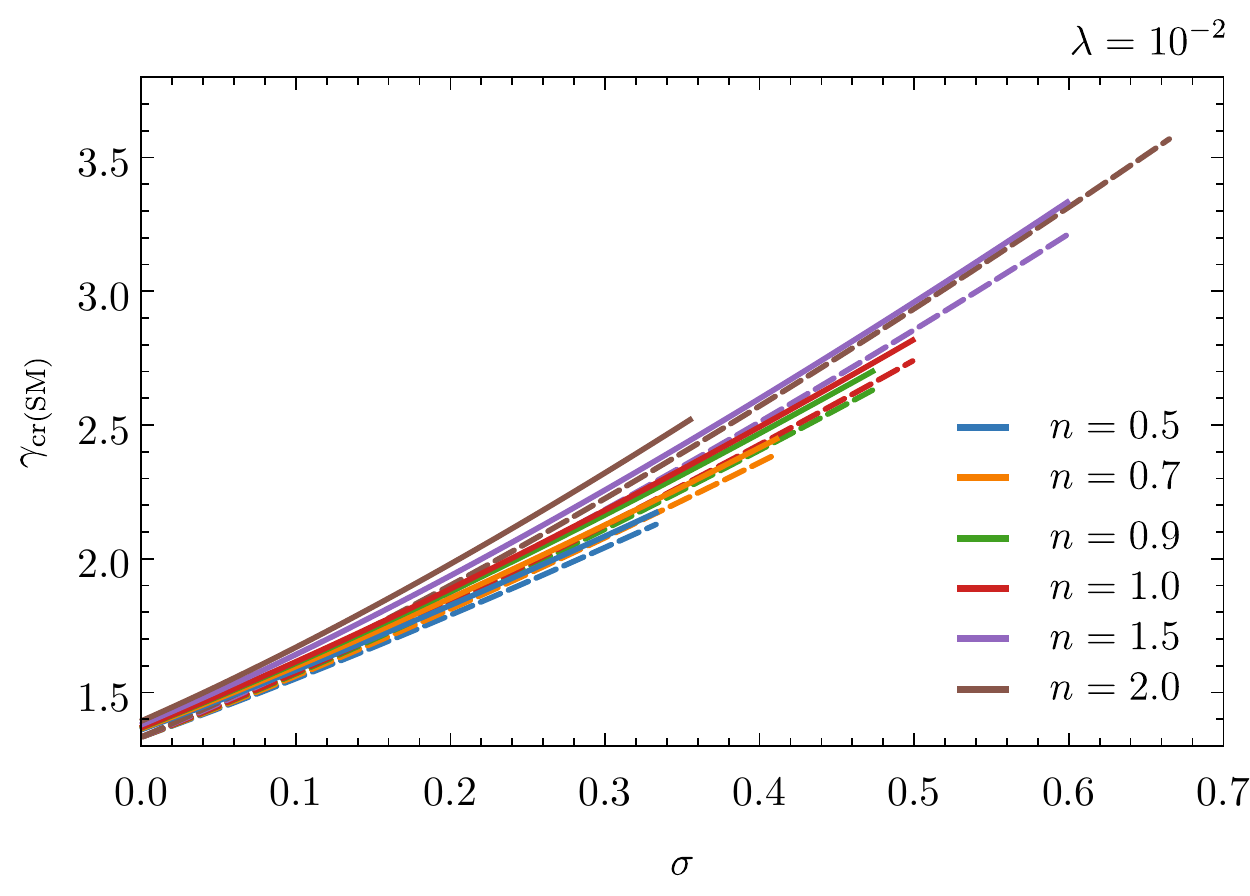}\\
    \centering \includegraphics[width=0.44\linewidth]{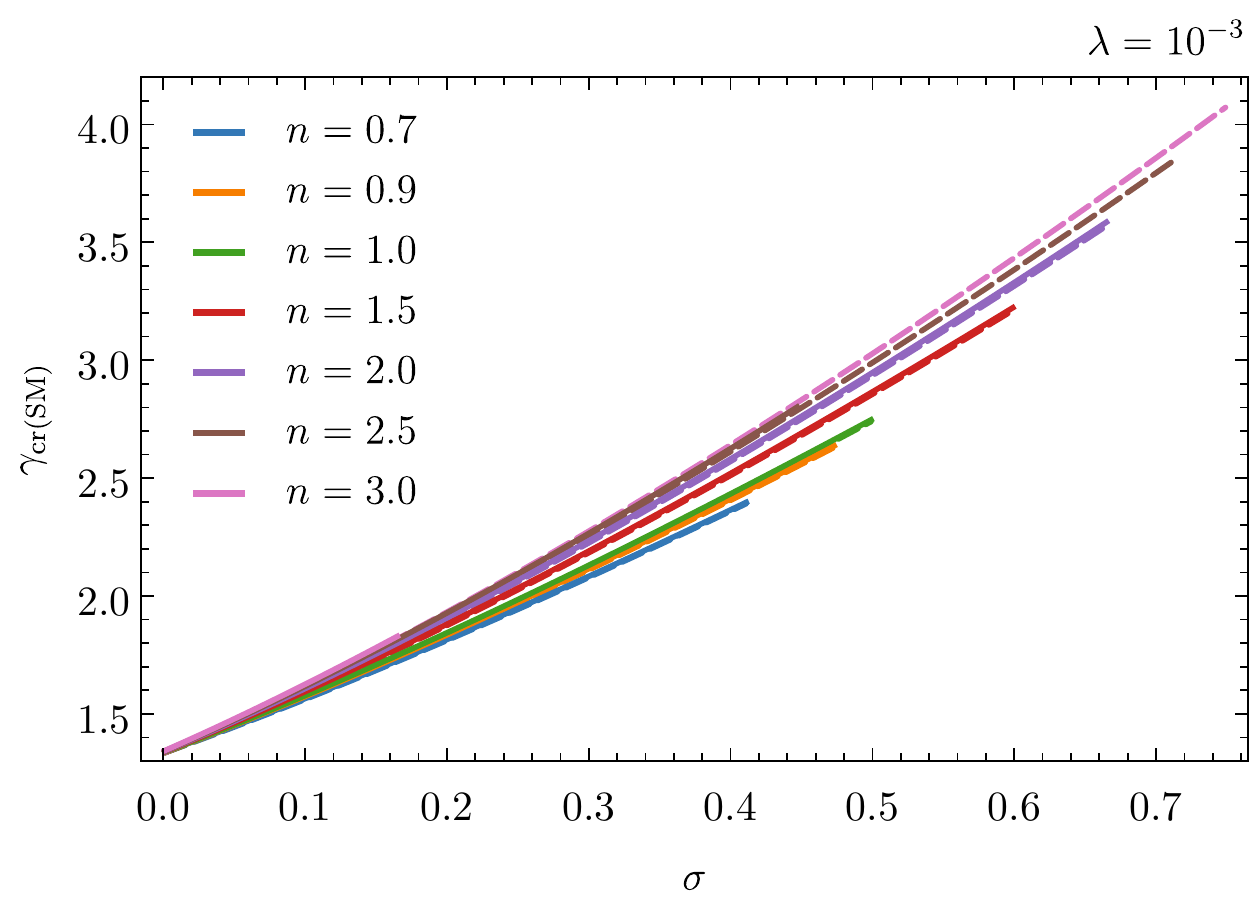} \quad \includegraphics[width=0.44\linewidth]{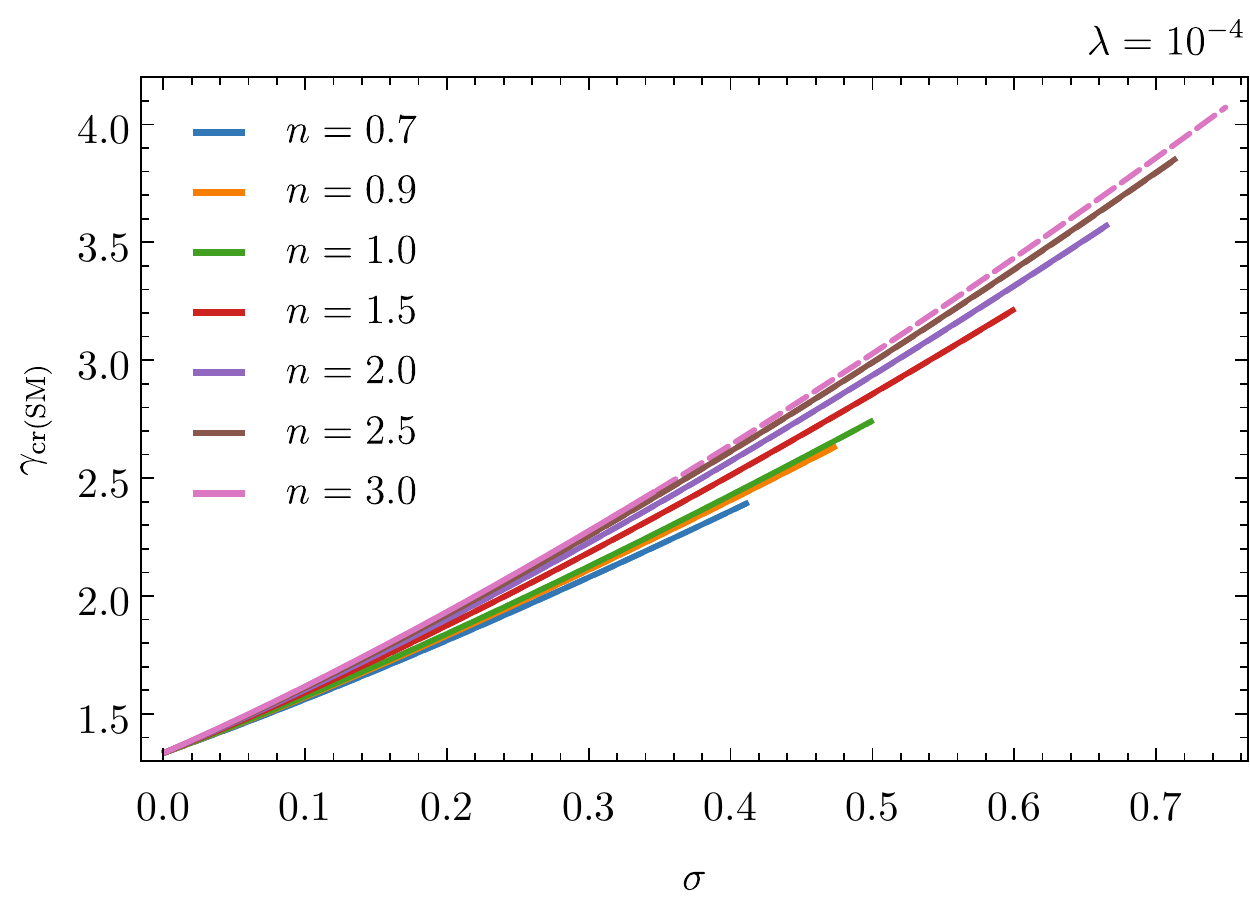}
    \caption{\label{fig2}Critical adiabatic index $\gamma_{\mathrm{cr}}$, for the onset of instability, for polytropic spheres in the range $0.5\leq\, n \leq\,3$, for the index $\lambda \in [10^{-4}, 10^{-1}]$. The dashed lines (in same color) indicate the corresponding values of $\gamma_{\mathrm{cr}}$ with $\lambda = 0$. Note that the critical adiabatic index increases from its corresponding value with $\lambda = 0$.}
\end{figure*}

As a second step in our analysis, we determined the critical adiabatic index $\gamma_{\mathrm{cr}}$ for the onset of instability, as a function of $\sigma$, for several values of the indexes $(n, \lambda)$. We show our results in Fig.~\ref{fig2}, which were obtained via the shooting method (see Sec.~\ref{sect:5A}).  For comparison we also plotted the values of $\gamma_{\mathrm{cr}}$ for each corresponding polytropic configuration with $\lambda = 0$. For large values of $\lambda$, for instance, $\lambda = 10^{-2}$ and $\lambda = 10^{-1}$ we found that the values of $\gamma_{\mathrm{cr}}$ increase with respect to their values with $\lambda = 0$. These results indicate that large values of $\lambda$ tend to destabilize the polytropic spheres.  Of particular interest is the case $\lambda = 10^{-1}$ which shows that in the Newtonian limit, when $\sigma\to\,0$, the value of $\gamma_{\mathrm{cr}}$ deviates from the expected value $\gamma_\mathrm{N} = 4/3$.

\begin{figure*}[ht]
    \centering \includegraphics[width=0.44\linewidth]{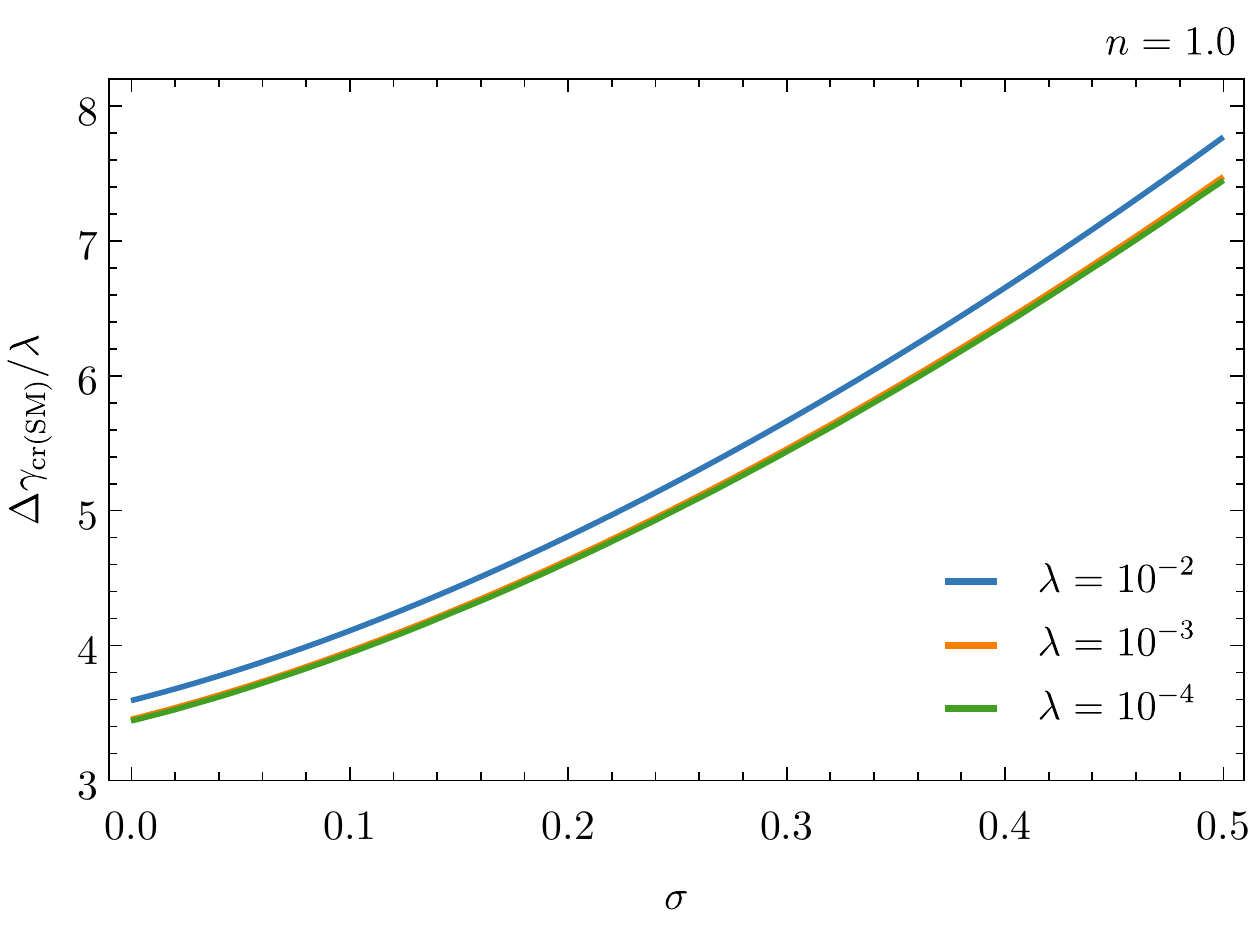} \quad \includegraphics[width=0.44\linewidth]{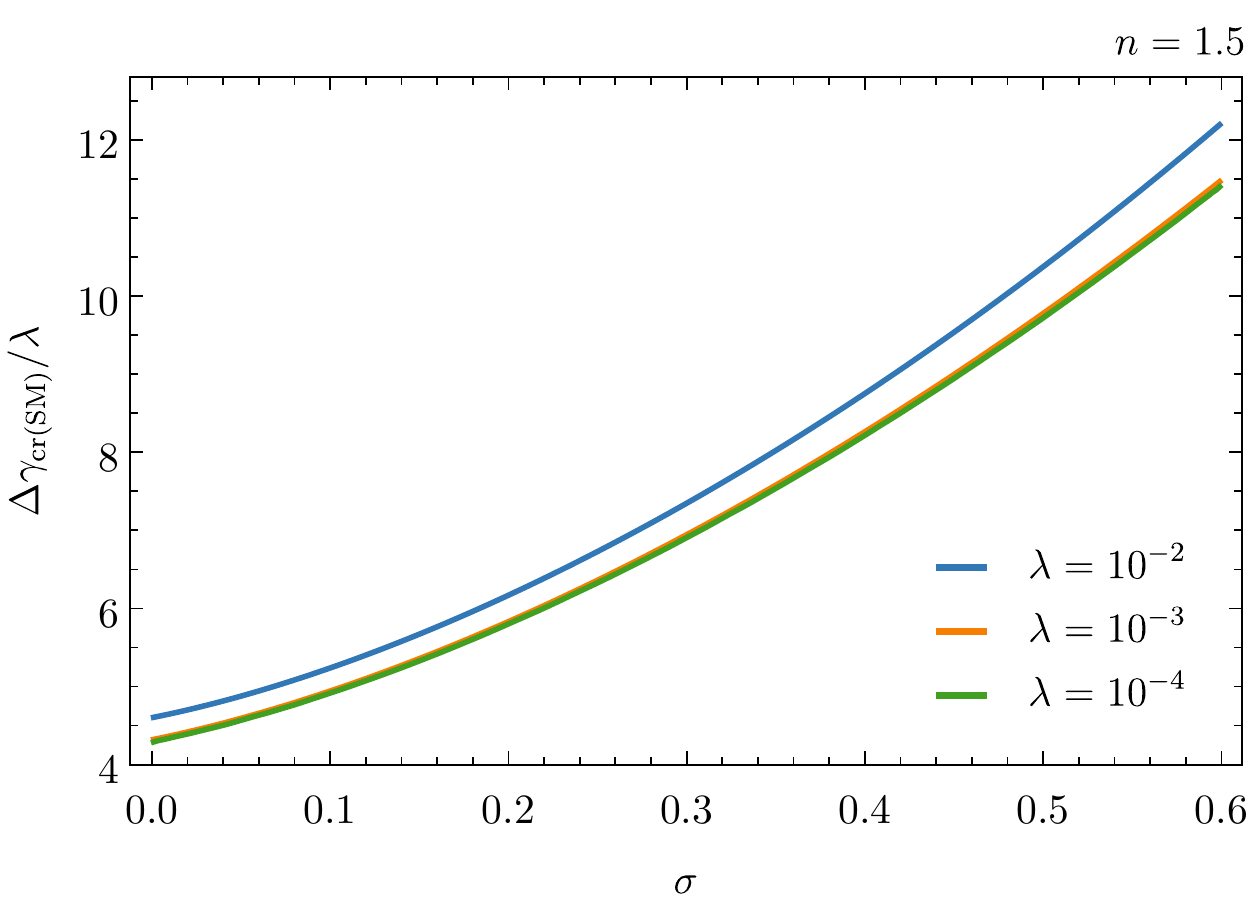}\\
    \centering \includegraphics[width=0.44\linewidth]{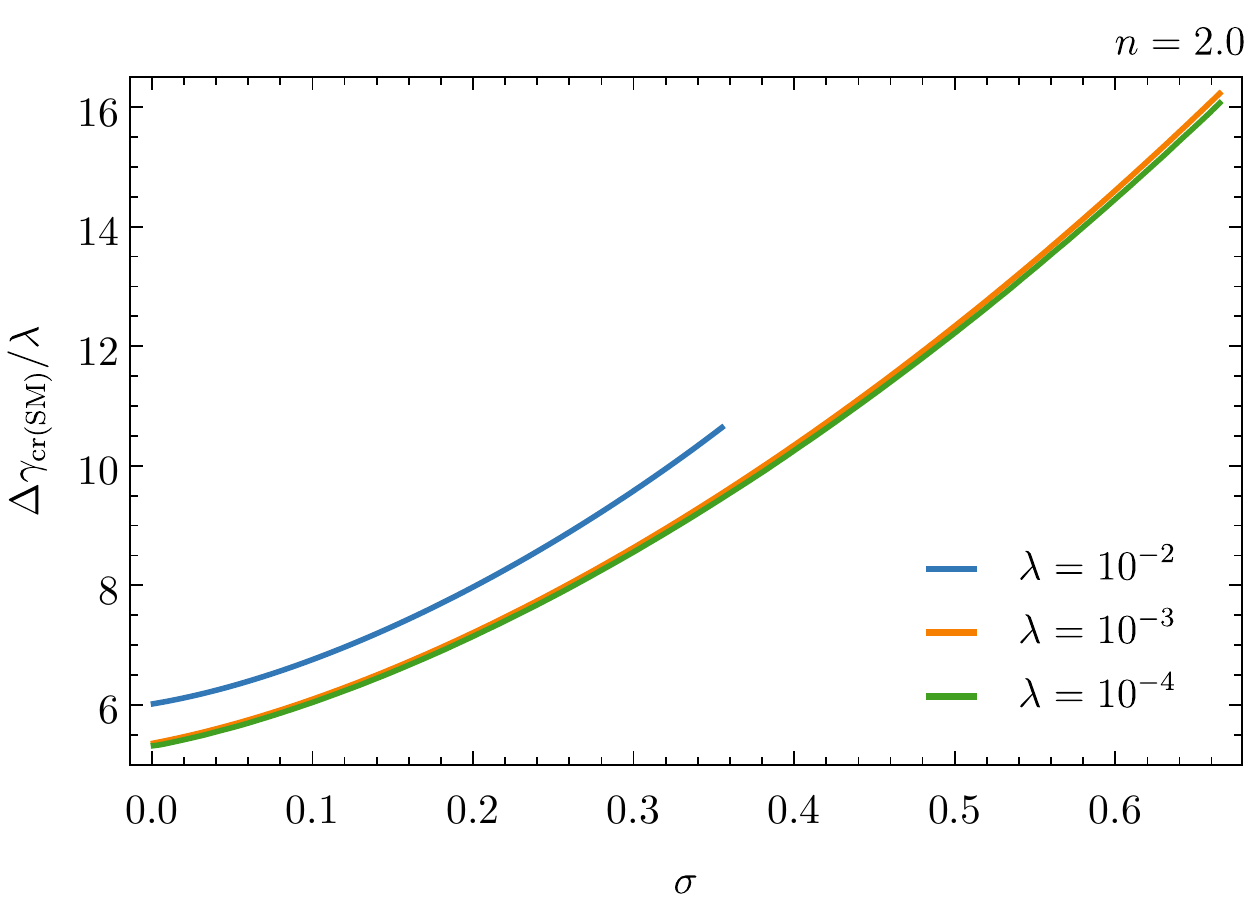} \quad \includegraphics[width=0.44\linewidth]{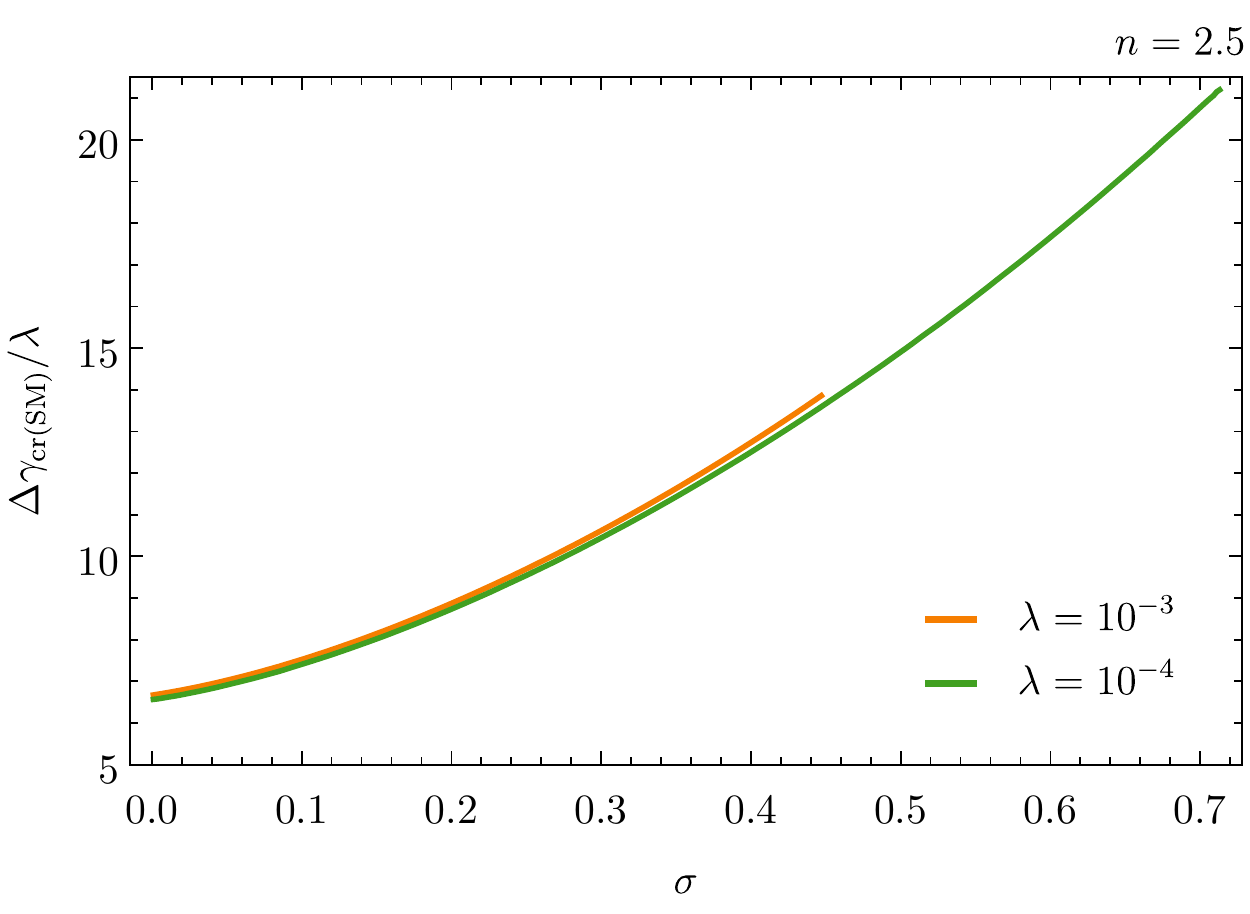}
    \caption{\label{fig3} Differences of the critical adiabatic index $\gamma_{\mathrm{cr}}$, for the polytropes $n = \{1.0, 1.5, 2.0, 2.5\}$, of the values for $\lambda \in [10^{-6},10^{-3}]$ from their corresponding values with $\lambda = 0$. We have ‘normalized' the differences by dividing by the corresponding value of $\lambda$.}
\end{figure*}

We will be interested in the differences of certain general quantities $q$ with $\lambda_{i} \neq 0$, from its corresponding value with $\lambda = 0$
\begin{equation}\label{difference}
    \Delta q \equiv q^{(\lambda_i)} - q^{(\lambda_0)},
\end{equation}
where $\lambda_i$ denotes $\lambda \in \{10^{-6}, 10^{-4}, 10^{-3}, 10^{-2}\}$, and $\lambda_0$ indicates $\lambda = 0$. In Fig.~\ref{fig3}, we show our results for the differences of $\gamma_{\mathrm{cr}}$ for $\lambda \in [10^{-6}, 10^{-2}]$, for the polytropes $n \in \{1.0, 1.5, 2.0, 2.5\}$. We have normalized the differences by dividing by the corresponding value of $\lambda$. Note that the curves of the differences show the same qualitative behavior.

\begin{figure*}[h]
    \begin{minipage}{\linewidth}
   \centering \includegraphics[width=0.3\linewidth]{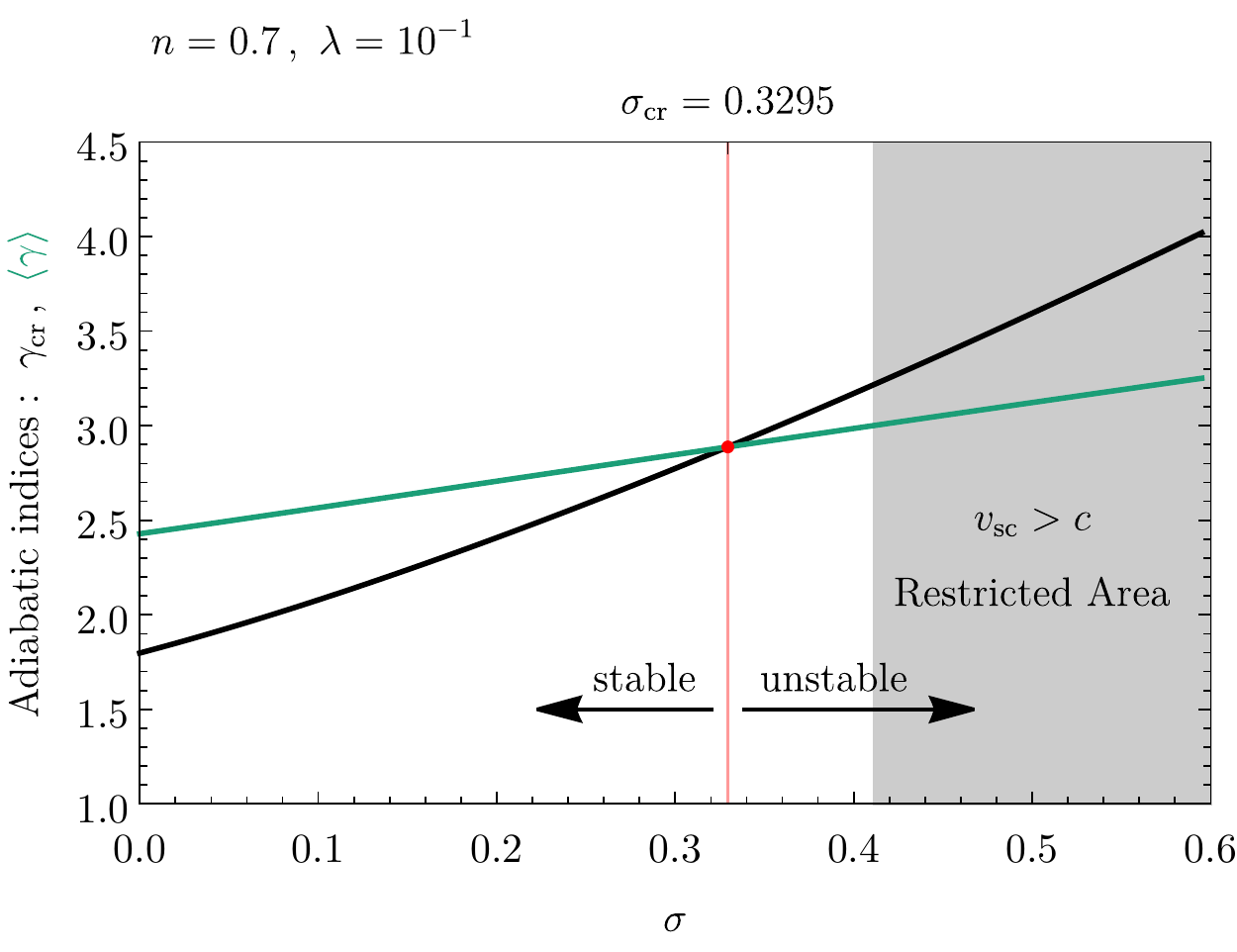}\quad\includegraphics[width=0.3\linewidth]{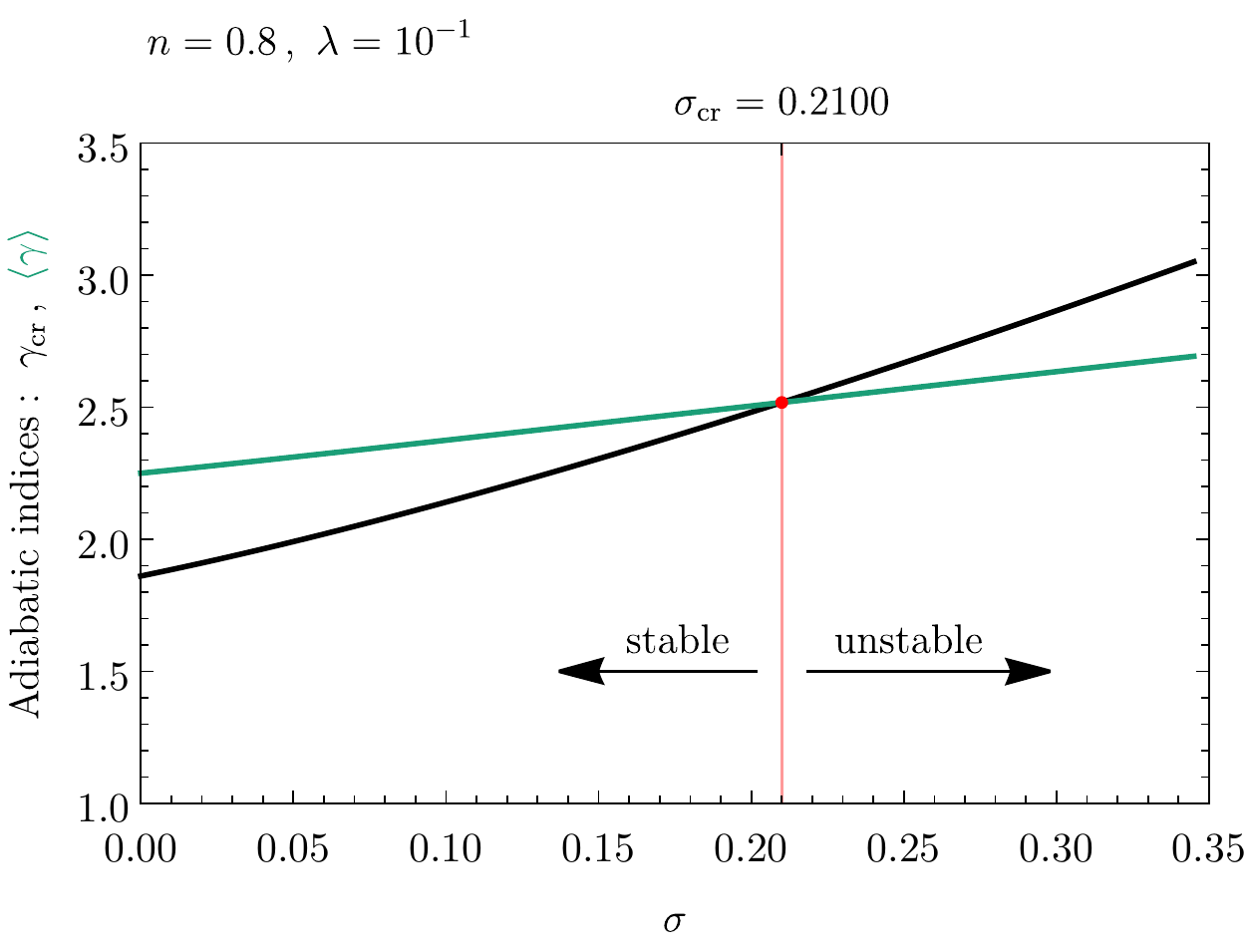}\quad\includegraphics[width=0.3\linewidth]{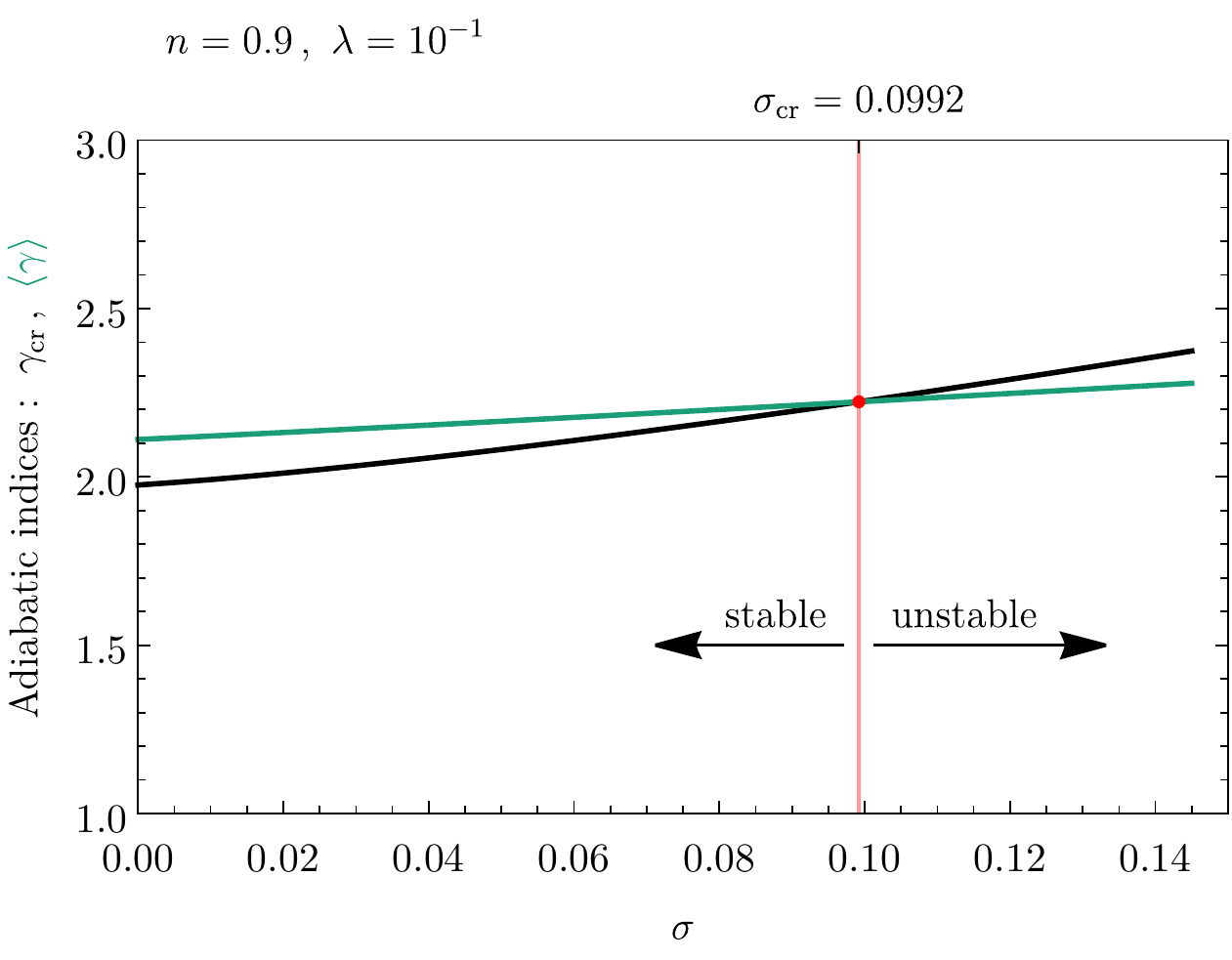}\\
    \end{minipage}\mbox{}\\[2mm]
    \centering \rule{10cm}{0.4pt} \\[2mm]
    \begin{minipage}{0.3\linewidth}
        \hbox to \linewidth{$n = 1.0$\hfill}
        \includegraphics[width=\linewidth]{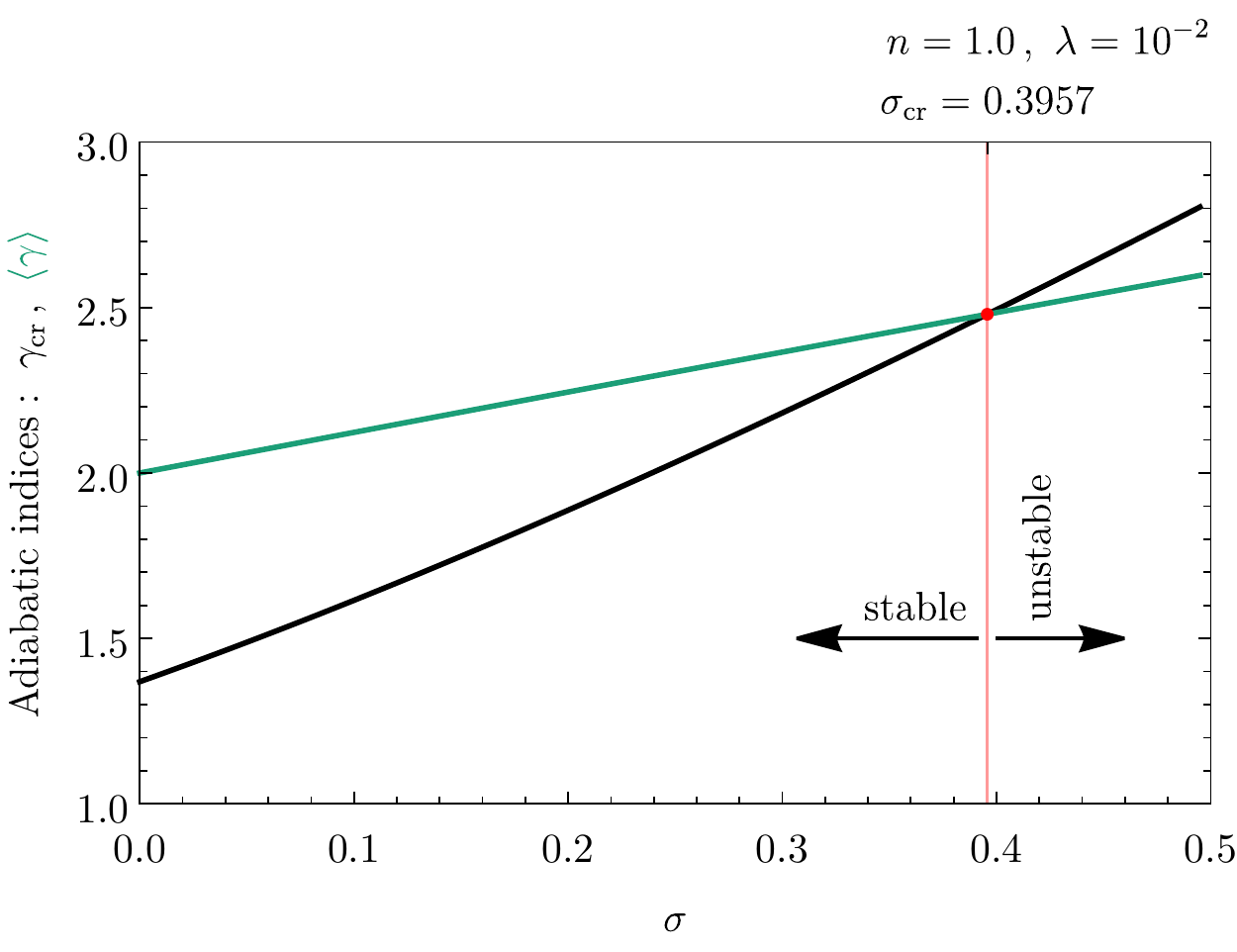}\\
        \includegraphics[width=\linewidth]{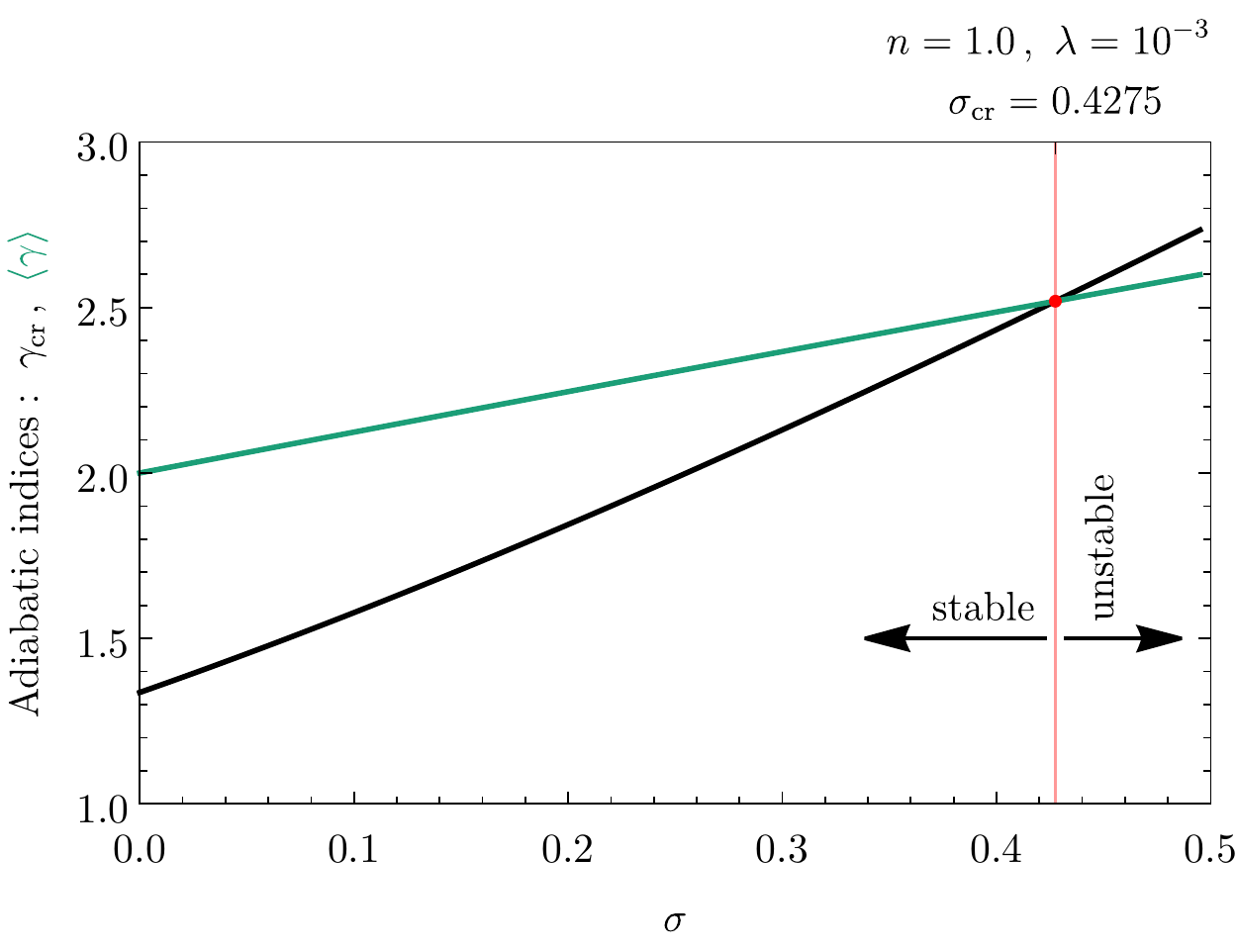}\\
        \includegraphics[width=\linewidth]{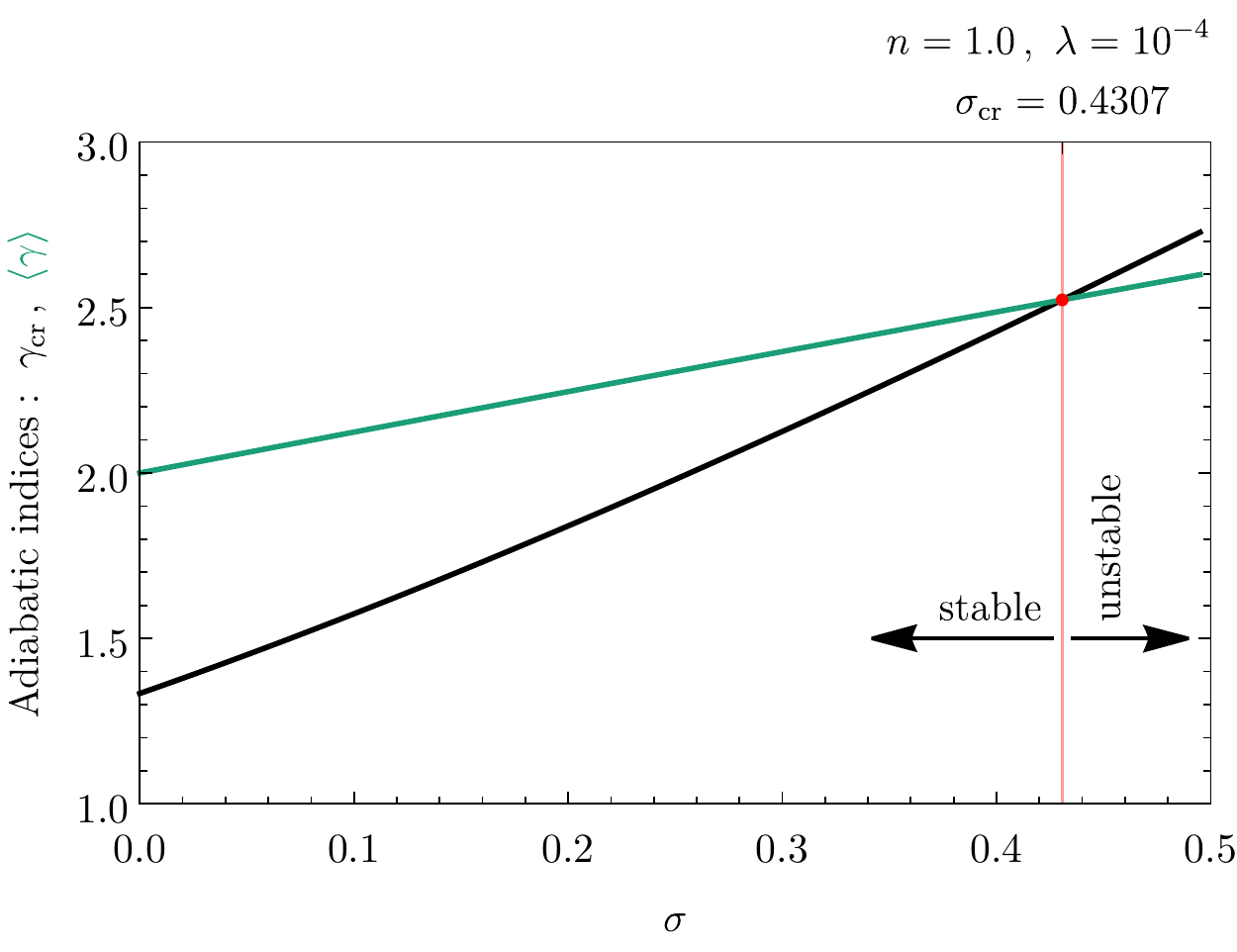}
    \end{minipage}
    \begin{minipage}{0.01\linewidth}
      \centering \rule{0.4pt}{10cm}
    \end{minipage}
    \begin{minipage}{0.3\linewidth}
         \hbox to \linewidth{$n = 1.5$\hfill}
        \includegraphics[width=\linewidth]{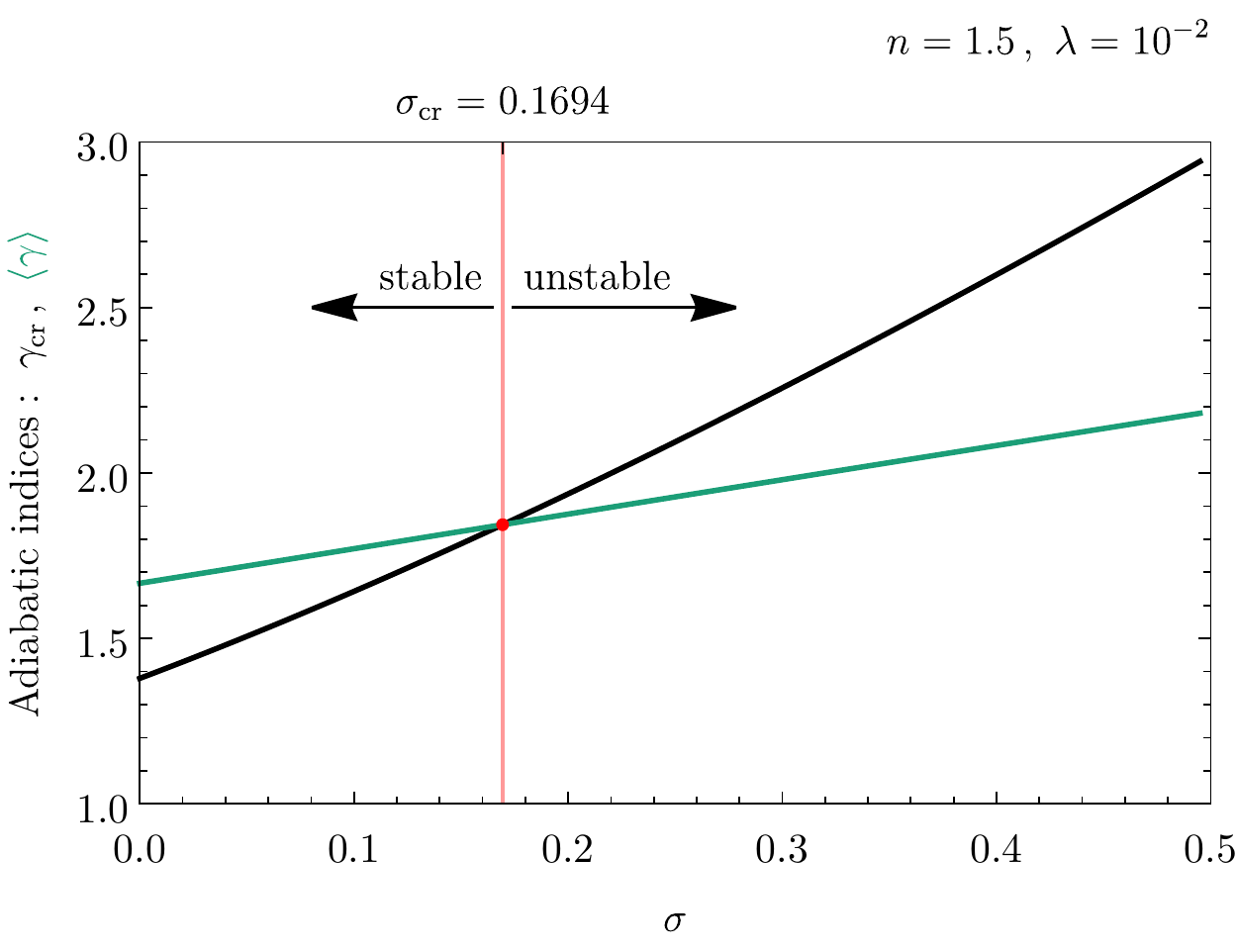}\\
        \includegraphics[width=\linewidth]{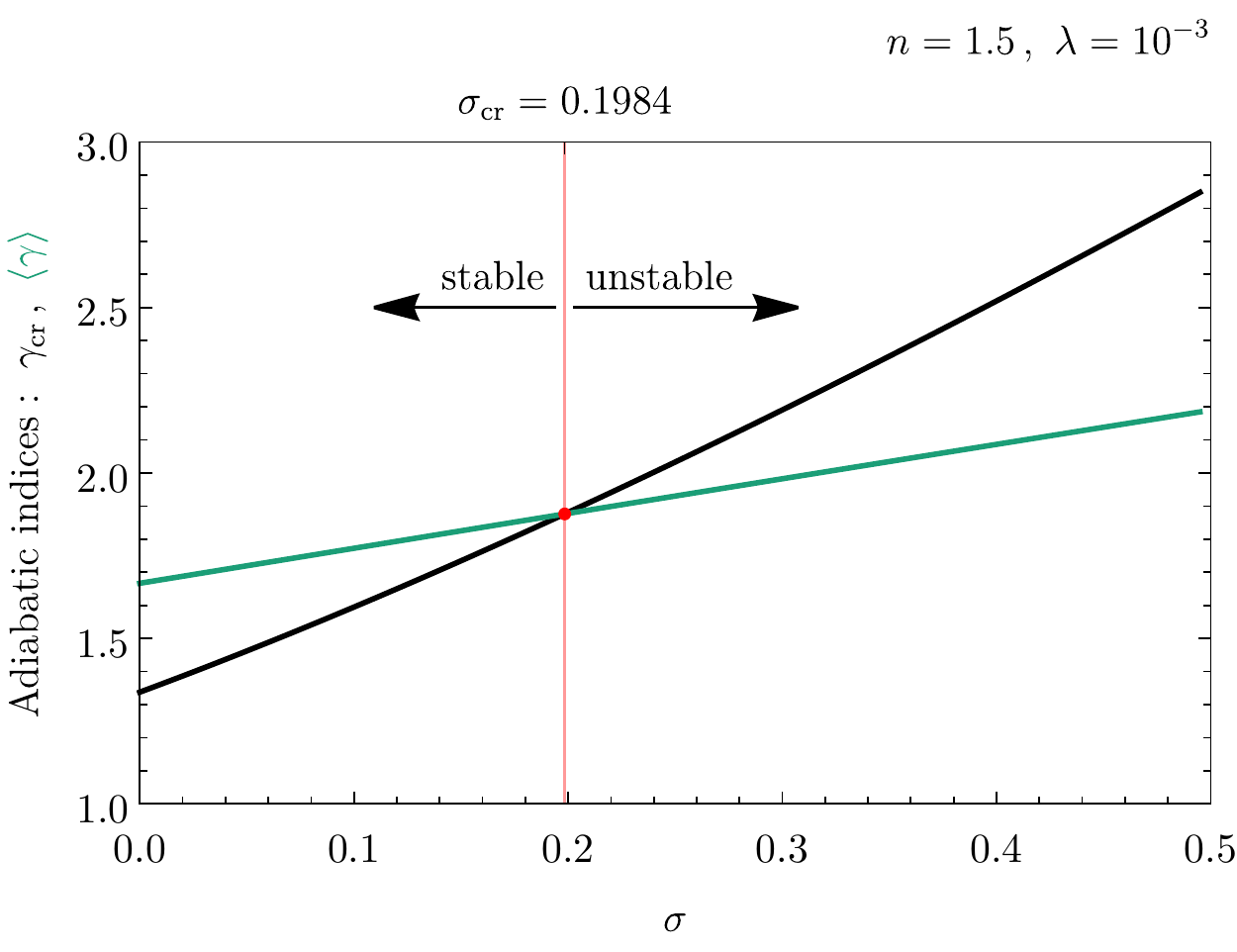}\\
        \includegraphics[width=\linewidth]{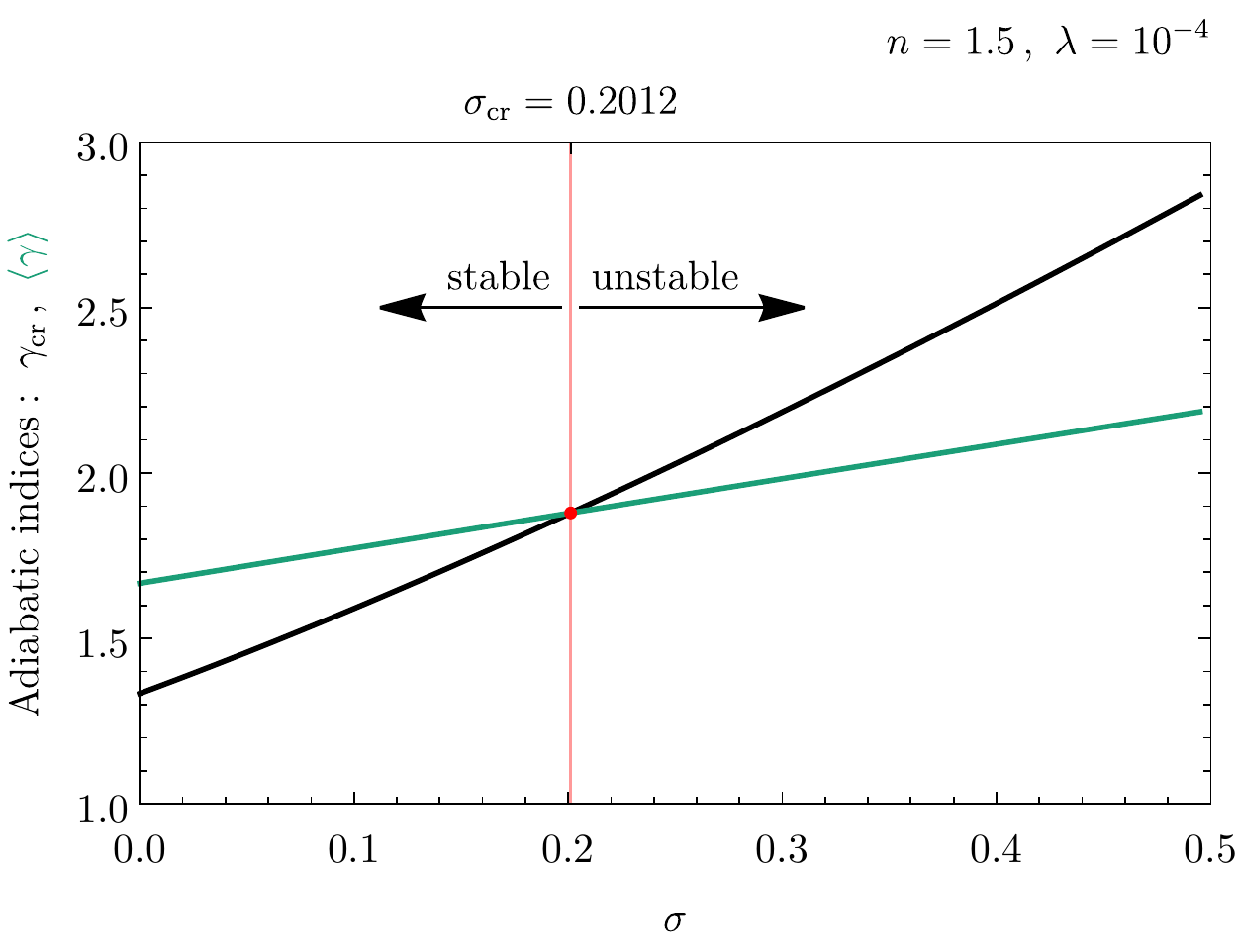}
    \end{minipage}
    \begin{minipage}{0.01\linewidth}
      \centering \rule{0.4pt}{10cm}
    \end{minipage}
    \begin{minipage}{0.3\linewidth}
         \hbox to \linewidth{$n = 2.0$\hfill}
        \includegraphics[width=\linewidth]{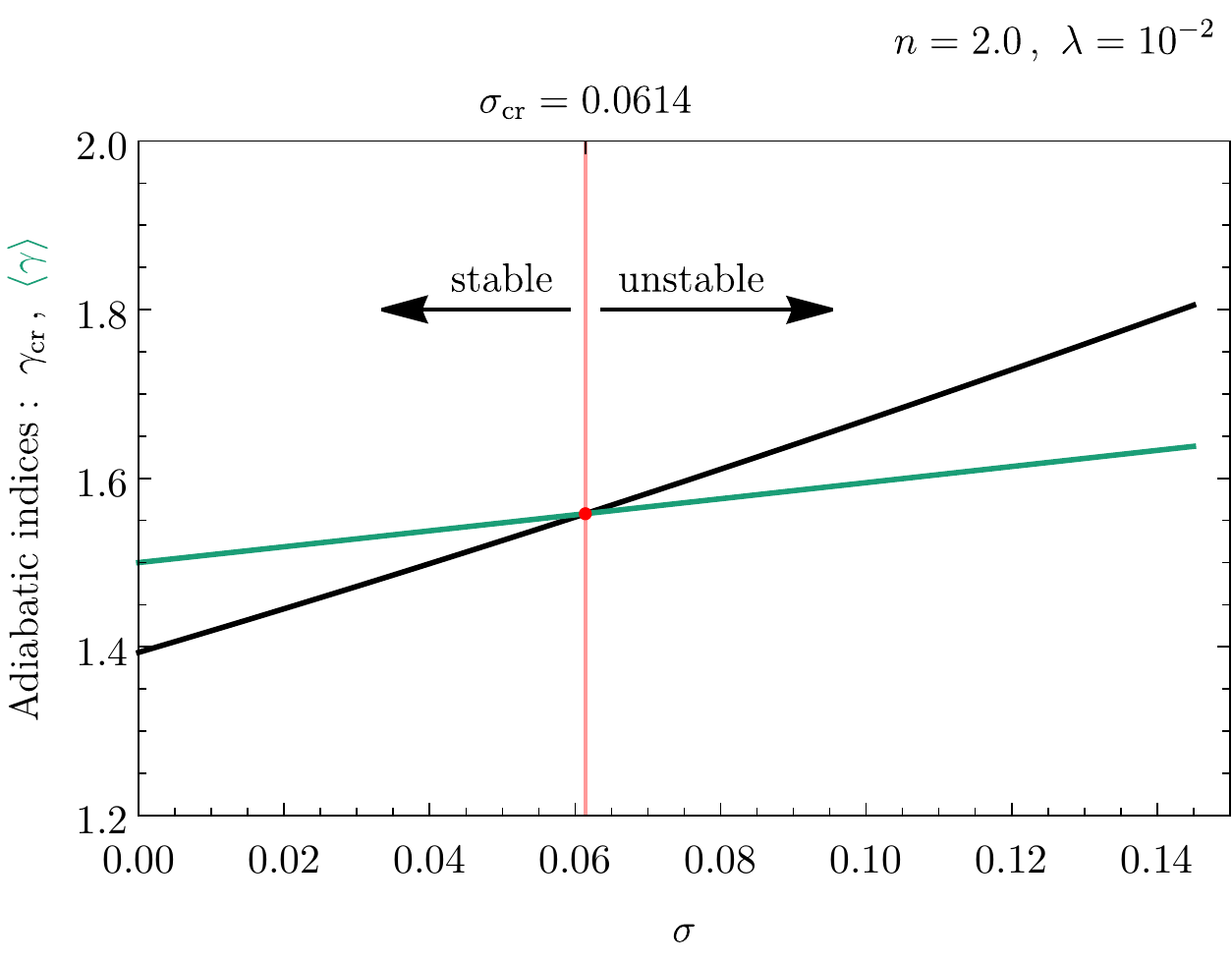}\\
        \includegraphics[width=\linewidth]{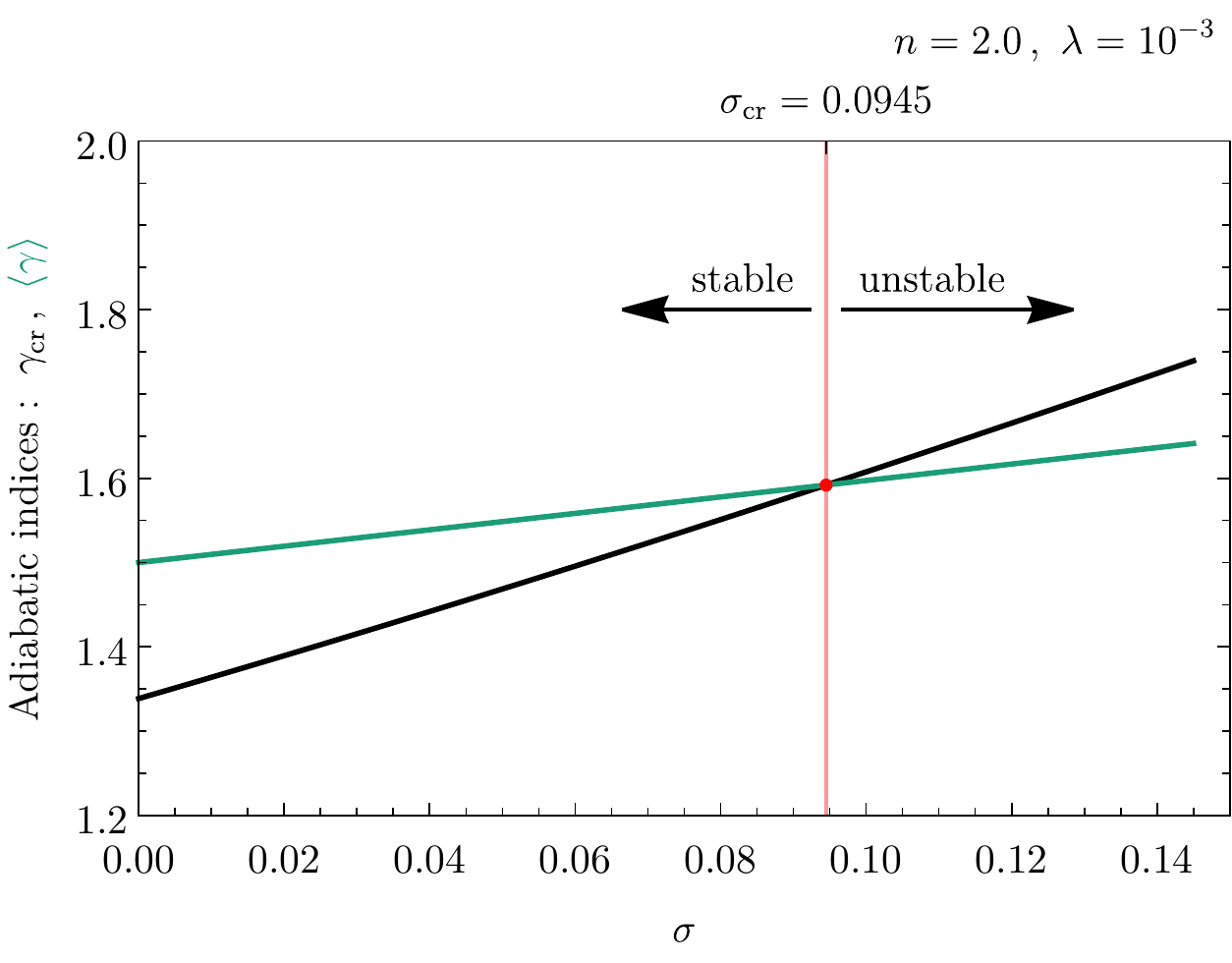}\\
        \includegraphics[width=\linewidth]{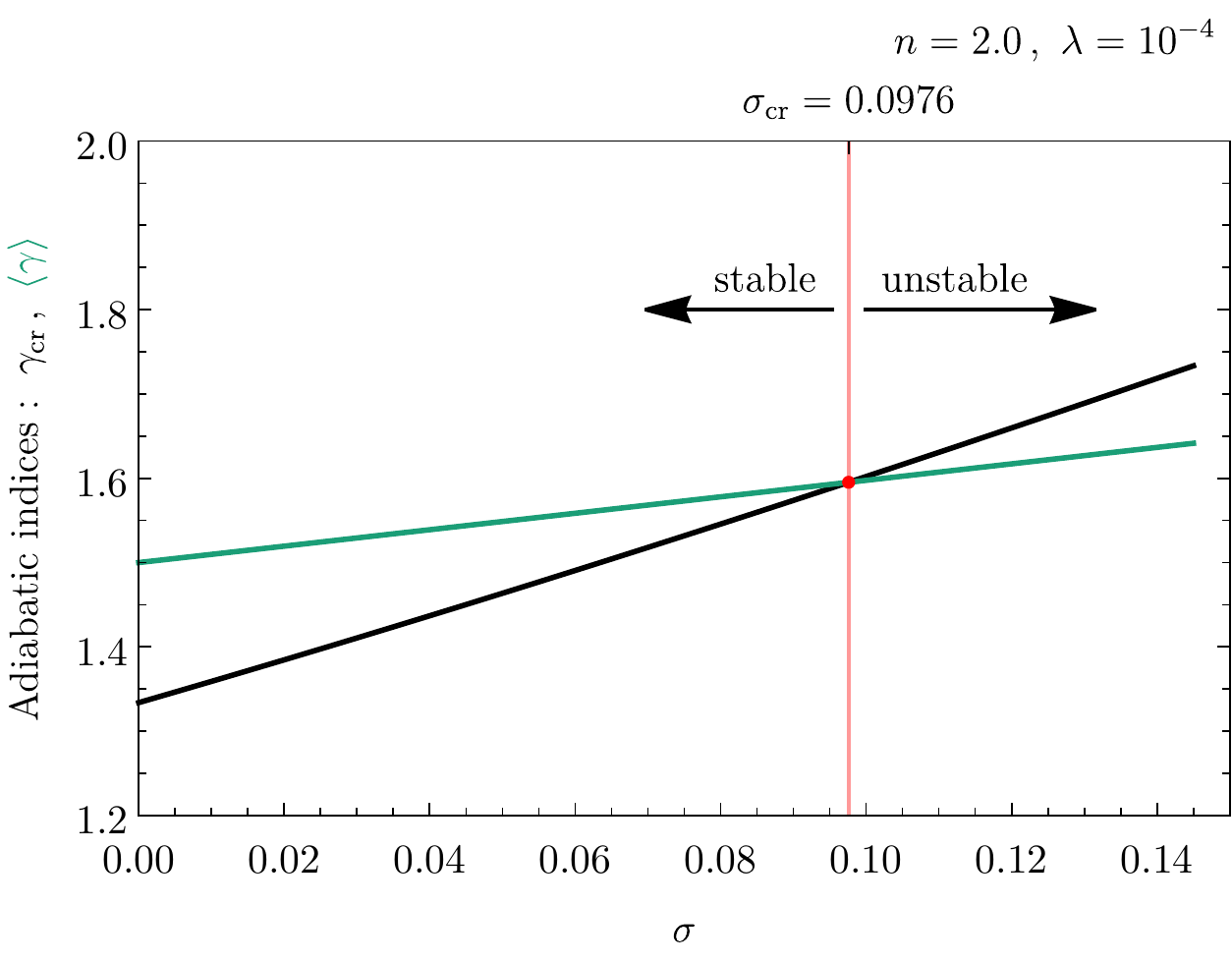}
    \end{minipage}
    \caption{\label{fig4} The stability domain as determined by comparison of the effective adiabatic index (green line) with the critical adiabatic index (black line) for polytropic spheres. The values of $\gamma_{\mathrm{cr}}$ were computed via the shooting method. The red line separates the stable from the unstable region given by the condition $\langle\gamma\rangle > \gamma_{\mathrm{cr}}$. The intersection point indicates the maximum permitted value $\sigma_{\mathrm{cr}}$ for stability. Note the role of the parameter $\lambda$ on $\sigma_\mathrm{cr}$.}
\end{figure*}

The stability domain, given by the condition $\langle\gamma\rangle > \gamma_{\mathrm{cr}}$, in dependence on the parameter $\lambda$ is shown in Fig.~\ref{fig4} for several representatives values of $n$. The effective $\langle\gamma\rangle$ was computed from Eq.~\eqref{effective} by using the results obtained from the shooting method. We also determined the critical values of $\sigma$ by using the trial functions given in Eq.~\eqref{Chandratrial}. In this case, we computed $\gamma_{\mathrm{cr}}$ from Eq.~\eqref{gammacr} and then we determined the effective adiabatic index from Eq.~\eqref{effective}. We summarize all of our results in Fig.~\ref{fig6}.

\begin{figure*}[h]
\begin{minipage}{\linewidth}
    \centering \includegraphics[width=0.3\linewidth]{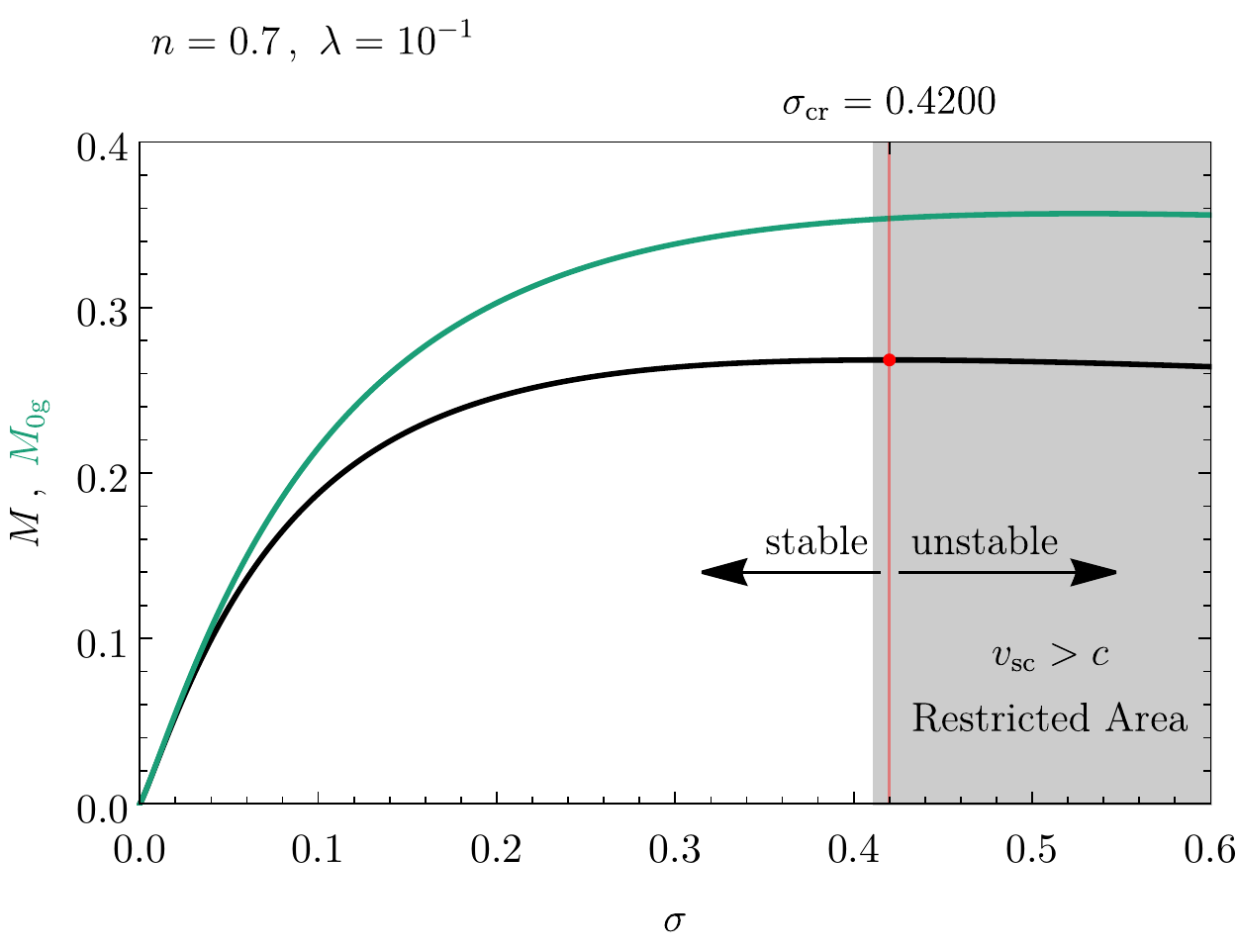}\quad\includegraphics[width=0.3\linewidth]{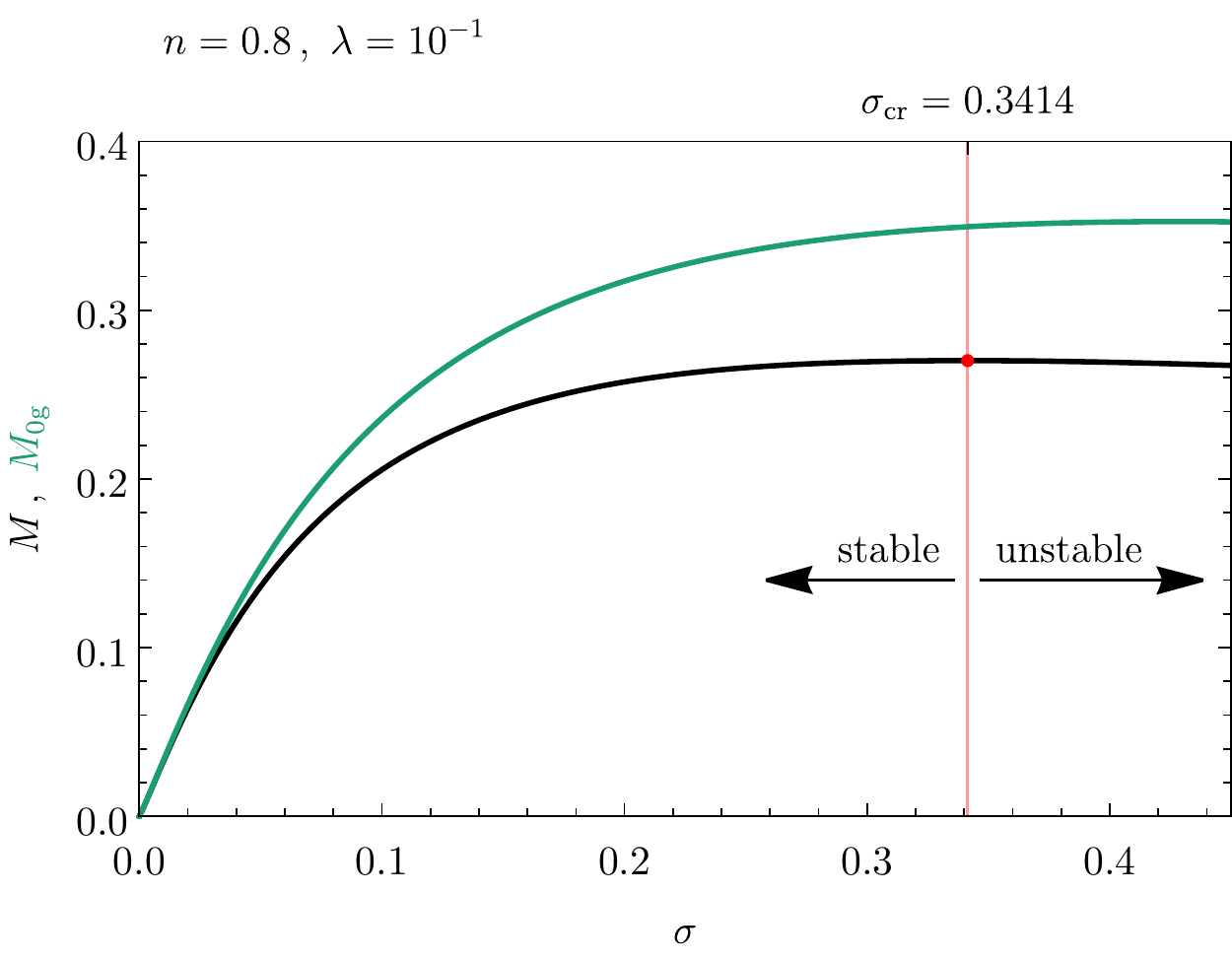}\quad\includegraphics[width=0.3\linewidth]{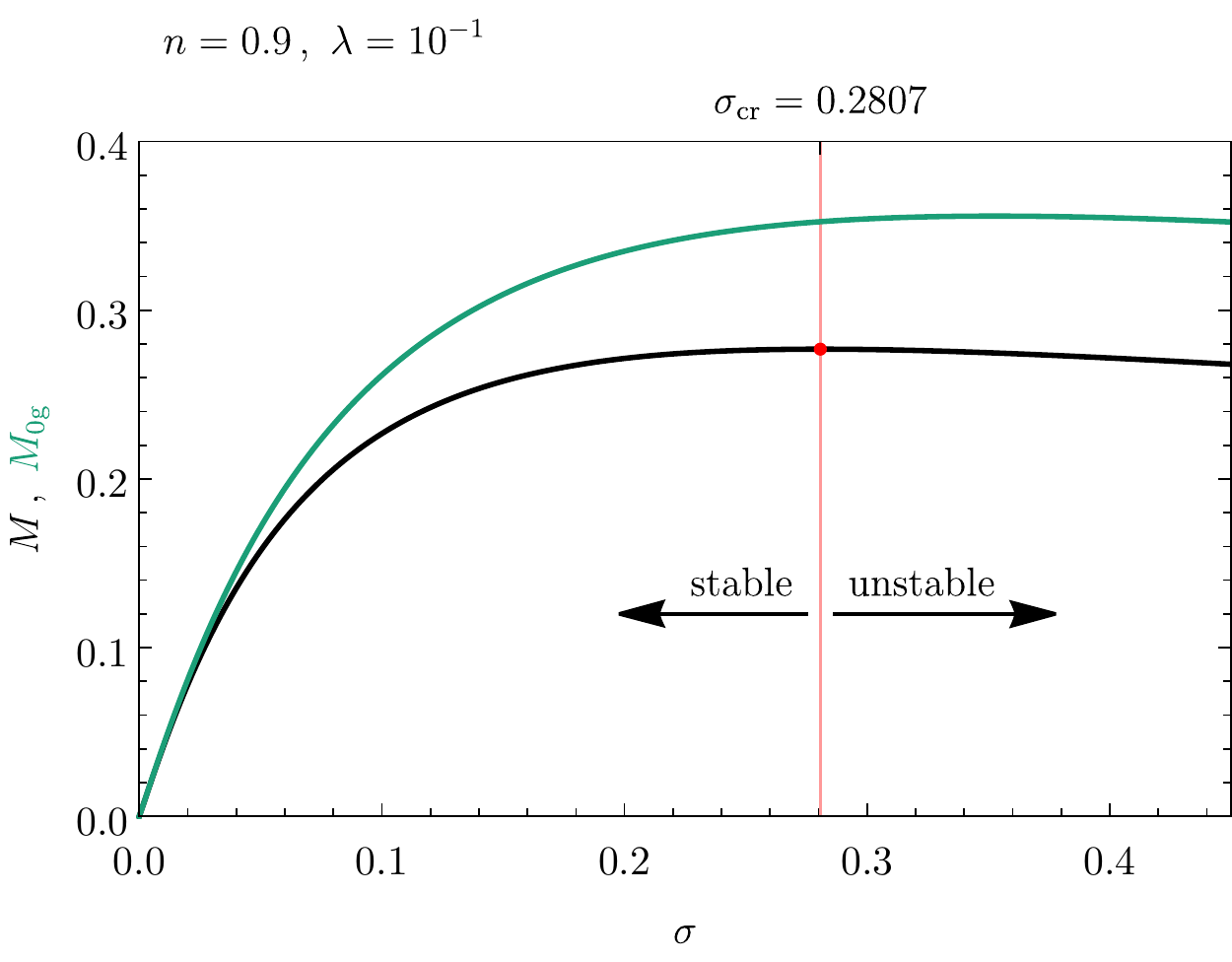}\\
    \end{minipage}\mbox{}\\[2mm]
    \centering \rule{10cm}{0.4pt} \\[2mm]
    \begin{minipage}{0.3\linewidth}
         \hbox to \linewidth{$n = 1.0$\hfill}
        \includegraphics[width=\linewidth]{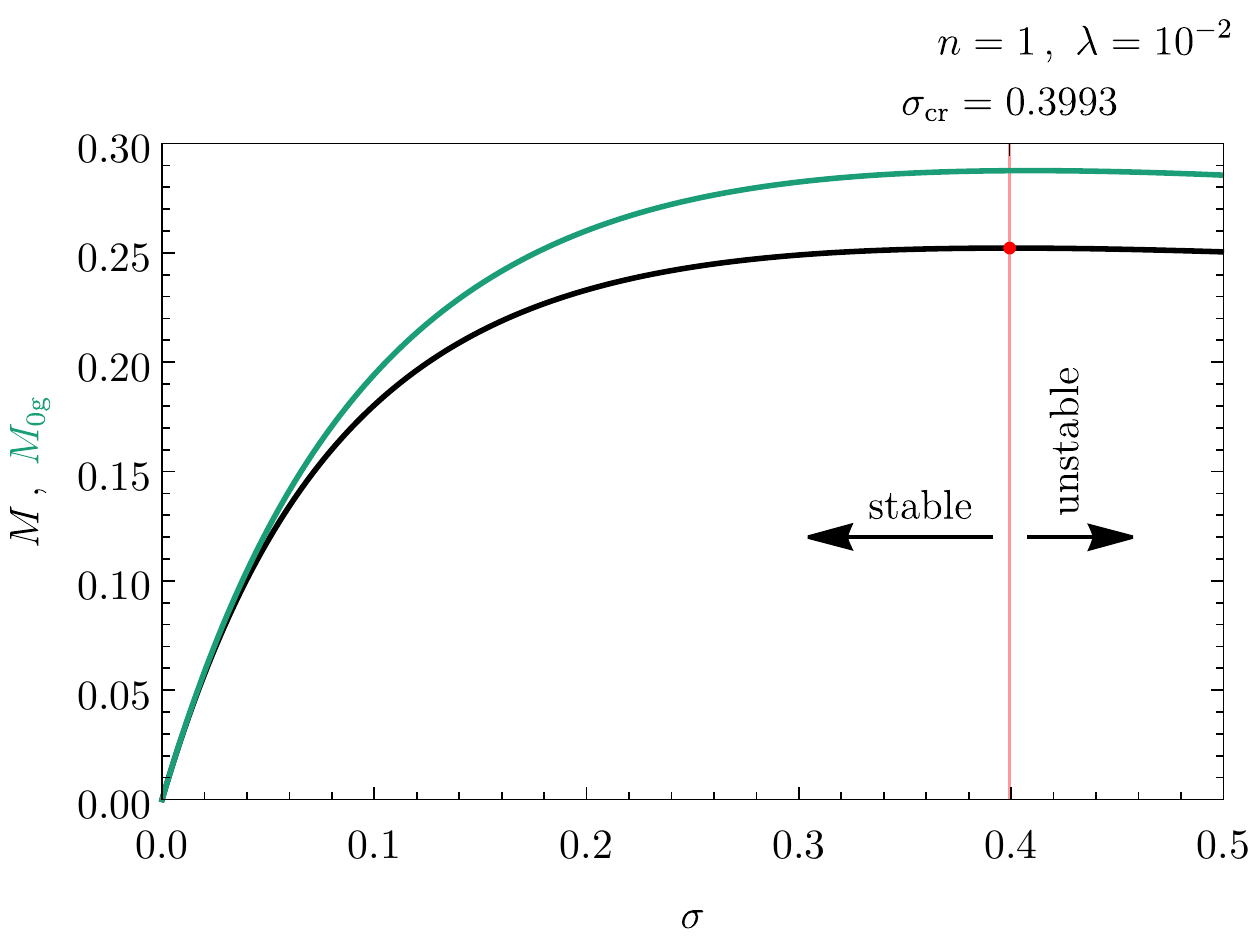}\\
        \includegraphics[width=\linewidth]{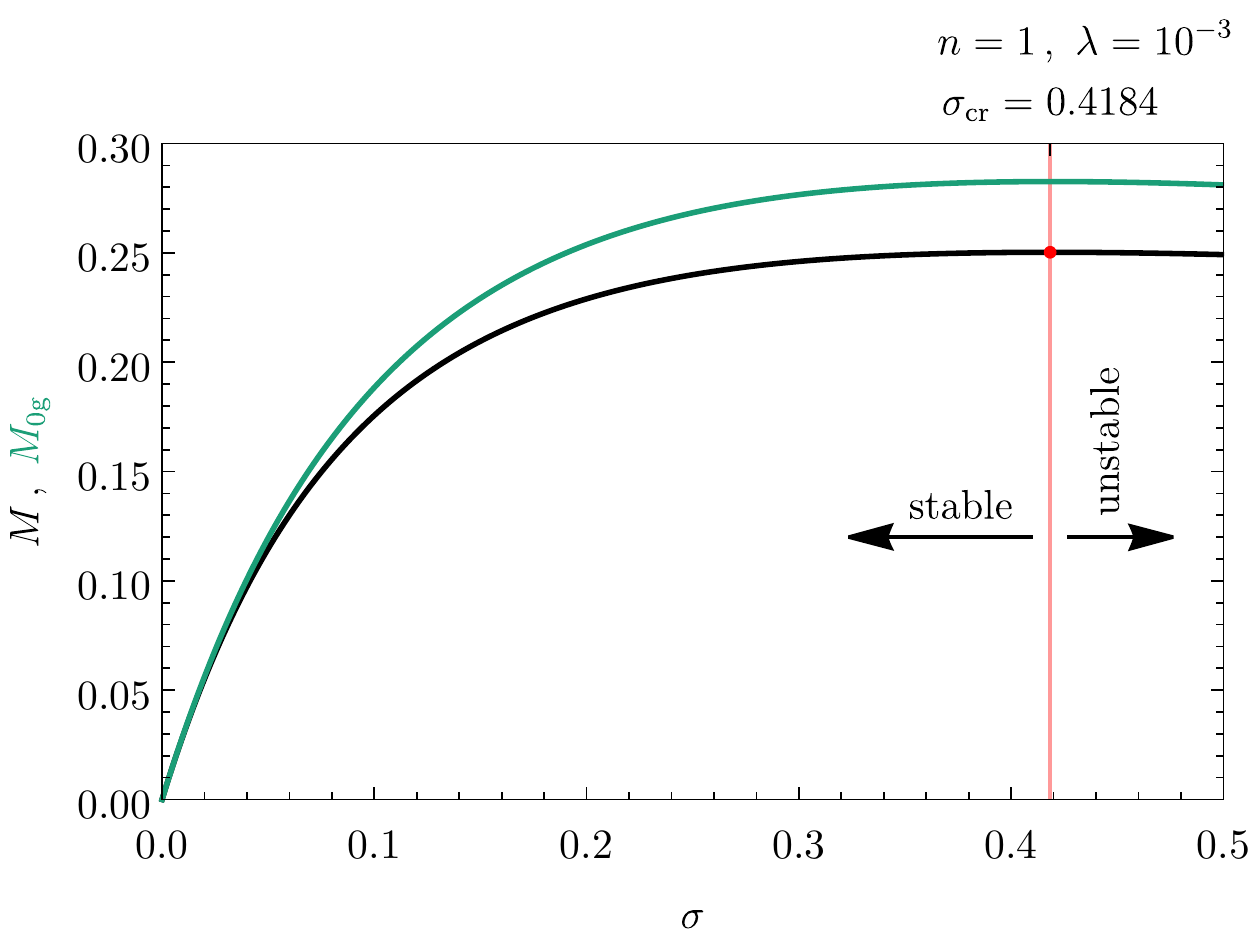}\\
        \includegraphics[width=\linewidth]{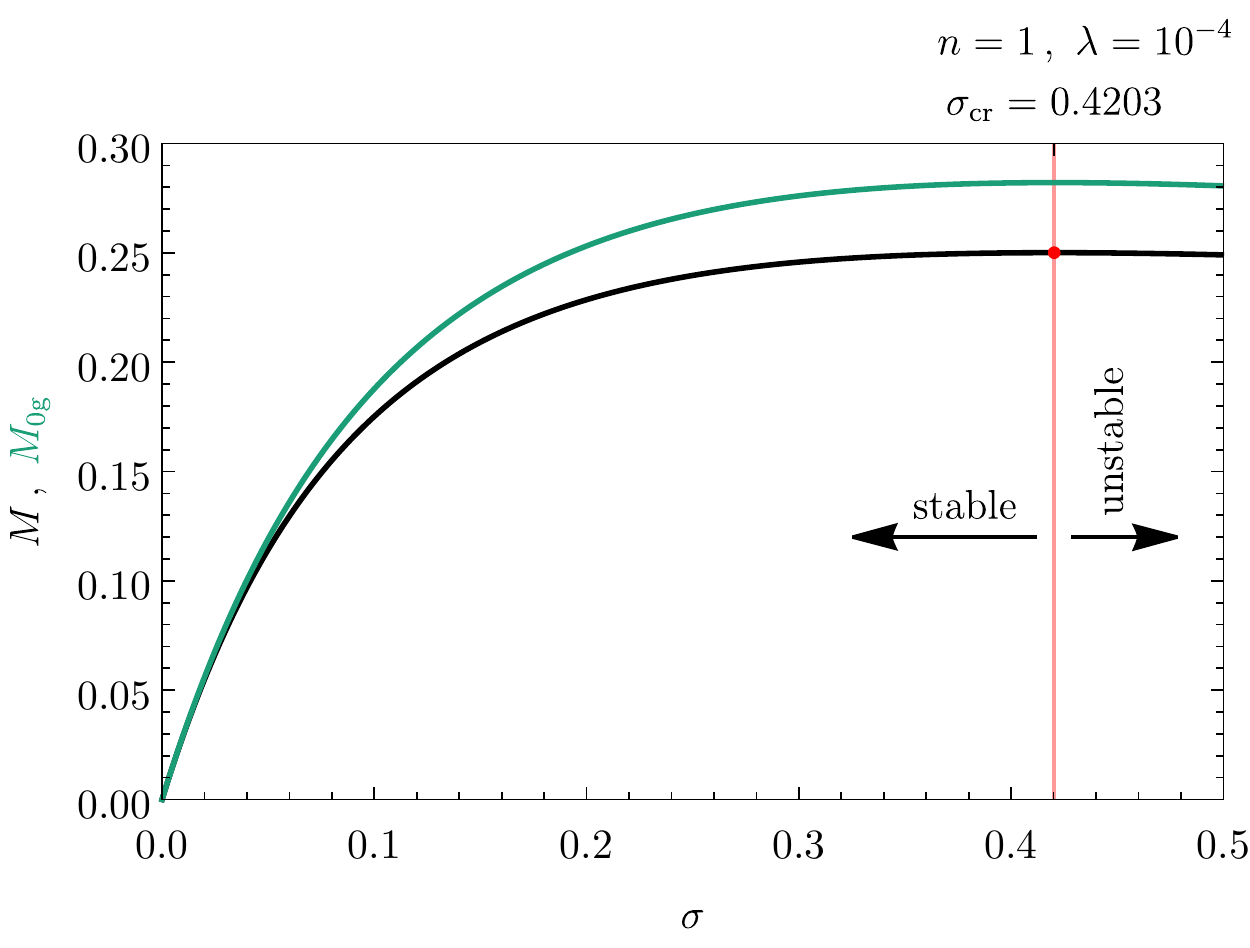}
    \end{minipage}
    \begin{minipage}{0.01\linewidth}
      \centering \rule{0.4pt}{10cm}
    \end{minipage}
    \begin{minipage}{0.3\linewidth}
         \hbox to \linewidth{$n = 1.5$\hfill}
        \includegraphics[width=\linewidth]{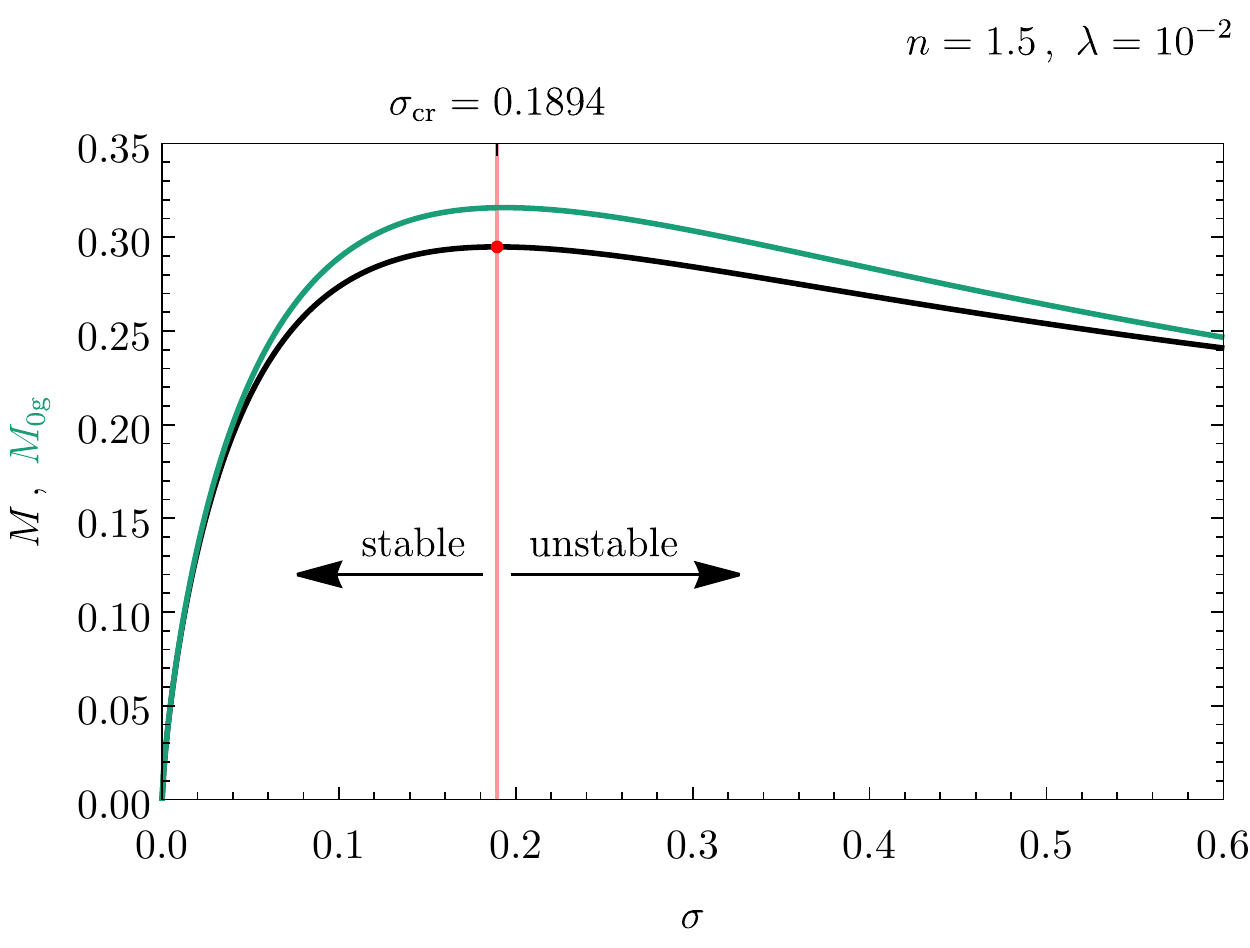}\\
        \includegraphics[width=\linewidth]{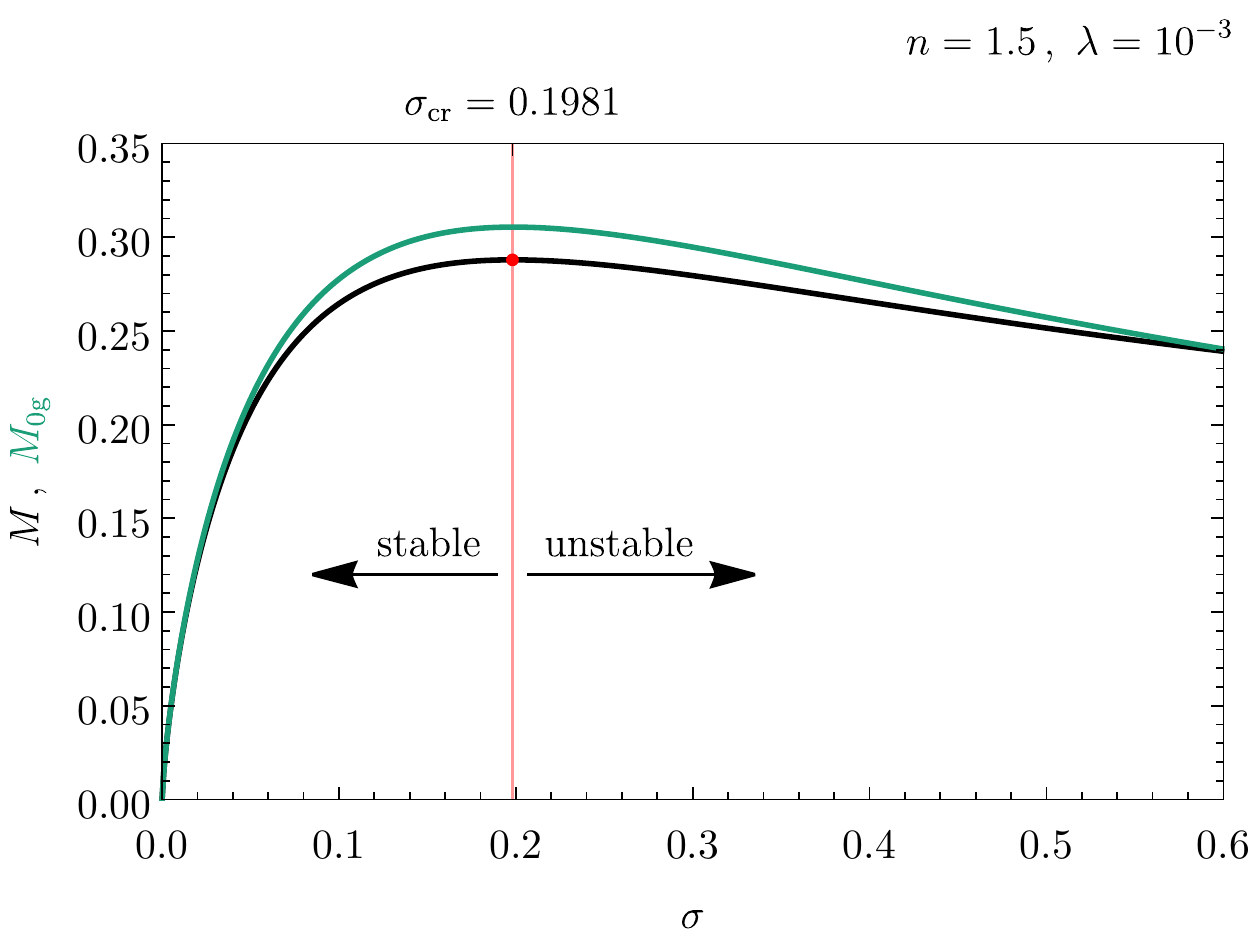}\\
        \includegraphics[width=\linewidth]{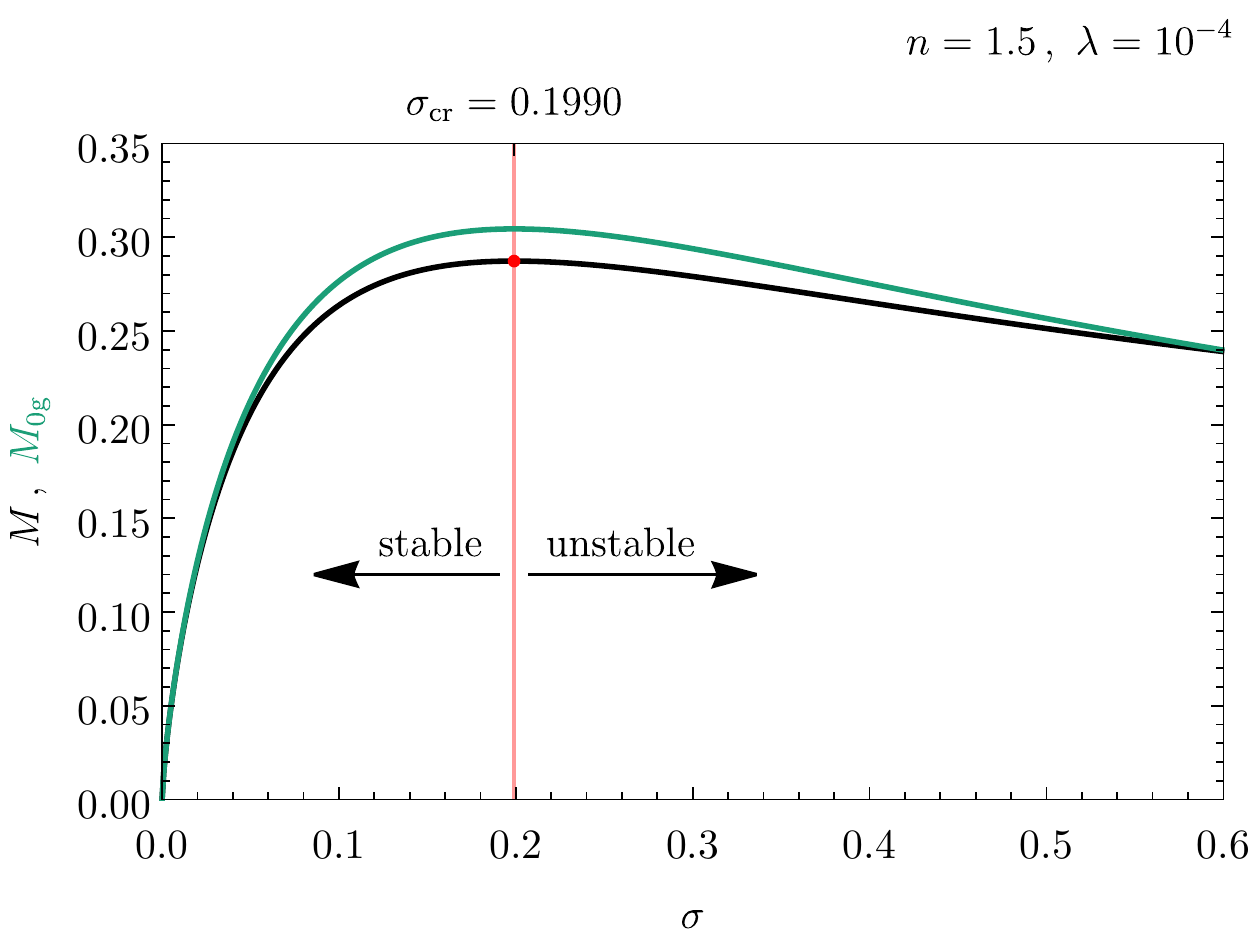}
    \end{minipage}
    \begin{minipage}{0.01\linewidth}
      \centering \rule{0.4pt}{10cm}
    \end{minipage}
    \begin{minipage}{0.3\linewidth}
         \hbox to \linewidth{$n = 2.0$\hfill}
        \includegraphics[width=\linewidth]{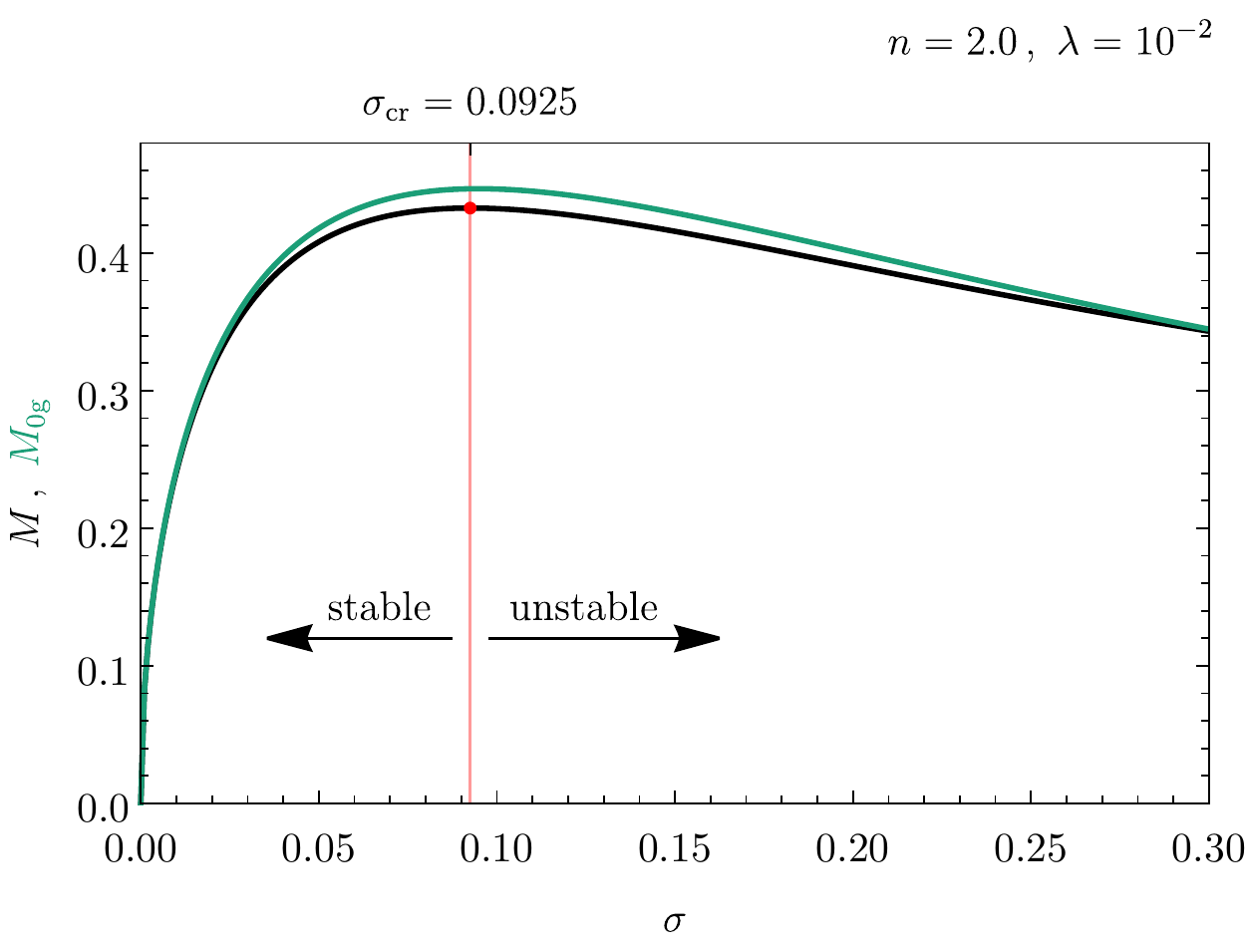}\\
        \includegraphics[width=\linewidth]{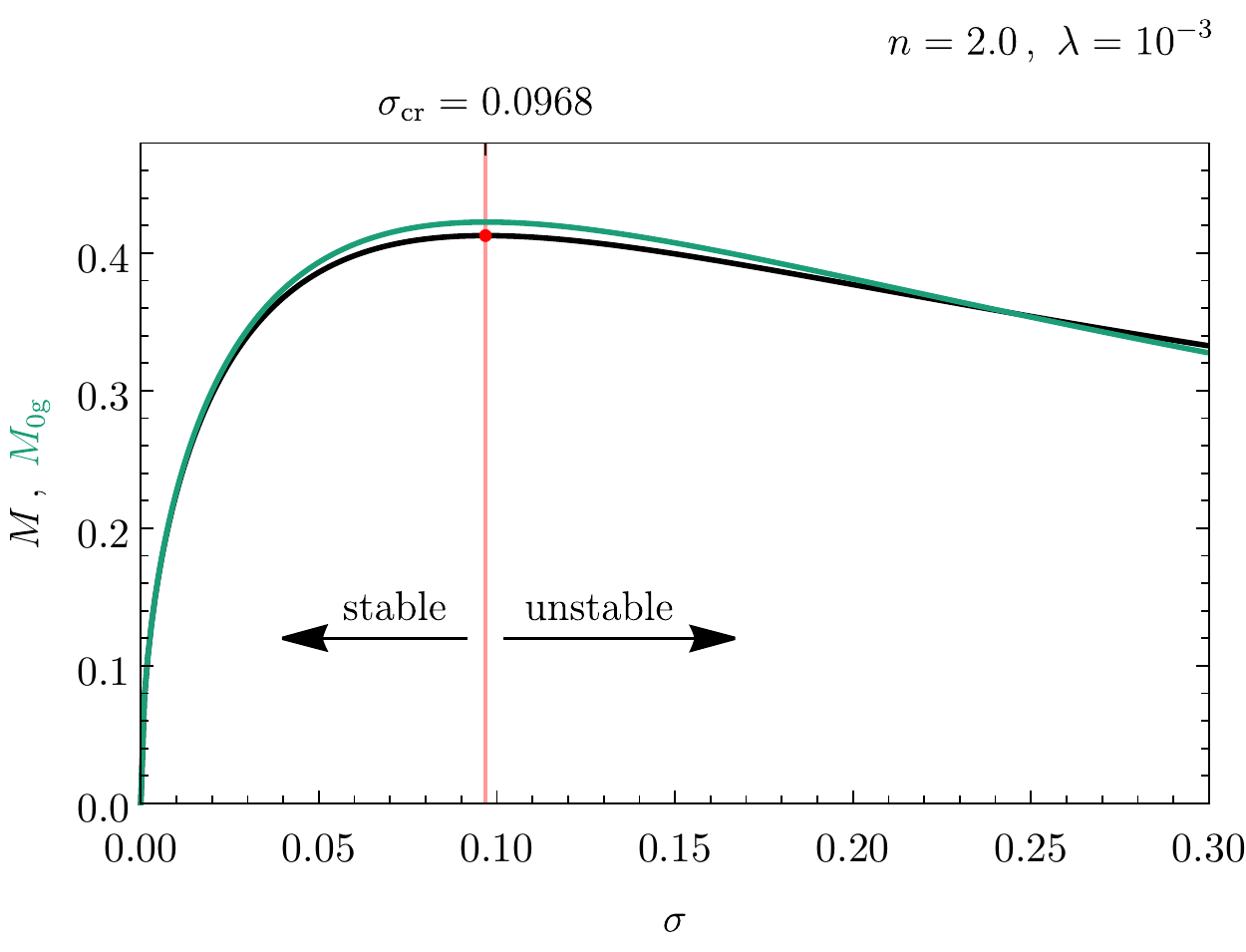}\\
        \includegraphics[width=\linewidth]{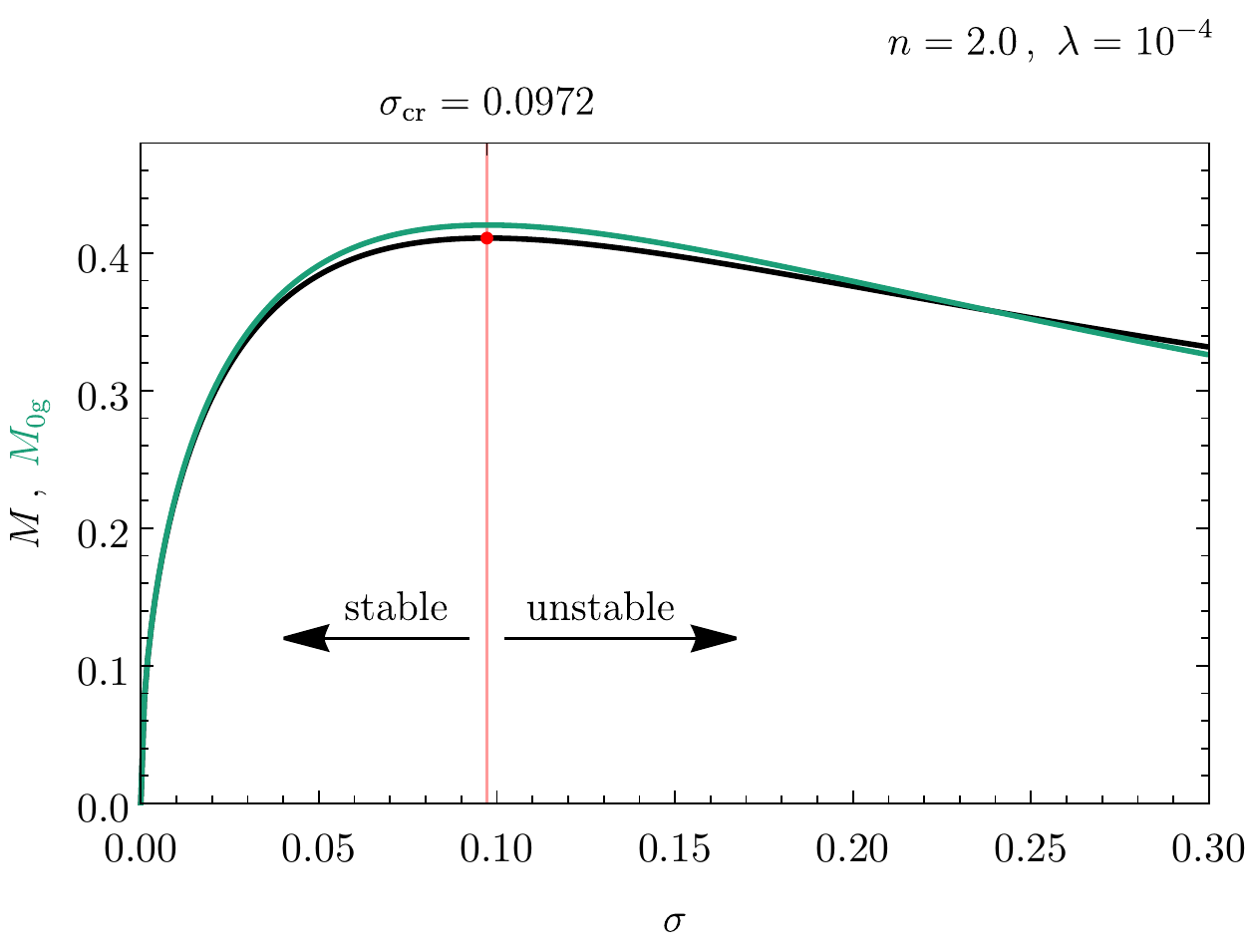}
    \end{minipage}
    \caption{\label{fig5} Profiles of the total mass (black line) and rest mass (green line), as a function of $\sigma$, for some polytropic spheres for different values of the vacuum constant index $\lambda$. The maximum of the curve for the total mass determines the critical value of $\sigma$ for stability; thus, it separates the stable and unstable regions.}
\end{figure*}

\subsection{Dynamical instability determined via the critical point method}
We follow the standard approach to examine the stability of the polytropic spheres using the energy considerations, or critical point method~\citep{Tooper:1964}. This analysis relies on the properties of static solutions to Einstein's equations. It is worthwhile to mention that static methods to study the stability of configurations might not be conclusive. Instabilities arising due to thermal effects may not be predicted by these methods, so one must turn to the full dynamical approach studied in the last section.

Substituting Eqs.~\eqref{sigma} and \eqref{polyx} in Eq.~\eqref{Mass} for the total mass $M$, we obtain

\begin{equation}
    M =  \frac{1}{\sqrt{4\pi}}(n + 1)^{3/2}K^{n/2}G^{-3/2}(\sigma\, c^2)^{(3 - n)/2} v(x_{1})\, ,
\end{equation}
where $K$ and $n$ are the parameters characterizing the polytrope (see Sect.~\ref{sect:2}). In our analysis we are considering configurations with $K$ and $n$ constants, therefore the total mass $M$ is proportional to $\sigma^{(3 - n)/2} v(x_{1})$ and the rest mass is proportional to $\sigma^{(3 - n)/2} v(x_{1})(E_{0\mathrm{g}}/E)$.

To study the stability, in Fig.~\ref{fig5}, we plot the total gravitational mass $M$ and the rest mass of baryons $M_{0\mathrm{g}}$ (‘preassembly mass'), given by Eq.~\eqref{M0g}, against the parameter $\sigma$. A necessary, but not sufficient, condition for stability is
\begin{equation}
    \frac{\dd M_{\mathrm{eq}}}{\dd \rho_\mathrm{c}} > 0\, ,
\end{equation}
where $M_{\mathrm{eq}}$ indicates the total mass at equilibrium. At the critical point, where
\begin{equation}
  \frac{\dd M_{\mathrm{eq}}}{\dd \rho_\mathrm{c}} = 0\, ,
\end{equation}
there is a change in stability due to the change in the sign of $\omega^2$ (see Sec.~\ref{sect:4}). Therefore, the critical point where the total mass $M$ has a maximum indicates the critical value $\sigma_{\mathrm{cr}}$ for stability.

In Fig.~\ref{fig5}, we show some profiles of total (and rest) mass, as a function of $\sigma$, for different polytropes in the range $0.5 < n < 3$ for several values of the index $\lambda \in [10^{-4}, 10^{-1}]$. In the plots, we have also indicated the maximum of the curve $M$ which provides the critical parameter $\sigma_\mathrm{cr}$, thus separating the stable from the unstable region.

\begin{figure*}[h]
    \centering \includegraphics[width=0.4\linewidth]{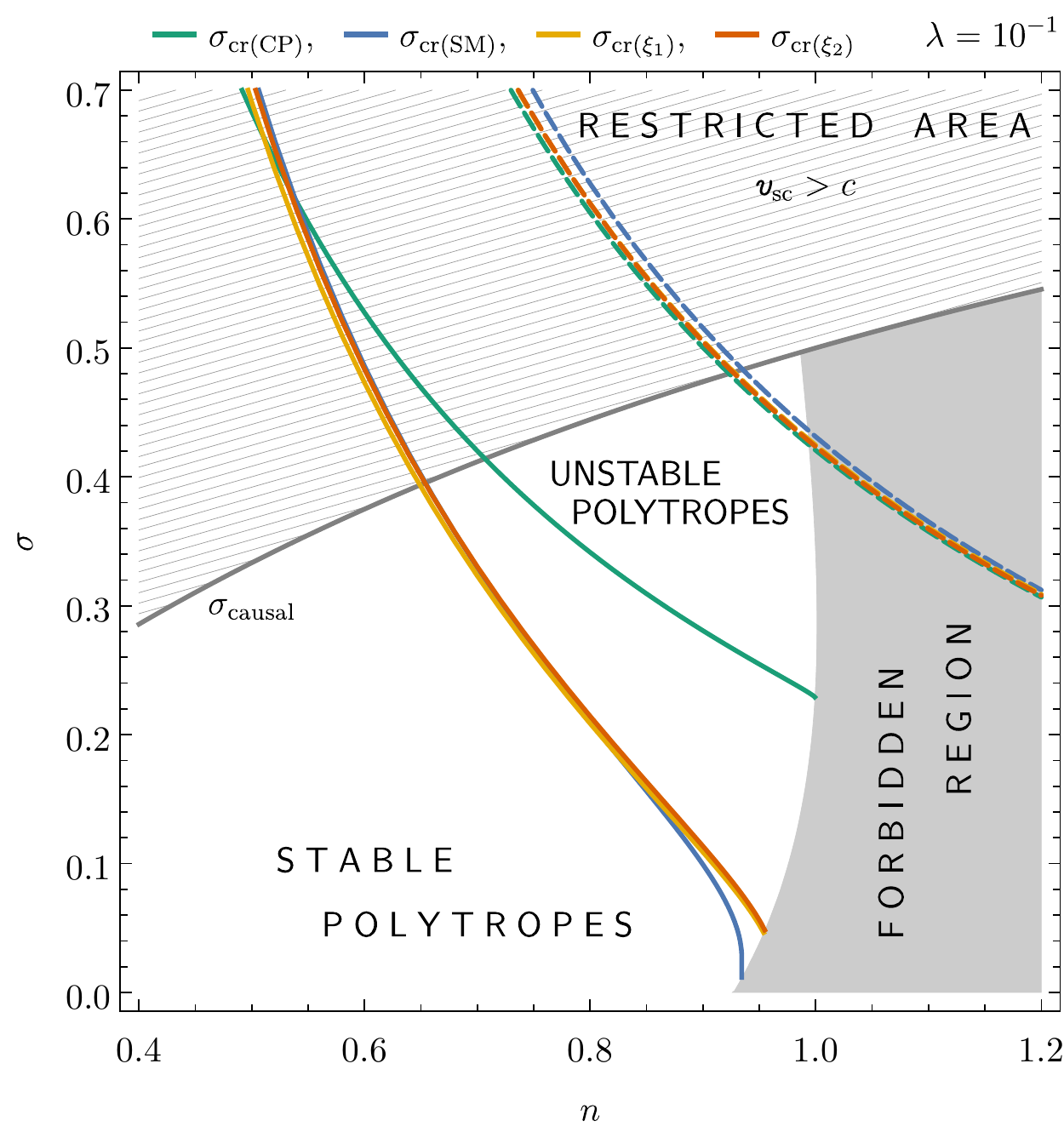} \quad \includegraphics[width=0.4\linewidth]{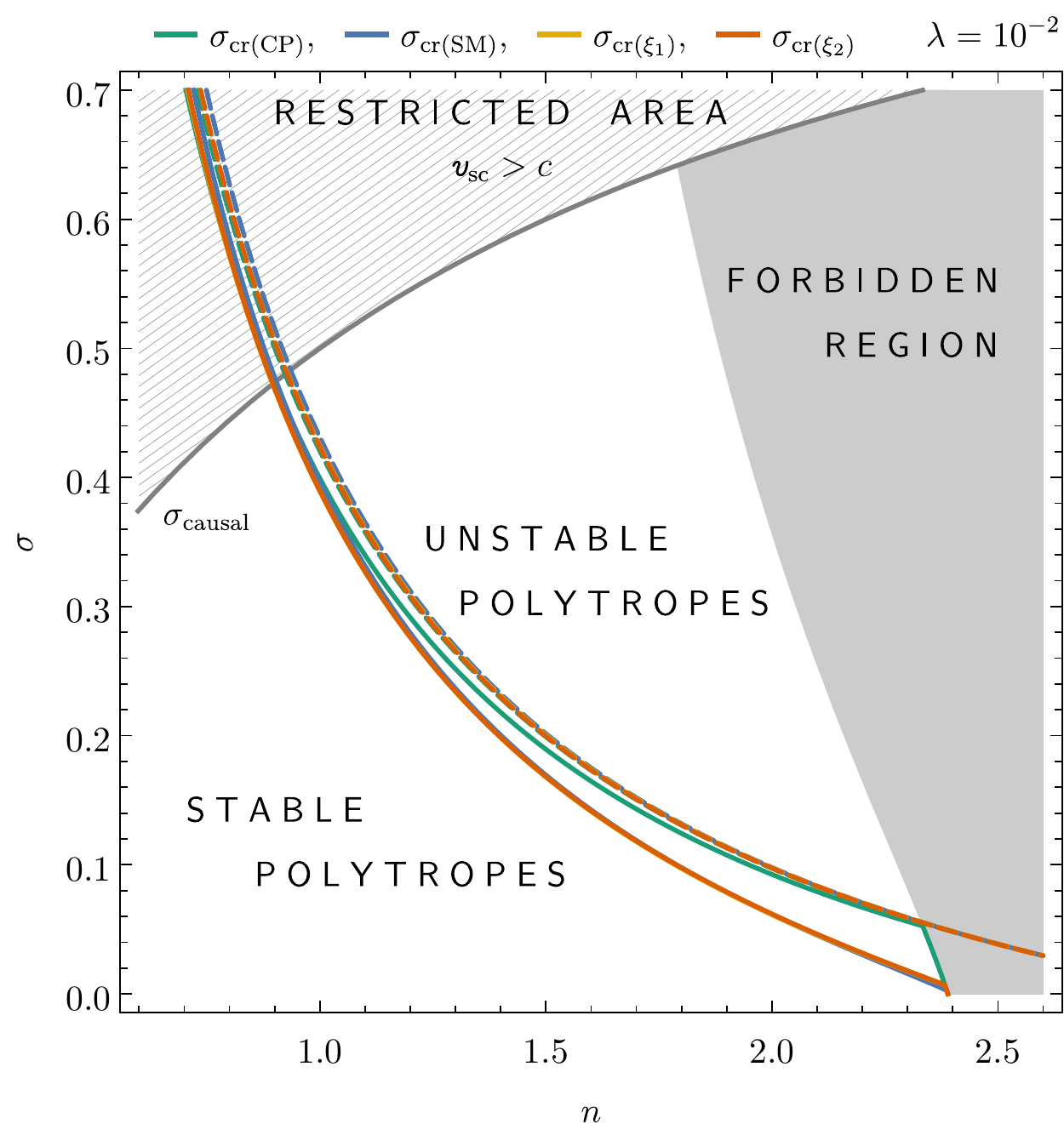}\\
    \centering \includegraphics[width=0.4\linewidth]{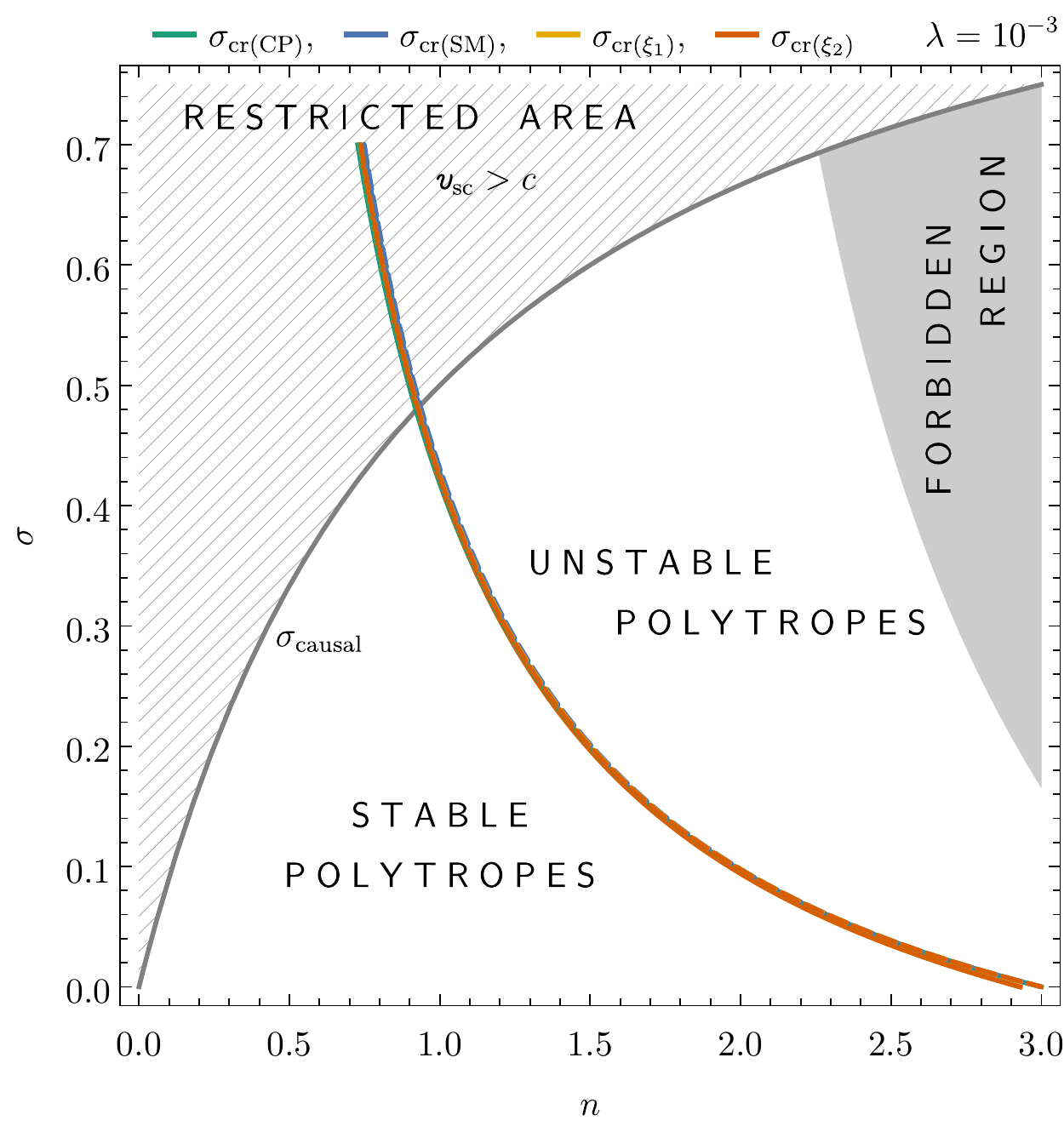} \quad \includegraphics[width=0.4\linewidth]{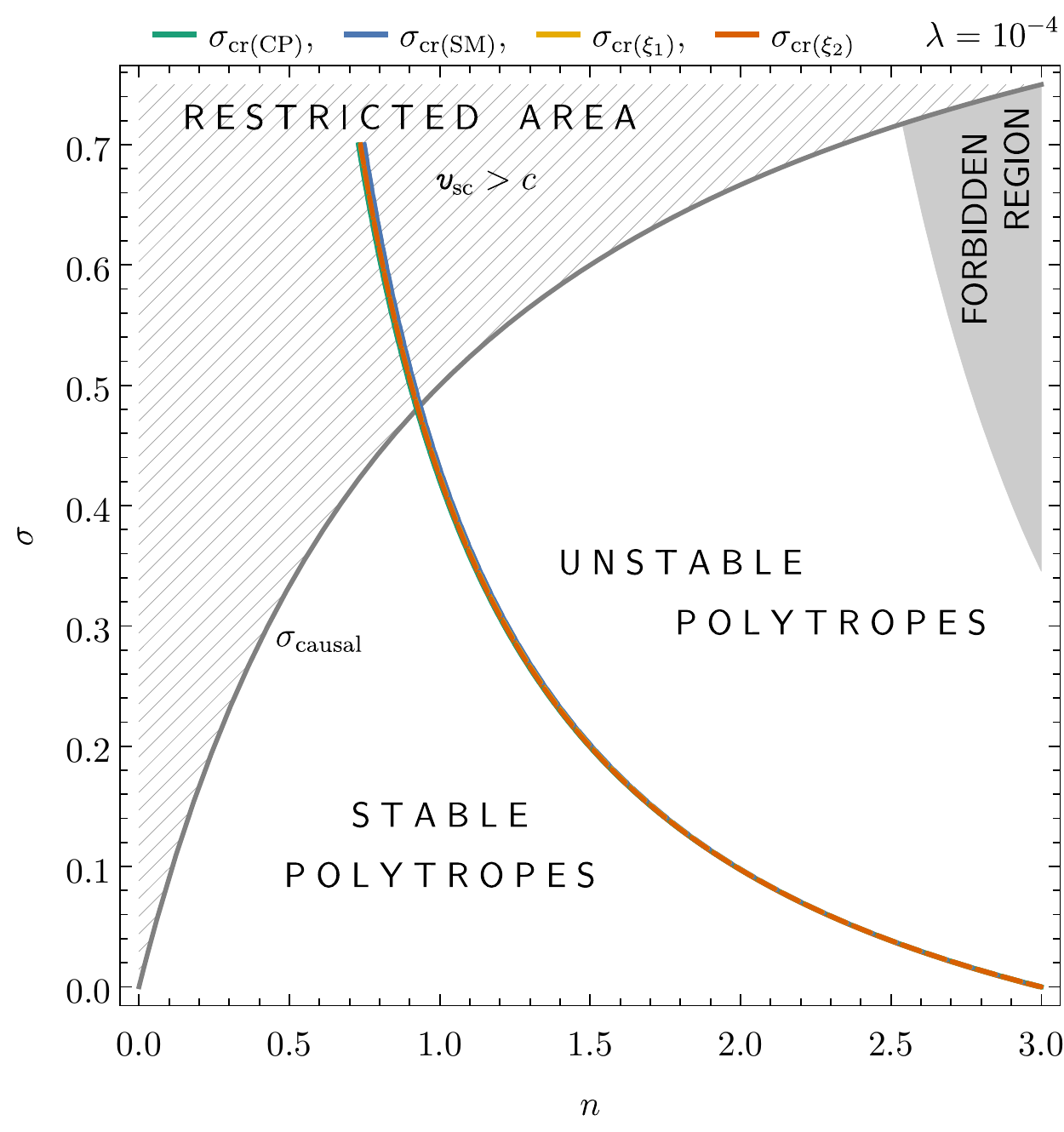}
    \caption{\label{fig6} Critical values of the relativity parameter $\sigma_{\mathrm{cr}}$, as a function of the polytropic index $n$, for the vacuum constant index $\lambda \in [10^{-4}, 10^{-1}]$. The forbidden region (gray background) corresponds to the region where the TOV equations do not give physically acceptable configurations for $\lambda\neq\,0$. Here we show the results obtained by using the CP method together with the results provided by Chandrasekhar's approach via the SM and the trial functions $\xi_1$ and $\xi_2$. The dashed lines (same color) indicate the corresponding values of $\sigma_\mathrm{cr}$ with $\lambda = 0$. Note that large values of $\lambda$, for instance, $\lambda = 10^{-2}$ and $\lambda = 10^{-1}$, lower the critical value $\sigma_\mathrm{cr}$ with respect to the corresponding value with $\lambda = 0$. Moreover, for these same values of $\lambda$, the critical point method and Chandrasekhar's dynamical approach predict different values of $\sigma_\mathrm{cr}$. For values of $\lambda < 10^{-4}$, its influence on the radial stability is practically negligible.}
\end{figure*}

The main result of our analysis is displayed in Fig.~\ref{fig6} where we determine the stable and unstable regions in the $n\mbox{--}\sigma$ parameter space, for several values of the vacuum constant index $\lambda\in[10^{-4}, 10^{-1}]$. In the plot, we present the results for the critical values of the parameter $\sigma_\mathrm{cr}$ as obtained from the critical point (CP) method, and those obtained from Chandrasekhar's method which were computed numerically using the shooting method (SM) (see Fig.~\ref{fig4}) and the trial functions [see Eq.~\eqref{Chandratrial}]. For comparison, we have also included the $\sigma_\mathrm{cr}$ values for the corresponding configurations with $\lambda=0$.

\begin{figure*}[h]
    \centering \includegraphics[width=0.4\linewidth]{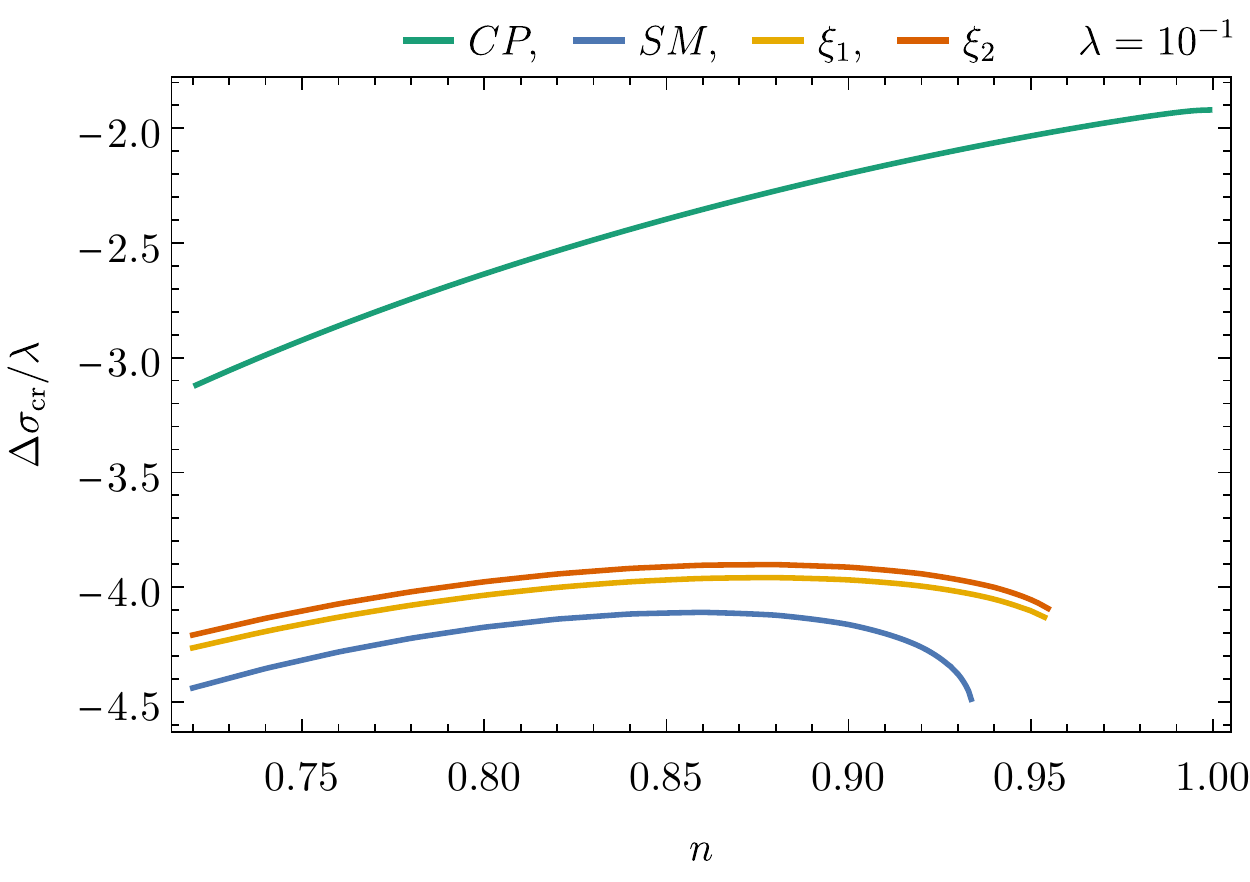} \quad \includegraphics[width=0.4\linewidth]{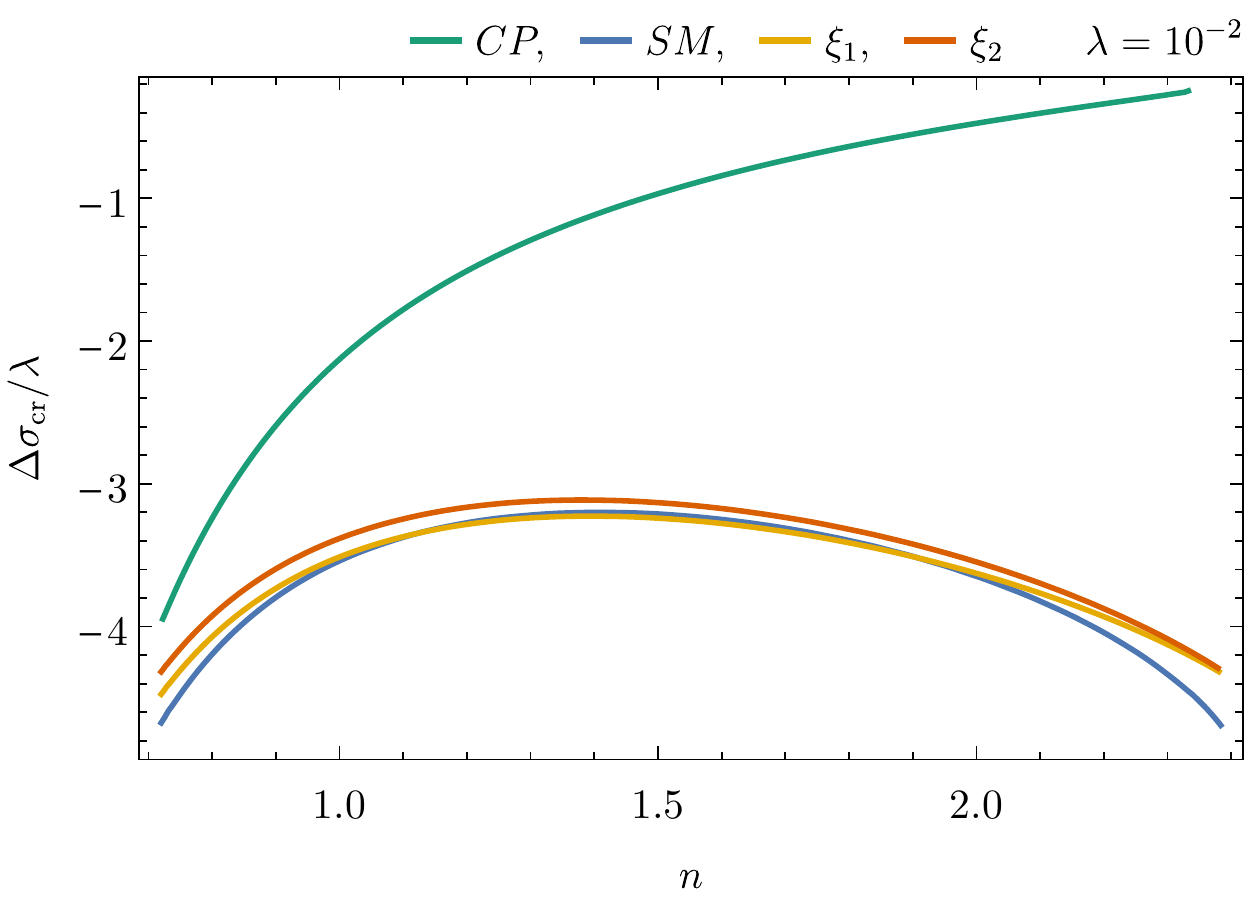}\\
    \centering \includegraphics[width=0.4\linewidth]{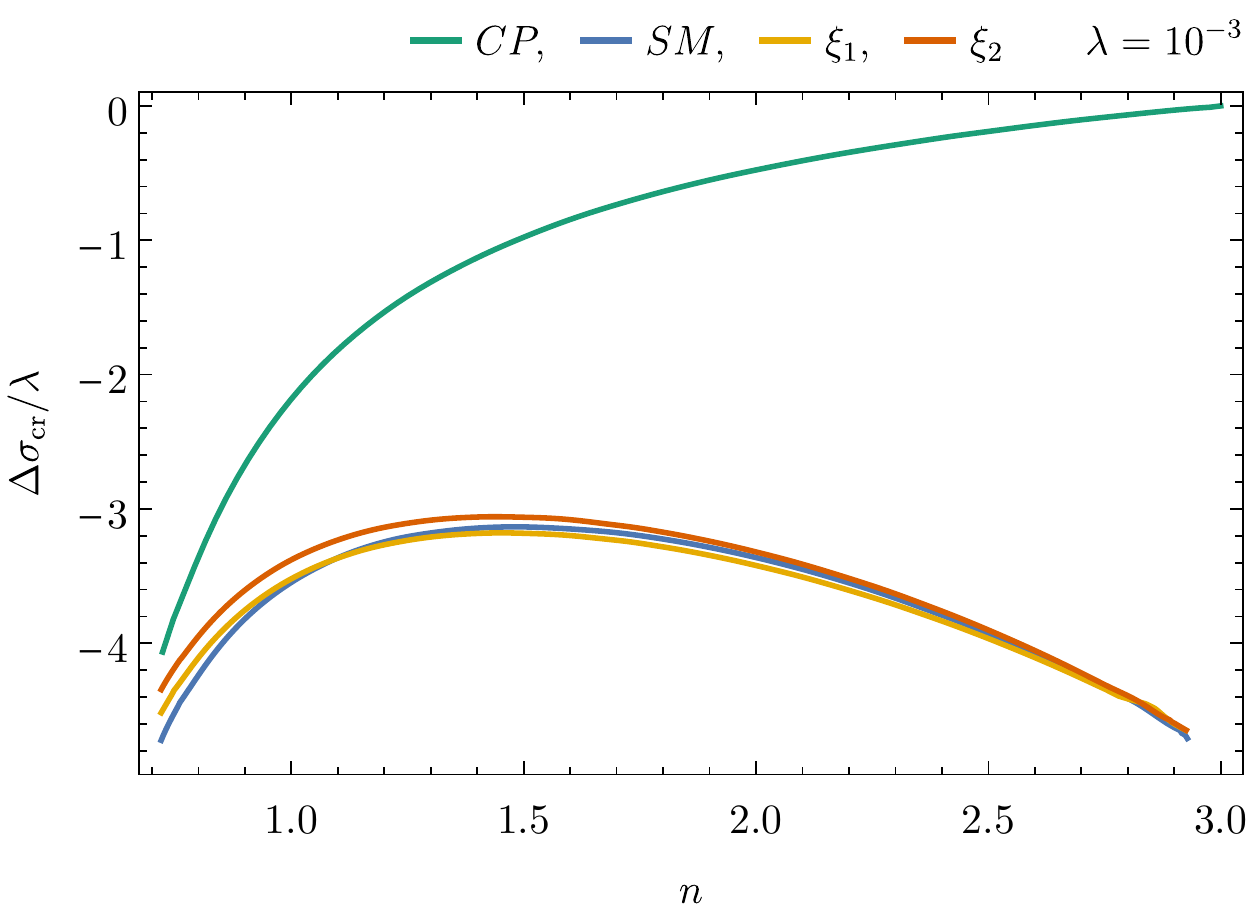} \quad \includegraphics[width=0.4\linewidth]{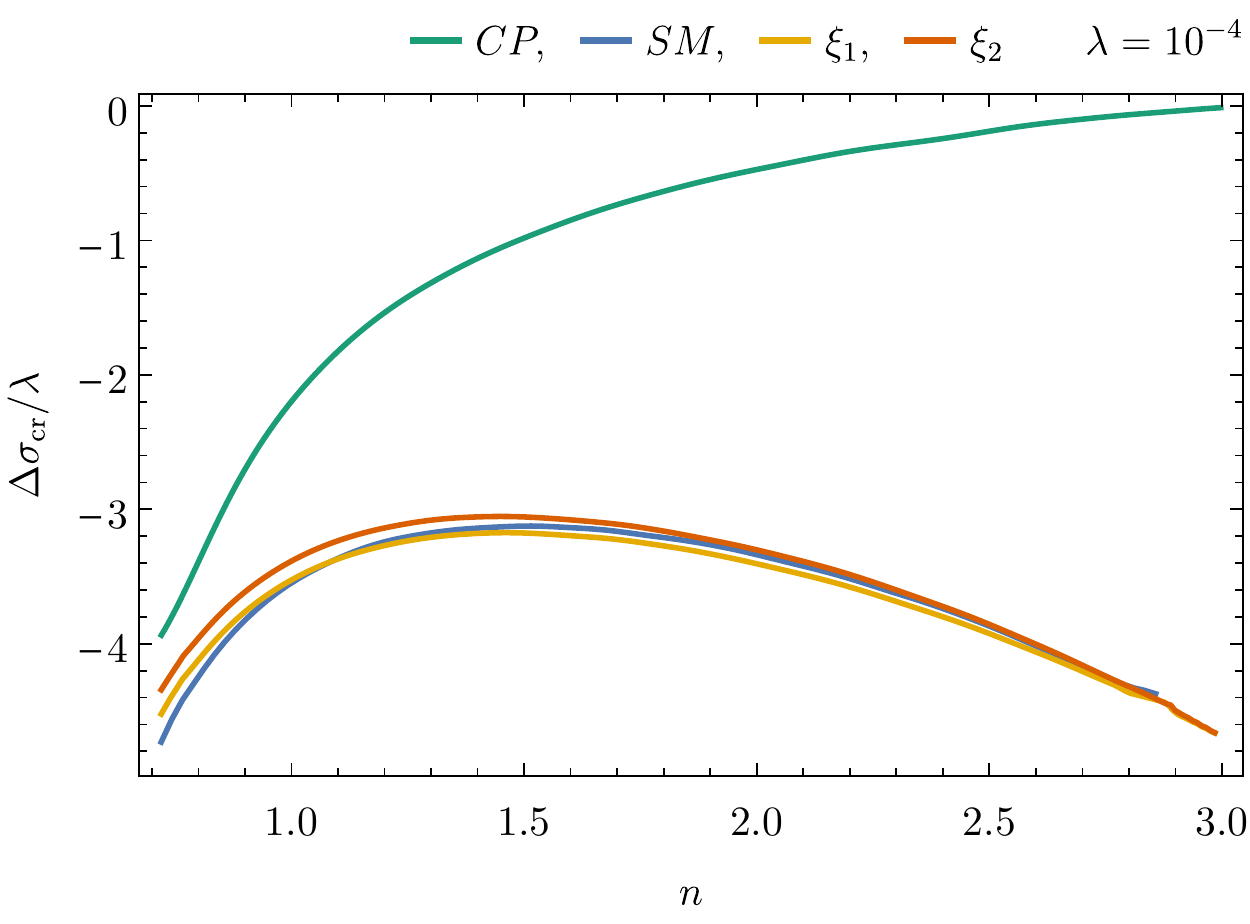}
    \caption{\label{fig7} Differences of the critical parameter $\sigma_{\mathrm{cr}}$, as a function of $n$, between the values for $\lambda \in [10^{-4},10^{-1}]$ and their corresponding values for $\lambda = 0$. Note that as $\lambda$ increases the differences between the values predicted by the CP method and Chandrasekhar's approach also increase.}
\end{figure*}

A first thing to notice is that for large values of the index $\lambda$, in particular $\lambda=10^{-2}$ and $\lambda=10^{-1}$, the values of $\sigma_\mathrm{cr}$ \emph{decrease} relative to the case with vanishing $\lambda$. We show the corresponding differences, as defined in Eq.~\eqref{difference}, in Fig.~\ref{fig7} for $\lambda \in [10^{-4},10^{-1}]$. Note that the differences are proportional, in order of magnitude, to the corresponding value of the index $\lambda$. These results are connected with those in Fig.~\ref{fig2} and the fact that large values of $\lambda$ tend to destabilize the polytropic spheres.

\begin{figure*}[h]
    \centering \includegraphics[width=0.4\linewidth]{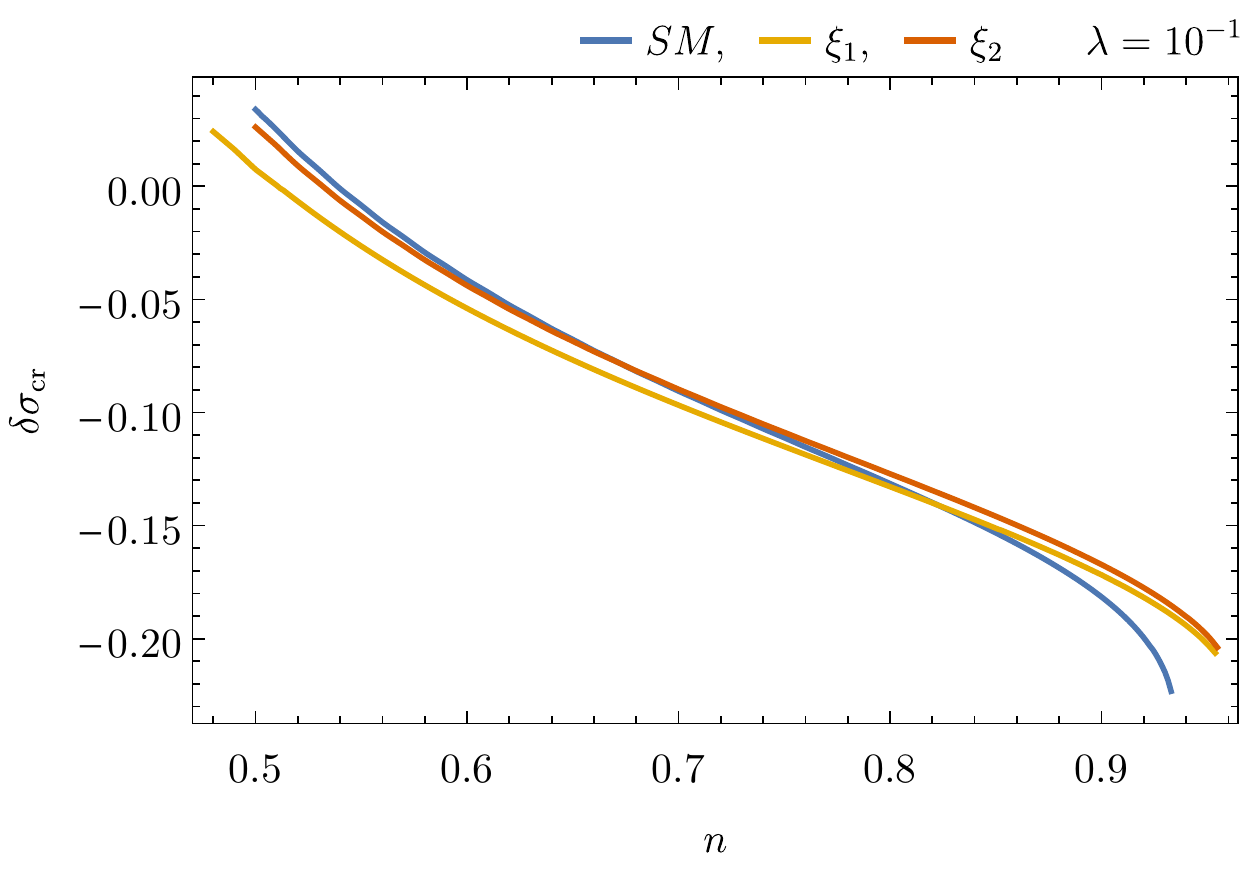} \quad \includegraphics[width=0.4\linewidth]{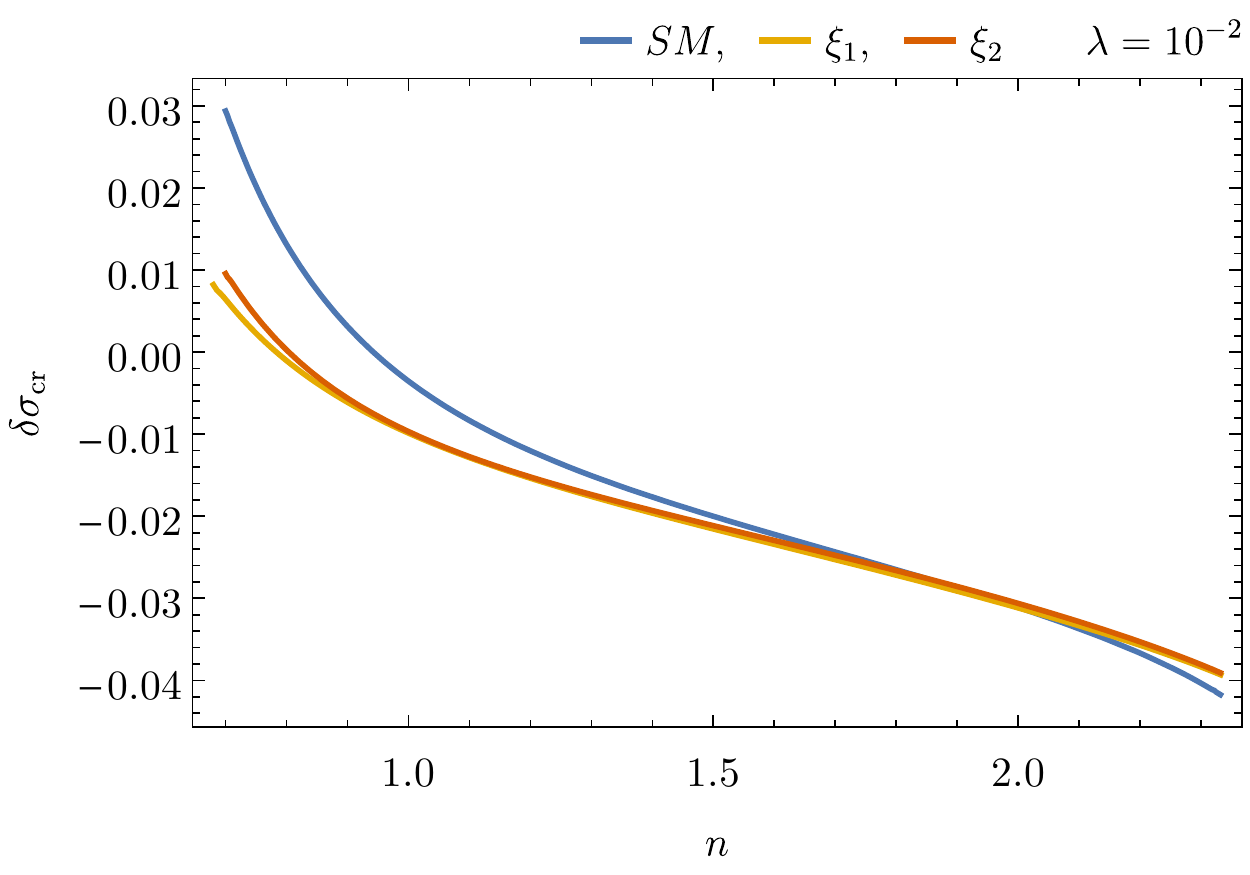}\\
    \centering \includegraphics[width=0.4\linewidth]{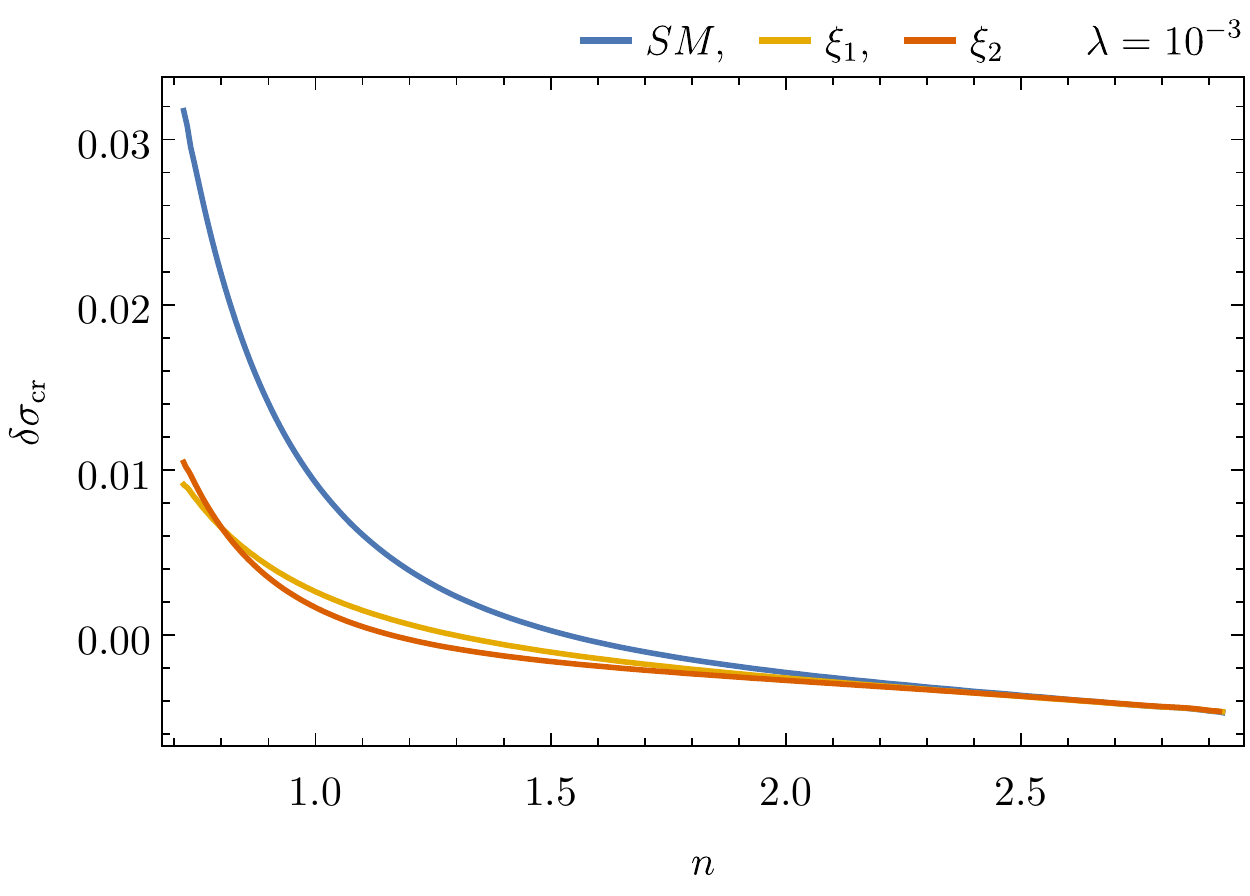} \quad \includegraphics[width=0.4\linewidth]{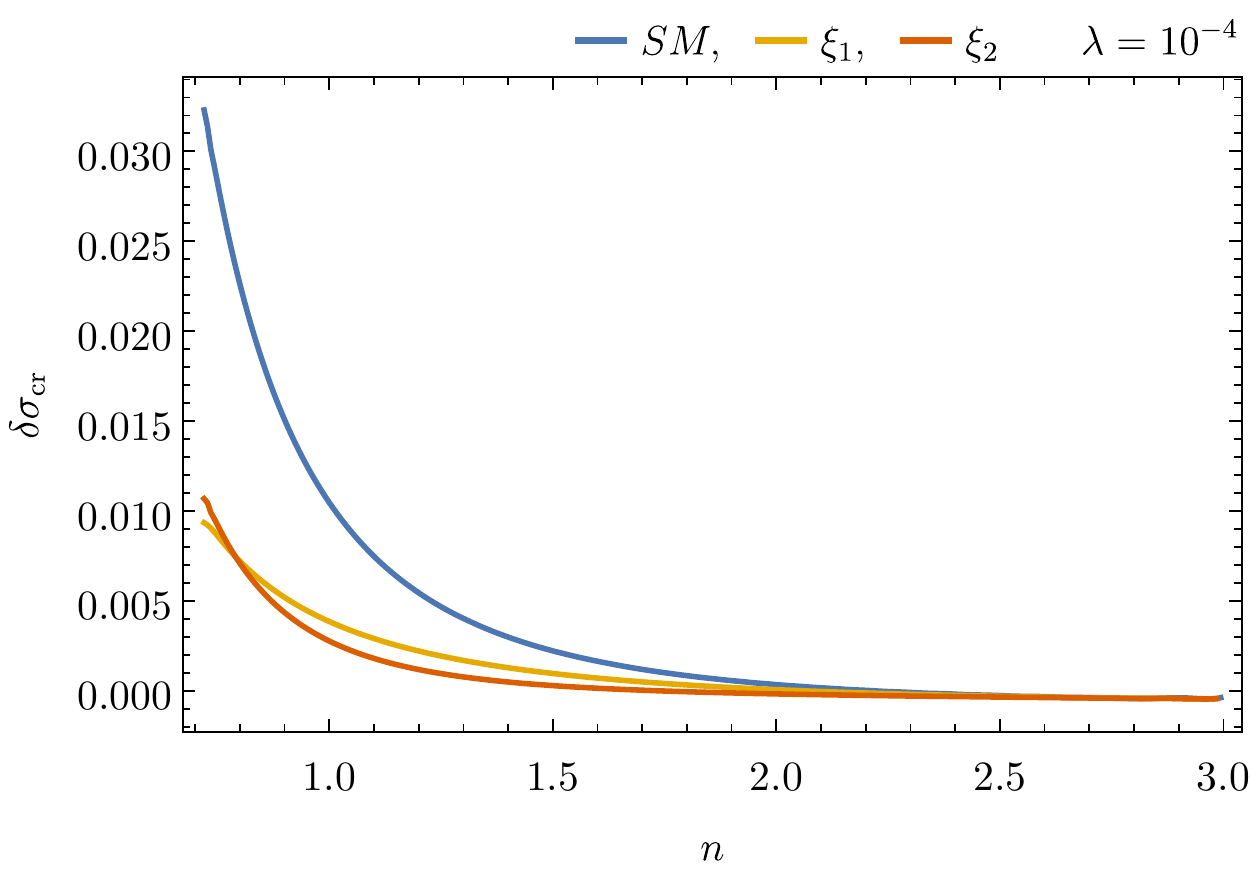}
\caption{\label{fig8} Differences $\delta\sigma_{\mathrm{cr}}$ of the critical parameter $\sigma_{\mathrm{cr}}$, as a function of the family of parameters $(n,\lambda)$, as determined via the CP method ($\sigma_{\mathrm{cr}}^{\mathrm{CP}}$) and Chandrasekhar's radial oscillations approach ($\sigma_{\mathrm{cr}}^{\mathrm{Ch}}$) via the SM and the trial functions $\xi_{1}$ and $\xi_{2}$. Note that for large values of $\lambda$ the differences grow as $n$ increases. For lower values of $\lambda$, the bigger differences are found in the low $n$ regime and tend to zero as $n \to 3$.}
\end{figure*}

Remarkably, we found that for large values of $\lambda$ the values of $\sigma_\mathrm{cr}$ obtained by using the critical point method \emph{differ} from those determined via the Chandrasekhar's dynamical approach. In Fig.~\ref{fig8}, we show the corresponding differences between both methods, as a function of $n$, for $\lambda \in [10^{-4},10^{-1}]$.  Note that for the cases $\lambda = 10^{-2}$ and $\lambda = 10^{-1}$, the differences increase with $n$. On the other hand, for lower values of $\lambda$, for instance, $\lambda = 10^{-4}$, the bigger differences are found in the regime of small $n$ and tend to zero as $n\to 3$. Note that these results are closely similar to those depicted in our preceding paper~\cite{Hladik:2020xfw}[Fig.~8] for the case $\lambda = 0$.

\section{Discussion}\label{sect:6}
In this paper we have investigated the role of the cosmological constant $\Lambda$ in the dynamical stability of relativistic polytropes by using two different approaches, namely, the energetic or critical point method and the infinitesimal radial oscillations method. Using Chandrasekhar's pulsation equation, we found that large values of $\lambda$ rise the critical adiabatic index $\gamma_{\mathrm{cr}}$ relative to their corresponding values for $\lambda = 0$. Thus, the cosmological constant tends to destabilize the polytropes.

Our results clearly show that the critical point method and the theory of radial oscillations predict different values of the critical parameter $\sigma_{\mathrm{cr}}$, for nonzero $\lambda$. The nature of this discrepancy might be attained to the different physical approach adopted in each method. Energy considerations are based on static solutions to Einstein's equations. Meanwhile Chandrasekhar's method considers a linearized analysis of time-dependent  perturbations on the given equilibrium configuration. Our results show that large values of the cosmological parameter $\lambda$ enhance this difference.

Finally, we would like to remark that the role of the vacuum energy on the radial stability of polytropic spheres becomes relevant for the parameter $\lambda$ sufficiently large---it is negligible for $\lambda$ smaller than $10^{-4}$ and becomes significant for $\lambda$ comparable to $10^{-1}$.


\begin{acknowledgements}
The authors acknowledge the support of the Institute of Physics and its Research Centre for Theoretical Physics and Astrophysics at the Silesian University in Opava.
\end{acknowledgements}


\bibliographystyle{apsrev4-1.bst}
\bibliography{main}

\begin{thebibliography}{45}%
\makeatletter
\providecommand \@ifxundefined [1]{%
 \@ifx{#1\undefined}
}%
\providecommand \@ifnum [1]{%
 \ifnum #1\expandafter \@firstoftwo
 \else \expandafter \@secondoftwo
 \fi
}%
\providecommand \@ifx [1]{%
 \ifx #1\expandafter \@firstoftwo
 \else \expandafter \@secondoftwo
 \fi
}%
\providecommand \natexlab [1]{#1}%
\providecommand \enquote  [1]{``#1''}%
\providecommand \bibnamefont  [1]{#1}%
\providecommand \bibfnamefont [1]{#1}%
\providecommand \citenamefont [1]{#1}%
\providecommand \href@noop [0]{\@secondoftwo}%
\providecommand \href [0]{\begingroup \@sanitize@url \@href}%
\providecommand \@href[1]{\@@startlink{#1}\@@href}%
\providecommand \@@href[1]{\endgroup#1\@@endlink}%
\providecommand \@sanitize@url [0]{\catcode `\\12\catcode `\$12\catcode
  `\&12\catcode `\#12\catcode `\^12\catcode `\_12\catcode `\%12\relax}%
\providecommand \@@startlink[1]{}%
\providecommand \@@endlink[0]{}%
\providecommand \url  [0]{\begingroup\@sanitize@url \@url }%
\providecommand \@url [1]{\endgroup\@href {#1}{\urlprefix }}%
\providecommand \urlprefix  [0]{URL }%
\providecommand \Eprint [0]{\href }%
\providecommand \doibase [0]{http://dx.doi.org/}%
\providecommand \selectlanguage [0]{\@gobble}%
\providecommand \bibinfo  [0]{\@secondoftwo}%
\providecommand \bibfield  [0]{\@secondoftwo}%
\providecommand \translation [1]{[#1]}%
\providecommand \BibitemOpen [0]{}%
\providecommand \bibitemStop [0]{}%
\providecommand \bibitemNoStop [0]{.\EOS\space}%
\providecommand \EOS [0]{\spacefactor3000\relax}%
\providecommand \BibitemShut  [1]{\csname bibitem#1\endcsname}%
\let\auto@bib@innerbib\@empty
\bibitem [{\citenamefont {Stuchl{\'\i}k}(2005)}]{Stuchlik:2005dv}%
  \BibitemOpen
  \bibfield  {author} {\bibinfo {author} {\bibfnamefont {Z.}~\bibnamefont
  {Stuchl{\'\i}k}},\ }\href {\doibase 10.1142/S0217732305016865} {\bibfield
  {journal} {\bibinfo  {journal} {Mod. Phys. Lett.}\ }\textbf {\bibinfo
  {volume} {A20}},\ \bibinfo {pages} {561} (\bibinfo {year} {2005})},\ \Eprint
  {http://arxiv.org/abs/0804.2266} {arXiv:0804.2266 [astro-ph]} \BibitemShut
  {NoStop}%
\bibitem [{\citenamefont {Stuchl{\'\i}k}\ \emph {et~al.}(2020)\citenamefont
  {Stuchl{\'\i}k}, \citenamefont {Kolo\v{s}}, \citenamefont {Kov\'{a}\v{r}},
  \citenamefont {Slan{\'y}},\ and\ \citenamefont
  {Tursunov}}]{Stuchlik:2020rls}%
  \BibitemOpen
  \bibfield  {author} {\bibinfo {author} {\bibfnamefont {Z.}~\bibnamefont
  {Stuchl{\'\i}k}}, \bibinfo {author} {\bibfnamefont {M.}~\bibnamefont
  {Kolo\v{s}}}, \bibinfo {author} {\bibfnamefont {J.}~\bibnamefont
  {Kov\'{a}\v{r}}}, \bibinfo {author} {\bibfnamefont {P.}~\bibnamefont
  {Slan{\'y}}}, \ and\ \bibinfo {author} {\bibfnamefont {A.}~\bibnamefont
  {Tursunov}},\ }\href {\doibase 10.3390/universe6020026} {\bibfield  {journal}
  {\bibinfo  {journal} {Universe}\ }\textbf {\bibinfo {volume} {6}},\ \bibinfo
  {pages} {26} (\bibinfo {year} {2020})}\BibitemShut {NoStop}%
\bibitem [{\citenamefont {Ade}\ \emph {et~al.}(2016)\citenamefont {Ade} \emph
  {et~al.}}]{Ade:2015xua}%
  \BibitemOpen
  \bibfield  {author} {\bibinfo {author} {\bibfnamefont {P.~A.~R.}\
  \bibnamefont {Ade}} \emph {et~al.} (\bibinfo {collaboration} {Planck}),\
  }\href {\doibase 10.1051/0004-6361/201525830} {\bibfield  {journal} {\bibinfo
   {journal} {Astron. Astrophys.}\ }\textbf {\bibinfo {volume} {594}},\
  \bibinfo {pages} {A13} (\bibinfo {year} {2016})},\ \Eprint
  {http://arxiv.org/abs/1502.01589} {arXiv:1502.01589 [astro-ph.CO]}
  \BibitemShut {NoStop}%
\bibitem [{\citenamefont {{Stuchl{\'\i}k}}(1983)}]{Stuchlik:1983BAC}%
  \BibitemOpen
  \bibfield  {author} {\bibinfo {author} {\bibfnamefont {Z.}~\bibnamefont
  {{Stuchl{\'\i}k}}},\ }\href@noop {} {\bibfield  {journal} {\bibinfo
  {journal} {Bull. Astron. Inst. Czechoslov.}\ }\textbf {\bibinfo {volume}
  {34}},\ \bibinfo {pages} {129} (\bibinfo {year} {1983})}\BibitemShut
  {NoStop}%
\bibitem [{\citenamefont {{Stuchl{\'\i}k}}(1984)}]{Stuchlik:1984BAC}%
  \BibitemOpen
  \bibfield  {author} {\bibinfo {author} {\bibfnamefont {Z.}~\bibnamefont
  {{Stuchl{\'\i}k}}},\ }\href@noop {} {\bibfield  {journal} {\bibinfo
  {journal} {Bull. Astron. Inst. Czechoslov.}\ }\textbf {\bibinfo {volume}
  {35}},\ \bibinfo {pages} {205} (\bibinfo {year} {1984})}\BibitemShut
  {NoStop}%
\bibitem [{\citenamefont {{Stuchl\'{\i}k}}\ and\ \citenamefont
  {{Hled\'{\i}k}}(1999)}]{Stuchlik:1999qk}%
  \BibitemOpen
  \bibfield  {author} {\bibinfo {author} {\bibfnamefont {Z.}~\bibnamefont
  {{Stuchl\'{\i}k}}}\ and\ \bibinfo {author} {\bibfnamefont {S.}~\bibnamefont
  {{Hled\'{\i}k}}},\ }\href@noop {} {\bibfield  {journal} {\bibinfo  {journal}
  {Phys. Rev.}\ }\textbf {\bibinfo {volume} {D60}},\ \bibinfo {pages} {044006}
  (\bibinfo {year} {1999})}\BibitemShut {NoStop}%
\bibitem [{\citenamefont {Stuchl\'{\i}k}\ and\ \citenamefont
  {Slan{\'y}}(2004)}]{Stuchlik:2004dt}%
  \BibitemOpen
  \bibfield  {author} {\bibinfo {author} {\bibfnamefont {Z.}~\bibnamefont
  {Stuchl\'{\i}k}}\ and\ \bibinfo {author} {\bibfnamefont {P.}~\bibnamefont
  {Slan{\'y}}},\ }\href {\doibase 10.1103/PhysRevD.69.064001} {\bibfield
  {journal} {\bibinfo  {journal} {Phys. Rev.}\ }\textbf {\bibinfo {volume}
  {D69}},\ \bibinfo {pages} {064001} (\bibinfo {year} {2004})},\ \Eprint
  {http://arxiv.org/abs/gr-qc/0307049} {arXiv:gr-qc/0307049 [gr-qc]}
  \BibitemShut {NoStop}%
\bibitem [{\citenamefont {Faraoni}(2016)}]{Far:2016:Uni}%
  \BibitemOpen
  \bibfield  {author} {\bibinfo {author} {\bibfnamefont {V.}~\bibnamefont
  {Faraoni}},\ }\href {\doibase 10.1016/j.dark.2015.11.001} {\bibfield
  {journal} {\bibinfo  {journal} {Phys. Dark Univ.}\ }\textbf {\bibinfo
  {volume} {11}},\ \bibinfo {pages} {11} (\bibinfo {year} {2016})},\ \Eprint
  {http://arxiv.org/abs/1508.00475} {arXiv:1508.00475 [gr-qc]} \BibitemShut
  {NoStop}%
\bibitem [{\citenamefont {Stuchl\'{\i}k}\ \emph {et~al.}(2018)\citenamefont
  {Stuchl\'{\i}k}, \citenamefont {Charbul\'{a}k},\ and\ \citenamefont
  {Schee}}]{Stuchlik:2018qyz}%
  \BibitemOpen
  \bibfield  {author} {\bibinfo {author} {\bibfnamefont {Z.}~\bibnamefont
  {Stuchl\'{\i}k}}, \bibinfo {author} {\bibfnamefont {D.}~\bibnamefont
  {Charbul\'{a}k}}, \ and\ \bibinfo {author} {\bibfnamefont {J.}~\bibnamefont
  {Schee}},\ }\href {\doibase 10.1140/epjc/s10052-018-5578-6} {\bibfield
  {journal} {\bibinfo  {journal} {Eur. Phys. J.}\ }\textbf {\bibinfo {volume}
  {C78}},\ \bibinfo {pages} {180} (\bibinfo {year} {2018})},\ \Eprint
  {http://arxiv.org/abs/1811.00072} {arXiv:1811.00072 [gr-qc]} \BibitemShut
  {NoStop}%
\bibitem [{\citenamefont {Stuchl{\'\i}k}\ \emph {et~al.}(2005)\citenamefont
  {Stuchl{\'\i}k}, \citenamefont {Slan{\'y}}, \citenamefont {{T\"{o}r\"{o}k}},\
  and\ \citenamefont {Abramowicz}}]{Stuchlik:2004wk}%
  \BibitemOpen
  \bibfield  {author} {\bibinfo {author} {\bibfnamefont {Z.}~\bibnamefont
  {Stuchl{\'\i}k}}, \bibinfo {author} {\bibfnamefont {P.}~\bibnamefont
  {Slan{\'y}}}, \bibinfo {author} {\bibfnamefont {G.}~\bibnamefont
  {{T\"{o}r\"{o}k}}}, \ and\ \bibinfo {author} {\bibfnamefont {M.~A.}\
  \bibnamefont {Abramowicz}},\ }\href {\doibase 10.1103/PhysRevD.71.024037}
  {\bibfield  {journal} {\bibinfo  {journal} {Phys. Rev.}\ }\textbf {\bibinfo
  {volume} {D71}},\ \bibinfo {pages} {024037} (\bibinfo {year} {2005})},\
  \Eprint {http://arxiv.org/abs/gr-qc/0411091} {arXiv:gr-qc/0411091 [gr-qc]}
  \BibitemShut {NoStop}%
\bibitem [{\citenamefont {{Stuchl{\'\i}k}}\ \emph {et~al.}(2000)\citenamefont
  {{Stuchl{\'\i}k}}, \citenamefont {{Slan{\'y}}},\ and\ \citenamefont
  {{Hled{\'\i}k}}}]{Stuchlik:2000AA}%
  \BibitemOpen
  \bibfield  {author} {\bibinfo {author} {\bibfnamefont {Z.}~\bibnamefont
  {{Stuchl{\'\i}k}}}, \bibinfo {author} {\bibfnamefont {P.}~\bibnamefont
  {{Slan{\'y}}}}, \ and\ \bibinfo {author} {\bibfnamefont {S.}~\bibnamefont
  {{Hled{\'\i}k}}},\ }\href@noop {} {\bibfield  {journal} {\bibinfo  {journal}
  {Astron. Astrophys.}\ }\textbf {\bibinfo {volume} {363}},\ \bibinfo {pages}
  {425} (\bibinfo {year} {2000})}\BibitemShut {NoStop}%
\bibitem [{\citenamefont {Slan{\'y}}\ and\ \citenamefont
  {Stuchl{\'\i}k}(2005)}]{Slany:2005vd}%
  \BibitemOpen
  \bibfield  {author} {\bibinfo {author} {\bibfnamefont {P.}~\bibnamefont
  {Slan{\'y}}}\ and\ \bibinfo {author} {\bibfnamefont {Z.}~\bibnamefont
  {Stuchl{\'\i}k}},\ }\href {\doibase 10.1088/0264-9381/22/17/019} {\bibfield
  {journal} {\bibinfo  {journal} {Class. Quant. Grav.}\ }\textbf {\bibinfo
  {volume} {22}},\ \bibinfo {pages} {3623} (\bibinfo {year}
  {2005})}\BibitemShut {NoStop}%
\bibitem [{\citenamefont {Stuchl{\'\i}k}\ and\ \citenamefont
  {Kov\'{a}\v{r}}(2008)}]{Stuchlik:2008ea}%
  \BibitemOpen
  \bibfield  {author} {\bibinfo {author} {\bibfnamefont {Z.}~\bibnamefont
  {Stuchl{\'\i}k}}\ and\ \bibinfo {author} {\bibfnamefont {J.}~\bibnamefont
  {Kov\'{a}\v{r}}},\ }\href {\doibase 10.1142/S021827180801373X} {\bibfield
  {journal} {\bibinfo  {journal} {Int. J. Mod. Phys.}\ }\textbf {\bibinfo
  {volume} {D17}},\ \bibinfo {pages} {2089} (\bibinfo {year} {2008})},\ \Eprint
  {http://arxiv.org/abs/0803.3641} {arXiv:0803.3641 [gr-qc]} \BibitemShut
  {NoStop}%
\bibitem [{\citenamefont {Stuchl{\'\i}k}\ \emph {et~al.}(2009)\citenamefont
  {Stuchl{\'\i}k}, \citenamefont {Slan{\'y}},\ and\ \citenamefont
  {Kovář}}]{Stuchlik:2009jv}%
  \BibitemOpen
  \bibfield  {author} {\bibinfo {author} {\bibfnamefont {Z.}~\bibnamefont
  {Stuchl{\'\i}k}}, \bibinfo {author} {\bibfnamefont {P.}~\bibnamefont
  {Slan{\'y}}}, \ and\ \bibinfo {author} {\bibfnamefont {J.}~\bibnamefont
  {Kovář}},\ }\href {\doibase 10.1088/0264-9381/26/21/215013} {\bibfield
  {journal} {\bibinfo  {journal} {Class. Quant. Grav.}\ }\textbf {\bibinfo
  {volume} {26}},\ \bibinfo {pages} {215013} (\bibinfo {year} {2009})},\
  \Eprint {http://arxiv.org/abs/0910.3184} {arXiv:0910.3184 [gr-qc]}
  \BibitemShut {NoStop}%
\bibitem [{\citenamefont {Stuchl{\'\i}k}\ and\ \citenamefont
  {Schee}(2011)}]{Stuchlik:2011zz}%
  \BibitemOpen
  \bibfield  {author} {\bibinfo {author} {\bibfnamefont {Z.}~\bibnamefont
  {Stuchl{\'\i}k}}\ and\ \bibinfo {author} {\bibfnamefont {J.}~\bibnamefont
  {Schee}},\ }\href {\doibase 10.1088/1475-7516/2011/09/018} {\bibfield
  {journal} {\bibinfo  {journal} {JCAP}\ }\textbf {\bibinfo {volume} {1109}},\
  \bibinfo {pages} {018} (\bibinfo {year} {2011})}\BibitemShut {NoStop}%
\bibitem [{\citenamefont {{Stuchl{\'\i}k}}\ and\ \citenamefont
  {{Schee}}(2012)}]{Stuchlik:2012}%
  \BibitemOpen
  \bibfield  {author} {\bibinfo {author} {\bibfnamefont {Z.}~\bibnamefont
  {{Stuchl{\'\i}k}}}\ and\ \bibinfo {author} {\bibfnamefont {J.}~\bibnamefont
  {{Schee}}},\ }\href {\doibase 10.1142/S0218271812500319} {\bibfield
  {journal} {\bibinfo  {journal} {Int. J. .Mod. Phys.}\ }\textbf {\bibinfo
  {volume} {D21}},\ \bibinfo {eid} {1250031} (\bibinfo {year}
  {2012})}\BibitemShut {NoStop}%
\bibitem [{\citenamefont {Schee}\ \emph {et~al.}(2013)\citenamefont {Schee},
  \citenamefont {Stuchl{\'\i}k},\ and\ \citenamefont
  {Petrásek}}]{Schee:2013wqa}%
  \BibitemOpen
  \bibfield  {author} {\bibinfo {author} {\bibfnamefont {J.}~\bibnamefont
  {Schee}}, \bibinfo {author} {\bibfnamefont {Z.}~\bibnamefont
  {Stuchl{\'\i}k}}, \ and\ \bibinfo {author} {\bibfnamefont {M.}~\bibnamefont
  {Petrásek}},\ }\href {\doibase 10.1088/1475-7516/2013/12/026} {\bibfield
  {journal} {\bibinfo  {journal} {JCAP}\ }\textbf {\bibinfo {volume} {1312}},\
  \bibinfo {pages} {026} (\bibinfo {year} {2013})},\ \Eprint
  {http://arxiv.org/abs/1312.0817} {arXiv:1312.0817 [astro-ph.GA]} \BibitemShut
  {NoStop}%
\bibitem [{\citenamefont {Stuchl{\'\i}k}\ \emph {et~al.}(2016)\citenamefont
  {Stuchl{\'\i}k}, \citenamefont {Hled{\'\i}k},\ and\ \citenamefont
  {Novotn\'{y}}}]{Stuchlik:2016xiq}%
  \BibitemOpen
  \bibfield  {author} {\bibinfo {author} {\bibfnamefont {Z.}~\bibnamefont
  {Stuchl{\'\i}k}}, \bibinfo {author} {\bibfnamefont {S.}~\bibnamefont
  {Hled{\'\i}k}}, \ and\ \bibinfo {author} {\bibfnamefont {J.}~\bibnamefont
  {Novotn\'{y}}},\ }\href {\doibase 10.1103/PhysRevD.94.103513} {\bibfield
  {journal} {\bibinfo  {journal} {Phys. Rev.}\ }\textbf {\bibinfo {volume}
  {D94}},\ \bibinfo {pages} {103513} (\bibinfo {year} {2016})},\ \Eprint
  {http://arxiv.org/abs/1611.05327} {arXiv:1611.05327 [gr-qc]} \BibitemShut
  {NoStop}%
\bibitem [{\citenamefont {Novotn\'{y}}\ \emph {et~al.}(2017)\citenamefont
  {Novotn\'{y}}, \citenamefont {Hlad{\'\i}k},\ and\ \citenamefont
  {Stuchl{\'\i}k}}]{Novotny:2017cep}%
  \BibitemOpen
  \bibfield  {author} {\bibinfo {author} {\bibfnamefont {J.}~\bibnamefont
  {Novotn\'{y}}}, \bibinfo {author} {\bibfnamefont {J.}~\bibnamefont
  {Hlad{\'\i}k}}, \ and\ \bibinfo {author} {\bibfnamefont {Z.}~\bibnamefont
  {Stuchl{\'\i}k}},\ }\href {\doibase 10.1103/PhysRevD.95.043009} {\bibfield
  {journal} {\bibinfo  {journal} {Phys. Rev.}\ }\textbf {\bibinfo {volume}
  {D95}},\ \bibinfo {pages} {043009} (\bibinfo {year} {2017})},\ \Eprint
  {http://arxiv.org/abs/1703.04604} {arXiv:1703.04604 [gr-qc]} \BibitemShut
  {NoStop}%
\bibitem [{\citenamefont {Stuchl{\'\i}k}\ \emph {et~al.}(2017)\citenamefont
  {Stuchl{\'\i}k}, \citenamefont {Schee}, \citenamefont {Toshmatov},
  \citenamefont {Hlad{\'\i}k},\ and\ \citenamefont
  {Novotn\'{y}}}]{Stuchlik:2017qiz}%
  \BibitemOpen
  \bibfield  {author} {\bibinfo {author} {\bibfnamefont {Z.}~\bibnamefont
  {Stuchl{\'\i}k}}, \bibinfo {author} {\bibfnamefont {J.}~\bibnamefont
  {Schee}}, \bibinfo {author} {\bibfnamefont {B.}~\bibnamefont {Toshmatov}},
  \bibinfo {author} {\bibfnamefont {J.}~\bibnamefont {Hlad{\'\i}k}}, \ and\
  \bibinfo {author} {\bibfnamefont {J.}~\bibnamefont {Novotn\'{y}}},\ }\href
  {\doibase 10.1088/1475-7516/2017/06/056} {\bibfield  {journal} {\bibinfo
  {journal} {JCAP}\ }\textbf {\bibinfo {volume} {1706}},\ \bibinfo {pages}
  {056} (\bibinfo {year} {2017})},\ \Eprint {http://arxiv.org/abs/1704.07713}
  {arXiv:1704.07713 [gr-qc]} \BibitemShut {NoStop}%
\bibitem [{\citenamefont {Konoplya}\ and\ \citenamefont
  {Zhidenko}(2011)}]{Konoplya:2011qq}%
  \BibitemOpen
  \bibfield  {author} {\bibinfo {author} {\bibfnamefont {R.~A.}\ \bibnamefont
  {Konoplya}}\ and\ \bibinfo {author} {\bibfnamefont {A.}~\bibnamefont
  {Zhidenko}},\ }\href {\doibase 10.1103/RevModPhys.83.793} {\bibfield
  {journal} {\bibinfo  {journal} {Rev. Mod. Phys.}\ }\textbf {\bibinfo {volume}
  {83}},\ \bibinfo {pages} {793} (\bibinfo {year} {2011})},\ \Eprint
  {http://arxiv.org/abs/1102.4014} {arXiv:1102.4014 [gr-qc]} \BibitemShut
  {NoStop}%
\bibitem [{\citenamefont {Toshmatov}\ and\ \citenamefont
  {Stuchl{\'\i}k}(2017)}]{Toshmatov:2017qrq}%
  \BibitemOpen
  \bibfield  {author} {\bibinfo {author} {\bibfnamefont {B.}~\bibnamefont
  {Toshmatov}}\ and\ \bibinfo {author} {\bibfnamefont {Z.}~\bibnamefont
  {Stuchl{\'\i}k}},\ }\href {\doibase 10.1140/epjp/i2017-11596-3} {\bibfield
  {journal} {\bibinfo  {journal} {Eur. Phys. J. Plus}\ }\textbf {\bibinfo
  {volume} {132}},\ \bibinfo {pages} {324} (\bibinfo {year} {2017})},\ \Eprint
  {http://arxiv.org/abs/1707.07419} {arXiv:1707.07419 [gr-qc]} \BibitemShut
  {NoStop}%
\bibitem [{\citenamefont {{Stuchl\'{\i}k}}(2000)}]{Stuchlik:2000xe}%
  \BibitemOpen
  \bibfield  {author} {\bibinfo {author} {\bibfnamefont {Z.}~\bibnamefont
  {{Stuchl\'{\i}k}}},\ }\href@noop {} {\bibfield  {journal} {\bibinfo
  {journal} {Acta Phys. Slov.}\ }\textbf {\bibinfo {volume} {50}},\ \bibinfo
  {pages} {219} (\bibinfo {year} {2000})},\ \Eprint
  {http://arxiv.org/abs/0803.2530} {arXiv:0803.2530 [gr-qc]} \BibitemShut
  {NoStop}%
\bibitem [{\citenamefont {Boehmer}(2004{\natexlab{a}})}]{Boehmer:2003iv}%
  \BibitemOpen
  \bibfield  {author} {\bibinfo {author} {\bibfnamefont {C.~G.}\ \bibnamefont
  {Boehmer}},\ }\href {\doibase 10.1088/0264-9381/21/4/025} {\bibfield
  {journal} {\bibinfo  {journal} {Class. Quant. Grav.}\ }\textbf {\bibinfo
  {volume} {21}},\ \bibinfo {pages} {1119} (\bibinfo {year}
  {2004}{\natexlab{a}})},\ \Eprint {http://arxiv.org/abs/gr-qc/0310058}
  {arXiv:gr-qc/0310058 [gr-qc]} \BibitemShut {NoStop}%
\bibitem [{\citenamefont {Boehmer}(2004{\natexlab{b}})}]{Boehmer:2003uz}%
  \BibitemOpen
  \bibfield  {author} {\bibinfo {author} {\bibfnamefont {C.~G.}\ \bibnamefont
  {Boehmer}},\ }\href {\doibase 10.1023/B:GERG.0000018088.69051.3b} {\bibfield
  {journal} {\bibinfo  {journal} {Gen. Rel. Grav.}\ }\textbf {\bibinfo {volume}
  {36}},\ \bibinfo {pages} {1039} (\bibinfo {year} {2004}{\natexlab{b}})},\
  \Eprint {http://arxiv.org/abs/gr-qc/0312027} {arXiv:gr-qc/0312027 [gr-qc]}
  \BibitemShut {NoStop}%
\bibitem [{\citenamefont {Axenides}\ \emph {et~al.}(2013)\citenamefont
  {Axenides}, \citenamefont {Georgiou},\ and\ \citenamefont
  {Roupas}}]{Axenides:2013hrq}%
  \BibitemOpen
  \bibfield  {author} {\bibinfo {author} {\bibfnamefont {M.}~\bibnamefont
  {Axenides}}, \bibinfo {author} {\bibfnamefont {G.}~\bibnamefont {Georgiou}},
  \ and\ \bibinfo {author} {\bibfnamefont {Z.}~\bibnamefont {Roupas}},\ }\href
  {\doibase 10.1016/j.nuclphysb.2013.02.003} {\bibfield  {journal} {\bibinfo
  {journal} {Nucl. Phys. B}\ }\textbf {\bibinfo {volume} {871}},\ \bibinfo
  {pages} {21} (\bibinfo {year} {2013})},\ \Eprint
  {http://arxiv.org/abs/1302.1977} {arXiv:1302.1977 [astro-ph.CO]} \BibitemShut
  {NoStop}%
\bibitem [{\citenamefont {Posada}\ and\ \citenamefont
  {Batic}(2014)}]{Posada:2013eqa}%
  \BibitemOpen
  \bibfield  {author} {\bibinfo {author} {\bibfnamefont {C.}~\bibnamefont
  {Posada}}\ and\ \bibinfo {author} {\bibfnamefont {D.}~\bibnamefont {Batic}},\
  }\href {\doibase 10.2478/s11534-014-0458-7} {\bibfield  {journal} {\bibinfo
  {journal} {Central Eur. J. Phys.}\ }\textbf {\bibinfo {volume} {12}},\
  \bibinfo {pages} {297} (\bibinfo {year} {2014})},\ \Eprint
  {http://arxiv.org/abs/1310.7894} {arXiv:1310.7894 [gr-qc]} \BibitemShut
  {NoStop}%
\bibitem [{\citenamefont {{Shapiro}}\ and\ \citenamefont
  {{Teukolsky}}(1983)}]{Shapiro:1983du}%
  \BibitemOpen
  \bibfield  {author} {\bibinfo {author} {\bibfnamefont {S.~L.}\ \bibnamefont
  {{Shapiro}}}\ and\ \bibinfo {author} {\bibfnamefont {S.~A.}\ \bibnamefont
  {{Teukolsky}}},\ }\href@noop {} {\emph {\bibinfo {title} {{Black holes, white
  dwarfs, and neutron stars: The physics of compact objects}}}}\ (\bibinfo
  {publisher} {Wiley},\ \bibinfo {address} {New York},\ \bibinfo {year}
  {1983})\BibitemShut {NoStop}%
\bibitem [{\citenamefont {{Alvarez-Castillo}}\ and\ \citenamefont
  {{Blaschke}}(2017)}]{Alvarez-Castillo:2017qki}%
  \BibitemOpen
  \bibfield  {author} {\bibinfo {author} {\bibfnamefont {D.~E.}\ \bibnamefont
  {{Alvarez-Castillo}}}\ and\ \bibinfo {author} {\bibfnamefont {D.~B.}\
  \bibnamefont {{Blaschke}}},\ }\href@noop {} {\bibfield  {journal} {\bibinfo
  {journal} {Phys. Rev.}\ }\textbf {\bibinfo {volume} {C96}},\ \bibinfo {pages}
  {045809} (\bibinfo {year} {2017})},\ \Eprint
  {http://arxiv.org/abs/1703.02681} {arXiv:1703.02681 [nucl-th]} \BibitemShut
  {NoStop}%
\bibitem [{\citenamefont {{Zel'dovich}}\ and\ \citenamefont
  {{Novikov}}(1971)}]{Zeldovich:1971}%
  \BibitemOpen
  \bibfield  {author} {\bibinfo {author} {\bibfnamefont {Y.~B.}\ \bibnamefont
  {{Zel'dovich}}}\ and\ \bibinfo {author} {\bibfnamefont {I.~D.}\ \bibnamefont
  {{Novikov}}},\ }\href@noop {} {\emph {\bibinfo {title} {{Relativistic
  Astrophysics. Vol.1: Stars and Relativity}}}}\ (\bibinfo  {publisher}
  {University of Chicago Press},\ \bibinfo {address} {Chicago},\ \bibinfo
  {year} {1971})\BibitemShut {NoStop}%
\bibitem [{\citenamefont {{Tooper}}(1964)}]{Tooper:1964}%
  \BibitemOpen
  \bibfield  {author} {\bibinfo {author} {\bibfnamefont {R.~F.}\ \bibnamefont
  {{Tooper}}},\ }\href@noop {} {\bibfield  {journal} {\bibinfo  {journal}
  {Astrophys. J.}\ }\textbf {\bibinfo {volume} {140}},\ \bibinfo {pages} {434}
  (\bibinfo {year} {1964})}\BibitemShut {NoStop}%
\bibitem [{\citenamefont {Chandrasekhar}(1964)}]{Chandrasekhar:1964zz}%
  \BibitemOpen
  \bibfield  {author} {\bibinfo {author} {\bibfnamefont {S.}~\bibnamefont
  {Chandrasekhar}},\ }\href@noop {} {\bibfield  {journal} {\bibinfo  {journal}
  {Astrophys. J.}\ }\textbf {\bibinfo {volume} {140}},\ \bibinfo {pages} {417}
  (\bibinfo {year} {1964})},\ \bibinfo {note} {[Erratum: Astrophys.
  J.140,1342(1964)]}\BibitemShut {NoStop}%
\bibitem [{\citenamefont {Gleiser}\ and\ \citenamefont
  {Watkins}(1989)}]{Gleiser:1988ih}%
  \BibitemOpen
  \bibfield  {author} {\bibinfo {author} {\bibfnamefont {M.}~\bibnamefont
  {Gleiser}}\ and\ \bibinfo {author} {\bibfnamefont {R.}~\bibnamefont
  {Watkins}},\ }\href {\doibase 10.1016/0550-3213(89)90627-5} {\bibfield
  {journal} {\bibinfo  {journal} {Nucl.\ Phys.\ B}\ }\textbf {\bibinfo {volume}
  {319}},\ \bibinfo {pages} {733} (\bibinfo {year} {1989})}\BibitemShut
  {NoStop}%
\bibitem [{\citenamefont {Moustakidis}(2017)}]{Moustakidis:2016ndw}%
  \BibitemOpen
  \bibfield  {author} {\bibinfo {author} {\bibfnamefont {C.~C.}\ \bibnamefont
  {Moustakidis}},\ }\href {\doibase 10.1007/s10714-017-2232-9} {\bibfield
  {journal} {\bibinfo  {journal} {Gen.\ Rel.\ Grav.}\ }\textbf {\bibinfo
  {volume} {49}},\ \bibinfo {pages} {68} (\bibinfo {year} {2017})},\ \Eprint
  {http://arxiv.org/abs/1612.01726} {arXiv:1612.01726 [gr-qc]} \BibitemShut
  {NoStop}%
\bibitem [{\citenamefont {Posada}\ and\ \citenamefont
  {Chirenti}(2019)}]{Posada:2018goy}%
  \BibitemOpen
  \bibfield  {author} {\bibinfo {author} {\bibfnamefont {C.}~\bibnamefont
  {Posada}}\ and\ \bibinfo {author} {\bibfnamefont {C.}~\bibnamefont
  {Chirenti}},\ }\href {\doibase 10.1088/1361-6382/ab0526} {\bibfield
  {journal} {\bibinfo  {journal} {Class. Quant. Grav.}\ }\textbf {\bibinfo
  {volume} {36}},\ \bibinfo {pages} {065004} (\bibinfo {year} {2019})},\
  \Eprint {http://arxiv.org/abs/1811.09589} {arXiv:1811.09589 [gr-qc]}
  \BibitemShut {NoStop}%
\bibitem [{\citenamefont {{Stuchl{\'\i}k}}\ and\ \citenamefont
  {{Hled{\'\i}k}}(2005)}]{Stuchlik:2005rag}%
  \BibitemOpen
  \bibfield  {author} {\bibinfo {author} {\bibfnamefont {Z.}~\bibnamefont
  {{Stuchl{\'\i}k}}}\ and\ \bibinfo {author} {\bibfnamefont {S.}~\bibnamefont
  {{Hled{\'\i}k}}},\ }in\ \href@noop {} {\emph {\bibinfo {booktitle}
  {Proceedings of RAGtime 6/7: Workshops on black holes and neutron stars}}},\
  \bibinfo {editor} {edited by\ \bibinfo {editor} {\bibfnamefont
  {S.}~\bibnamefont {{Hled{\'\i}k}}}\ and\ \bibinfo {editor} {\bibfnamefont
  {Z.}~\bibnamefont {{Stuchl{\'\i}k}}}}\ (\bibinfo {address} {Silesian
  University in Opava},\ \bibinfo {year} {2005})\ p.\ \bibinfo {pages}
  {209}\BibitemShut {NoStop}%
\bibitem [{\citenamefont {Boehmer}\ and\ \citenamefont
  {Harko}(2005)}]{Boehmer:2005kk}%
  \BibitemOpen
  \bibfield  {author} {\bibinfo {author} {\bibfnamefont {C.~G.}\ \bibnamefont
  {Boehmer}}\ and\ \bibinfo {author} {\bibfnamefont {T.}~\bibnamefont
  {Harko}},\ }\href {\doibase 10.1103/PhysRevD.71.084026} {\bibfield  {journal}
  {\bibinfo  {journal} {Phys. Rev.}\ }\textbf {\bibinfo {volume} {D71}},\
  \bibinfo {pages} {084026} (\bibinfo {year} {2005})},\ \Eprint
  {http://arxiv.org/abs/gr-qc/0504075} {arXiv:gr-qc/0504075 [gr-qc]}
  \BibitemShut {NoStop}%
\bibitem [{\citenamefont {Hlad{\'\i}k}\ \emph {et~al.}(2020)\citenamefont
  {Hlad{\'\i}k}, \citenamefont {Posada},\ and\ \citenamefont
  {Stuchl{\'\i}k}}]{Hladik:2020xfw}%
  \BibitemOpen
  \bibfield  {author} {\bibinfo {author} {\bibfnamefont {J.}~\bibnamefont
  {Hlad{\'\i}k}}, \bibinfo {author} {\bibfnamefont {C.}~\bibnamefont {Posada}},
  \ and\ \bibinfo {author} {\bibfnamefont {Z.}~\bibnamefont {Stuchl{\'\i}k}},\
  }\href {\doibase 10.1142/S0218271820500303} {\bibfield  {journal} {\bibinfo
  {journal} {{Int. J. Mod. Phys.}}\ }\textbf {\bibinfo {volume} {{D29}}},\
  \bibinfo {pages} {2050030} (\bibinfo {year} {{2020}})},\ \Eprint
  {http://arxiv.org/abs/2001.05999} {arXiv:2001.05999 [gr-qc]} \BibitemShut
  {NoStop}%
\bibitem [{\citenamefont {Tolman}(1934)}]{Tolman:1934}%
  \BibitemOpen
  \bibfield  {author} {\bibinfo {author} {\bibfnamefont {R.~C.}\ \bibnamefont
  {Tolman}},\ }\href@noop {} {\emph {\bibinfo {title} {{Relativity,
  Thermodynamics and Cosmology}}}}\ (\bibinfo  {publisher} {Clarendon Press},\
  \bibinfo {address} {Oxford},\ \bibinfo {year} {1934})\BibitemShut {NoStop}%
\bibitem [{\citenamefont {{Kottler}}(1918)}]{Kottler:1918AnP}%
  \BibitemOpen
  \bibfield  {author} {\bibinfo {author} {\bibfnamefont {F.}~\bibnamefont
  {{Kottler}}},\ }\href {\doibase 10.1002/andp.19183611402} {\bibfield
  {journal} {\bibinfo  {journal} {Annalen der Physik}\ }\textbf {\bibinfo
  {volume} {361}},\ \bibinfo {pages} {401} (\bibinfo {year}
  {1918})}\BibitemShut {NoStop}%
\bibitem [{\citenamefont {{Misner}}\ \emph {et~al.}(1973)\citenamefont
  {{Misner}}, \citenamefont {{Thorne}},\ and\ \citenamefont
  {{Wheeler}}}]{Misner:1974qy}%
  \BibitemOpen
  \bibfield  {author} {\bibinfo {author} {\bibfnamefont {C.~W.}\ \bibnamefont
  {{Misner}}}, \bibinfo {author} {\bibfnamefont {K.~S.}\ \bibnamefont
  {{Thorne}}}, \ and\ \bibinfo {author} {\bibfnamefont {J.~A.}\ \bibnamefont
  {{Wheeler}}},\ }\href@noop {} {\emph {\bibinfo {title} {{Gravitation}}}}\
  (\bibinfo  {publisher} {Freeman},\ \bibinfo {address} {San Francisco},\
  \bibinfo {year} {1973})\BibitemShut {NoStop}%
\bibitem [{\citenamefont {{Merafina}}\ and\ \citenamefont
  {{Ruffini}}(1989)}]{Merafina:1989}%
  \BibitemOpen
  \bibfield  {author} {\bibinfo {author} {\bibfnamefont {M.}~\bibnamefont
  {{Merafina}}}\ and\ \bibinfo {author} {\bibfnamefont {R.}~\bibnamefont
  {{Ruffini}}},\ }\href@noop {} {\bibfield  {journal} {\bibinfo  {journal}
  {Astron. Astrophys.}\ }\textbf {\bibinfo {volume} {221}},\ \bibinfo {pages}
  {4} (\bibinfo {year} {1989})}\BibitemShut {NoStop}%
\bibitem [{\citenamefont {{Bardeen}}\ \emph {et~al.}(1966)\citenamefont
  {{Bardeen}}, \citenamefont {{Thorne}},\ and\ \citenamefont
  {{Meltzer}}}]{Bardeen:1966}%
  \BibitemOpen
  \bibfield  {author} {\bibinfo {author} {\bibfnamefont {J.~M.}\ \bibnamefont
  {{Bardeen}}}, \bibinfo {author} {\bibfnamefont {K.~S.}\ \bibnamefont
  {{Thorne}}}, \ and\ \bibinfo {author} {\bibfnamefont {D.~W.}\ \bibnamefont
  {{Meltzer}}},\ }\href@noop {} {\bibfield  {journal} {\bibinfo  {journal}
  {Astrophys. J.}\ }\textbf {\bibinfo {volume} {145}},\ \bibinfo {pages} {505}
  (\bibinfo {year} {1966})}\BibitemShut {NoStop}%
\bibitem [{\citenamefont {Press}\ \emph {et~al.}(1992)\citenamefont {Press},
  \citenamefont {Teukolsky}, \citenamefont {Vetterling},\ and\ \citenamefont
  {Flannery}}]{Press:1992zz}%
  \BibitemOpen
  \bibfield  {author} {\bibinfo {author} {\bibfnamefont {W.~H.}\ \bibnamefont
  {Press}}, \bibinfo {author} {\bibfnamefont {S.~A.}\ \bibnamefont
  {Teukolsky}}, \bibinfo {author} {\bibfnamefont {W.~T.}\ \bibnamefont
  {Vetterling}}, \ and\ \bibinfo {author} {\bibfnamefont {B.~P.}\ \bibnamefont
  {Flannery}},\ }\href@noop {} {\emph {\bibinfo {title} {{Numerical Recipes in
  C: The Art of Scientific Computing}}}},\ \bibinfo {edition} {2nd}\ ed.\
  (\bibinfo  {publisher} {Cambridge University Press},\ \bibinfo {address} {New
  York},\ \bibinfo {year} {1992})\BibitemShut {NoStop}%
\bibitem [{\citenamefont {{Kokkotas}}\ and\ \citenamefont
  {{Ruoff}}(2001)}]{Kokkotas:2000up}%
  \BibitemOpen
  \bibfield  {author} {\bibinfo {author} {\bibfnamefont {K.}~\bibnamefont
  {{Kokkotas}}}\ and\ \bibinfo {author} {\bibfnamefont {J.}~\bibnamefont
  {{Ruoff}}},\ }\href@noop {} {\bibfield  {journal} {\bibinfo  {journal}
  {Astron. Astrophys.}\ }\textbf {\bibinfo {volume} {366}},\ \bibinfo {pages}
  {565} (\bibinfo {year} {2001})},\ \Eprint
  {http://arxiv.org/abs/gr-qc/0011093} {arXiv:gr-qc/0011093 [gr-qc]}
  \BibitemShut {NoStop}%
\end{thebibliography}%

\end{document}